\documentclass[12pt, twoside]{book}

\usepackage[utf8]{inputenc}  
\usepackage[T1]{fontenc}  

\usepackage{lmodern}
\usepackage{microtype}
\usepackage{setspace}

\usepackage{ragged2e}
\usepackage{enumitem}
\usepackage[normalem]{ulem}

\usepackage{amsmath}
\usepackage{amssymb}
\usepackage{dsfont} 
\usepackage{mathtools}
\usepackage{bbold}
\usepackage{bm}

\usepackage[toc,page,title]{appendix}  
\usepackage{pdfpages}
\usepackage{changepage}
\usepackage{fancyhdr} 
\usepackage[toctitles,raggedright]{titlesec}
\usepackage{epigraph}  
\usepackage{chngcntr} 

\usepackage{float}  
\usepackage[section]{placeins}  
\usepackage[font=footnotesize]{caption}

\usepackage{standalone}  
\usepackage{import}  

\usepackage[disable,textsize=tiny]{todonotes}  
\usepackage[breakable]{tcolorbox}  
\usepackage[backend=bibtex,
style=phys, biblabel=brackets 
]{biblatex}
\newcommand{\refcolor}{blue}
\usepackage{tocloft}
\makeatletter
    \renewcommand{\@tocrmarg}{\@pnumwidth plus1fil} 
\makeatother
\usepackage[%
        colorlinks=true,urlcolor=\refcolor,linkcolor=\refcolor,citecolor=\refcolor]%
        {hyperref}  
 
\usepackage[outer=22mm,inner=25mm,bottom=30mm,b5paper]{geometry}

\assignpagestyle{\chapter}{empty}
\titleformat{\chapter}[display]
  {\raggedright\Huge\bfseries} 
  {\raggedleft\normalfont\itshape\Large{\chaptertitlename} \thechapter} 
  {0.2ex} 
  {\titlerule[0.4pt]\vspace{1.ex}} 
  [
  ] 

\newcommand{\kboltz}{k_\mathrm{B}}

\newcommand{\realnumbers}{\mathbb{R}}
\newcommand{\complexnumbers}{\mathbb{C}}
\renewcommand{\Re}{\mathrm{Re}}
\renewcommand{\Im}{\mathrm{Im}}
\newcommand{\transposesymb}{{\sf T}}
\newcommand{\transpose}[1]{#1^{\transposesymb}}
\newcommand{\commutator}[2]{[#1, #2]}
\newcommand{\dd}{\mathrm{d}}
\newcommand{\dv}[2][]{\frac{\dd #1}{\dd #2}}
\newcommand{\ddt}{\dv{t}}
\newcommand{\pdvD}[2][]{\frac{\partial #1}{\partial #2}}
\newcommand{\pdv}[2]{\partial_{#2}#1}
\newcommand{\bra}[1]{\langle #1 |}
\newcommand{\ket}[1]{| #1 \rangle}
\newcommand{\braket}[2]{\langle #1| #2 \rangle}
\newcommand{\ketbra}[1]{| #1 \rangle \langle #1|}
\newcommand{\bbra}[1]{\langle\langle #1 ||}
\newcommand{\kket}[1]{|| #1 \rangle\rangle}
\newcommand{\bbraket}[2]{\langle\langle #1|| #2 \rangle\rangle}

\newcommand{\avg}[1]{\langle #1 \rangle}
\newcommand{\avgT}[1]{\avg{#1}_{\currenttilt}}
\newcommand{\Var}{\textrm{Var}}
\newcommand{\abs}[1]{\lvert #1 \rvert}
\newcommand{\expsup}[1]{\text{e}^{#1}} 
\DeclarePairedDelimiter\floor{\lfloor}{\rfloor}

\newcommand{\ophat}[1]{\hat{#1}} 
\newcommand{\identityop}{\hat{\mathbb{1}}}

\newcommand{\hamilt}{H}
\newcommand{\invtemp}{\beta}
\newcommand{\crit}[1]{#1_{\mathrm{c}}}

\newcommand{\conf}{C}
\newcommand{\confset}{\Omega}
\newcommand{\confsetlen}{m}
\newcommand{\prob}{P}

\newcommand{\transitionratesymb}{w}
\newcommand{\transitionrate}[2]{\transitionratesymb_{#1 \to #2}}
\newcommand{\escaperatesymb}{r}
\newcommand{\escaperate}[1]{\escaperatesymb_{#1}}
\newcommand{\genmatmat}{\mathbb{W}} 
\newcommand{\genmat}{\ophat{\genmatmat}} 
\newcommand{\eigvectRnk}[1]{R_{#1}}
\newcommand{\eigvectLnk}[1]{L_{#1}}
\newcommand{\eigvectR}[1]{\ket{\eigvectRnk{#1}}}
\newcommand{\eigvectL}[1]{\bra{\eigvectLnk{#1}}}

\newcommand{\TEobservable}{A} 

\newcommand{\LDF}{I}
\newcommand{\sCGF}{\vartheta}
\newcommand{\DFE}{\sCGF}
\newcommand{\DPF}{Z}
\newcommand{\DPFwa}{\DPF_{\trajduration}(\biasingfield)}
\newcommand{\biasingfield}{\lambda} 
\newcommand{\currenttilt}{\lambda}
\newcommand{\escaperateT}[1]{\escaperate{#1}^{\currenttilt}}
\newcommand{\genmatT}{\genmat^{\currenttilt}}
\newcommand{\transitionrateT}[2]{\transitionrate{#1}{#2}^{\currenttilt}}
\newcommand{\probC}[1]{\prob_{\trajduration, #1}} 
\newcommand{\probT}{\prob^{\currenttilt}}

\newcommand{\probTvecttimenk}[1]{\probT_{#1, P_0}}
\newcommand{\probTvecttime}[1]{\ket{\probTvecttimenk{#1}}}
\newcommand{\probTvectstnk}{\probTvecttimenk{\text{ss}}}
\newcommand{\probTvectst}{\ket{\probTvectstnk}}

\newcommand{\eigvectTRnk}[1]{R^{\currenttilt}_{#1}}
\newcommand{\eigvectTLnk}[1]{L^{\currenttilt}_{#1}}
\newcommand{\eigvectTR}[1]{\ket{\eigvectTRnk{#1}}}
\newcommand{\eigvectTL}[1]{\bra{\eigvectTLnk{#1}}}
\newcommand{\Doobletter}{\mathrm{D}}
\newcommand{\escaperateD}[1]{\escaperate{#1}^{\currenttilt, \Doobletter}}
\newcommand{\genmatD}{\genmatT_{\Doobletter}}
\newcommand{\transitionrateD}[2]{\transitionrate{#1}{#2}^{\currenttilt, \Doobletter}}
\newcommand{\genmatDrate}{\genmatDrate_{\Doobletter}}

\newcommand{\eigvectDRnk}[1]{R^{\currenttilt}_{#1, \Doobletter}}
\newcommand{\eigvectDLnk}[1]{L^{\currenttilt}_{#1, \Doobletter}}
\newcommand{\eigvectDR}[1]{\ket{\eigvectDRnk{#1}}}
\newcommand{\eigvectDL}[1]{\bra{\eigvectDLnk{#1}}}
\newcommand{\probDphasevectnk}[1]{\Pi^{\currenttilt}_{#1}}
\newcommand{\probDphasevect}[1]{\ket{\probDphasevectnk{#1}}}
\newcommand{\probDphasevecttime}[1]{\ket{\probDphasevectnk{#1}(\timev)}}

\newcommand{\be}{\begin{equation}}
\newcommand{\ee}{\end{equation}}

\newcommand{\field}{E}
\newcommand{\nprtcls}{N}
\newcommand{\lattsize}{L}
\newcommand{\jumprate}{p}
\newcommand{\jumprateR}{\jumprate_{+}}
\newcommand{\jumprateL}{\jumprate_{-}}
\newcommand{\jumprateRL}{\jumprate_{\pm}}
\newcommand{\rateinL}{\alpha}
\newcommand{\rateoutR}{\beta}
\newcommand{\rateoutL}{\gamma}
\newcommand{\rateinR}{\delta}
\newcommand{\densL}{\dens_{L}}
\newcommand{\densR}{\dens_{R}}
\newcommand{\nstatesP}{r}

\newcommand{\occup}{n}
\newcommand{\mass}{\mu}
\newcommand{\intcurrent}{Q_{\trajduration}}

\newcommand{\current}{q}
\newcommand{\pos}{x}
\newcommand{\tv}{t}
\newcommand{\timev}{\tv} 
\newcommand{\dens}{\rho}
\newcommand{\densxt}{\dens(\pos,\tv)}
\newcommand{\diffcoef}{D}
\newcommand{\diffusivity}{\diffcoef} 
\newcommand{\mobility}{\sigma}
\newcommand{\meandens}{\dens_0}

\newcommand{\Zsymm}[1]{\mathbb{Z}_{#1}}
\newcommand{\symmop}{\hat{S}}
\newcommand{\symmeigval}[1]{\phi_{#1}}

\newcommand{\symmopR}{\symmop_{\frac{2\pi}{r}}}
\newcommand{\spinangle}{\varphi}
\newcommand{\meanmag}{\mathbf{m}}
\newcommand{\energytilt}{\lambda}
\newcommand{\trajduration}{\tau}

\newcommand{\traj}{\omega_{\trajduration}}
\newcommand{\currentcrit}{\crit{\current}}
\newcommand{\currenttiltcrit}{\crit{\lambda}}
\newcommand{\currenttiltcritmax}{\currenttiltcrit^{+}}
\newcommand{\currenttiltcritmin}{\currenttiltcrit^{-}}
\newcommand{\currenttiltcritminmax}{\currenttiltcrit^{\pm}}

\newcommand{\occupop}{\hat{\occup}}
\newcommand{\creationop}{\hat{\sigma}^{+}}
\newcommand{\destructop}{\hat{\sigma}^{-}}

\newcommand{\eigval}[1]{\theta_{#1}}

\newcommand{\eigvalT}[1]{\theta^{\currenttilt}_{#1}}

\newcommand{\eigvalD}[1]{\eigvalT{#1, \Doobletter}}

\newcommand{\confket}{\ket{\conf}}
\newcommand{\confbra}{\bra{\conf}}

\newcommand{\confpbra}{\bra{\confp}}
\newcommand{\flatvectnk}{-}
\newcommand{\flatvect}{\bra{\flatvectnk}}
\newcommand{\eigvectTLop}[1]{\hat{L}^{\currenttilt}_{#1}}

\newcommand{\probvecttimenk}[1]{\prob_{#1}}
\newcommand{\probvecttime}[1]{\ket{\probvecttimenk{#1}}}
\newcommand{\probvectt}{\probvecttime{t}}

\newcommand{\isep}{\mathrel{{.}\,{.}}\nobreak}

\newcommand{\Eorder}{m}
\newcommand{\KDopar}{z} 
\newcommand{\KDoparm}{\KDopar_{\Eorder}} 
\newcommand{\KDopararg}{\varphi} 
\newcommand{\KDoparargm}{\varphi_{\Eorder}}
\newcommand{\KDoparargmini}{\KDopararg_{\Eorder, 0}}
\newcommand{\E}{E}
\newcommand{\Ed}{\epsilon} 
\newcommand{\Epampl}{\lambda} 
\newcommand{\Epamplcritnom}{\lambda_{\mathrm{c}}} 
\newcommand{\Epamplcrit}[1]{\Epamplcritnom^{(#1)}}
\newcommand{\Epamplcritm}{\Epamplcrit{\Eorder}}
\newcommand{\twarg}{u}

\newcommand{\fcoef}[1]{C_{#1}} 
\newcommand{\fcoeft}[1]{\fcoef{#1}(\tv)} 
\newcommand{\wnumber}{k}
\newcommand{\binder}{U_4}

\newcommand{\clengthexp}{\nu}
\newcommand{\KDoparexp}{\beta}
\newcommand{\suscepexp}{\gamma}

\newcommand{\denspxt}{\delta\rho(\pos,\tv)}

\newcommand{\denstw}[1]{f_{#1}}

\newcommand{\minlevelTOC}{2}
\addtocontents{toc}{\protect\setcounter{tocdepth}{\minlevelTOC}}

\title{Rare events, time crystals and symmetry-breaking dynamical phase transitions}
\author{Rubén Hurtado Gutiérrez}

\begin{document} 

\frontmatter
    {
        \pagestyle{empty}

        \subinputfrom{titlepage/}{titlepage.tex}
        \clearpage

    \cleardoublepage
    }

   \tableofcontents

    \mainmatter

    \chapter{Introduction}  
    \label{chap:intro}
    {

\newcommand{\gencoord}{q}
\newcommand{\genmomen}{p}
\newcommand{\internalenergy}{U}
\newcommand{\energy}{E}
\newcommand{\energyN}{\energy_{\Npart}}
\newcommand{\energyPP}{h} 
\newcommand{\entropy}{S}
\newcommand{\entropyWA}{\entropy_{\Npart}(\internalenergy)} 
\newcommand{\entropyPP}{s} 
\newcommand{\entropyPPN}{\entropyPP_{\Npart}} 
\newcommand{\Nmicrostates}{\Omega}
\newcommand{\NmicrostatesWA}{\Nmicrostates_{\Npart}(\internalenergy)}
\newcommand{\totalNmicrostatesPP}{\Lambda} 
\newcommand{\temperature}{T}
\newcommand{\canonZ}{Z}
\newcommand{\canonZWA}{Z_{\Npart}(\invtemp)}
\newcommand{\helmholtzFE}{F}
\newcommand{\helmholtzFEWA}{\helmholtzFE_{\Npart}(\invtemp)}
\newcommand{\helmholtzFEPP}{f}
\newcommand{\massieupot}{\phi}
\newcommand{\volume}{V}
\newcommand{\Npart}{N}
\newcommand{\heatcapV}{C_{\volume}}
\newcommand{\probdens}{\rho}
\newcommand{\microstate}{\conf}
\newcommand{\trajj}{\omega} 
\newcommand{\dynobs}{A} 
\newcommand{\dynobsval}{a} 

\todo[inline]
{
    \textbf{Ideas chulas}

    - Al hablar de física estadística, hablamos del rol de las fluctuaciones microscópicas para dar lugar al mundo macro no fluctuante. Podemos recoger esto luego para preguntarnos al final por las fluctuaciones MACROSCÓPICAS.

    - El no equilibrio es mucho más complejo (negativo), pero también mucho más rico (positivo)

    - Podemos resolver bien tanto muy pocas como muchísimas partículas (en equilibrio). El problema surge para sistemas mesoscópicos

    - Creo que puede quedar bien que en algún momento reconozca que time-crystals parece que no pega y explica la conexión
}

\todo[inline]{out-of-equiibrium, con guiones}

\section{From the microscopic to the macroscopic world}

When we observe the world around us, it often gives us the impression of being a ``calm'' place.
When looking at the water inside a glass, we perceive it as a homogeneous and static liquid. 
This is true for many of the everyday objects we interact with, such as the walls of the room we are in, the wood of the table on which we might be reading or even the air we are breathing.
They all seem to be defined by a relatively small number of smooth and predictable average material properties \cite{jarzynski_2015}.
And, in fact, under our \emph{macroscopic} perception of the world, they indeed are.
However, if we ask ourselves about the underpinnings of this world, its building blocks, everything changes.
In a seemingly contradictory fashion, the foundations of this calm world lie in the hectic, vibrant and fluctuating realm of atoms and molecules found at the microscopic scale, which seems to be governed by a completely different set of rules and laws.
``How does the seemingly calm world we perceive emerge from the turmoil at the microscopic level?'' is the question addressed by the field of \emph{Statistical Mechanics}. 




In fact\todo{estaría bien conectar mejor}, Nature displays a deep hierarchic structure across different levels of description, each one characterized by its own particular observables and connected by a different set of physical laws.
For instance, the rules that explain the interaction between water molecules are remarkably different from the ones relating to the thermodynamic properties---such as temperature and pressure---of the water inside a glass, and again different than the ones controlling the dynamics of ocean currents.
As we ascend the hierarchy, the interaction between the vast number of degrees of freedom at each level generates new behaviors of increasing complexity. 
Statistical Mechanics aims to derive the ``effective'' laws describing these new phenomena from the fundamental laws at the bottom microscopic levels. 
While, in a strict sense, fundamental relations such as Schrödinger and Newton equations remain valid in the upper levels, they prove insufficient for a proper description of the new emergent phenomenologies.
In P. W. Anderson's words,  “Emergent properties are obedient to the laws of the more primitive level, but not conceptually consequent from that level” or more concisely ``More is different'' \cite{anderson_1972}.
\todo{Hablo de la equivalencia entre colectividades? Creo que no}
\todo{Creo que debería hablar un poco más de termo}
\todo{Resalta que la dinámica no juega ningún papel}
Rather than trying to solve the whole dynamical behavior at the microscopic level---an unfeasible task given the vast number of elements at this scale---, Statistical Mechanics employs a probabilistic approach, seeking the probability distribution of the microscopic states (microstate) compatible with each macroscopic state (macrostate) of the system.
In equilibrium systems---those in a stationary state in the absence of macroscopic fluxes of conserved quantities such as energy, momentum or mass---, this approach has attained extraordinary success in deriving the thermodynamics of systems from the laws governing their microscopic behavior.\todo{me gustaría explicar más de qué va esto antes de hablar de su éxito}
The cornerstone of this success lies in the \emph{ensemble theory} introduced by Gibbs \cite{pathria09a}.
This formalism represents the state of a system by a \emph{phase point}, $\conf = (\gencoord, \genmomen)$, where $\gencoord$ and $\genmomen$ denote its generalized coordinates and momenta. 
In this representation, an \emph{ensemble} is then defined as a ``swarm'' of infinitely many copies of the system, which traverse the \emph{phase space} according to the microscopic dynamics and physical constraints imposed on the system.
The spreading of the ensemble over the phase space gives rise to a stationary density distribution, $\prob(\gencoord, \genmomen)$.
This density function is the central focus of ensemble theory, serving as the perfect tool to perform the averages required to calculate the macroscopic properties\footnote{
    The use of the ensemble density function to perform averages is only valid in \emph{ergodic} systems, i.e., systems where the long-time average can be replaced with the ensemble average.
    While this property is expected to hold in most systems, an assumption known as \emph{ergodic hypthesis}, it has only been proved rigorously in a few simplified models.
    \todo[inline]{SI hay timempo, añadir que ergodico quiere decir que se explora todo el espacio de fases}
}.
According to the physical constraints imposed on the system, different ensembles are defined.

The \emph{microcanonical ensemble} is the one that describes equilibrium systems in which the macroscopic state is defined by a fixed number of particles $\Npart$, internal energy $\internalenergy$ and volume $\volume$ \footnote{
    For simplicity, in what follows we will assume a fixed density to eliminate the need for specifying the volume.
}.
In this ensemble, it is postulated that the probability is constant across all the microstates $\microstate$ compatible with the macrostate, and zero for the rest.
Therefore the distribution over the states with $\Npart$ particles is given by
\begin{equation}
    \prob^{\textrm{micro}}_{\internalenergy, \Npart}(\microstate)
    =
    \left\{
    \begin{array}[]{cc}
        \textrm{const}
        & \textrm{if}\; \energy(\microstate) = \internalenergy 
        \\
        0 
        & \textrm{else}
    \end{array}
    ,
    \right.
\end{equation}
where $\energy(\microstate)$ is the energy of the microstate $\microstate$.
The connection with thermodynamics in this ensemble is established by Boltzmann's equation, which relates the entropy of the system with the number of microstates $\NmicrostatesWA$ compatible with a given macrostate.
\begin{equation}
    \label{eq:entropy_intro}
    \entropyWA   
    =
    \kboltz \ln \NmicrostatesWA
    .
\end{equation}

However, fixing the energy is only possible in isolated systems, which are rarely found in Nature.
Therefore, it is useful to define a new ensemble in which the system in equilibrium interacts with a heat bath at temperature $\temperature$, allowing the system energy to fluctuate.
This defines the \emph{canonical ensemble}, in which the probability distribution of finding the sytem in the microstate $\microstate$ is given by the Gibbs distribution
\begin{equation}
    \label{eq:canon_ensemble}
    \prob^\textrm{canon}_{\invtemp, \Npart}(\microstate)
    =
    \frac{1}{\canonZWA} e^{-\invtemp\energy(\microstate)}
    ,
\end{equation}
with $\invtemp = 1/(\kboltz\temperature)$ denoting the inverse temperature.
The normalization factor in this equation, $\canonZWA = \sum_{\microstate} e^{-\invtemp\energy(\microstate)}$ is \emph{the canonical partition function}, which establishes the connection with thermodynamics in this ensemble.
Indeed, the Helmholtz free energy $\helmholtzFEWA$ of the macroscopic system is derived from the canonical partition function as
\todo{Creo que habría que añadir que también es útil en los cosos en los que queremos usar la micro por al equivalencia entre colectividades}
\begin{equation}
    \helmholtzFEWA
    =
    -\invtemp^{-1} \ln \canonZWA
\end{equation}

In addition, this ensemble offers insights into the significant role played by the microscopic fluctuations in the macroscopic description of a system.
An excellent example appears in the calculation of the heat capacity at constant volume, 
$
\heatcapV 
=
(\frac{\partial \internalenergy}{\partial \temperature})_{\volume, \Npart}
$. 
This macroscopic property can be readily measured from the microscopic fluctuations of the energy
\begin{equation}
    \heatcapV (\beta, \Npart)
    =
    \invtemp\temperature^{-1}\left(\avg{\energy^2}_{\beta, \Npart} - \avg{\energy}_{\beta, \Npart}^2\right)
    ,
\end{equation}
with $\avg{\cdot}_{\beta, \Npart}$ denoting the average with respect to the canonical distribution in Eq.~\eqref{eq:canon_ensemble}.
This is one of the many applications of the fluctuation-dissipation theorem \cite{callen51a,kubo66a}, which establishes the crucial link between the microscopic fluctuations and the thermodynamics of a system.
Indeed, Statistical Mechanics can also be seen as the theory that analyses the behavior of spontaneous fluctuations of physical systems \cite{chandler_1987}.


However, despite the undeniable success of Statistical Mechanics in the description of equilibrium systems, most of the phenomenology we find in Nature is out of equilibrium---characterized by the presence of net fluxes of conserved quantities, external forces and/or hysteretic behavior. 
From the gravitational collapse of a star to the inner workings of cells, nonequilibrium processes seem to be the rule rather than the exception. 
\todo{aquí me tendría que enrollar un poco más. Puedo poner una figura con muchos ejemplos}
In these systems, the central role played by the dynamics renders the development of a theory akin to equilibrium ensemble theory---connecting the macroscopic behavior with the microscopic invariants and constraints--- a formidable challenge.
Therefore, given our lack of \emph{a priori} knowledge of the underlying probability distribution of the microstates, we are compelled to consider thoroughly the full dynamics of the systems.
Today, the characterization and control of matter far from equilibrium remains one of the main challenges faced by Physics \cite{fleming_2008}.

\section{Large deviations and thermodynamics of trajectories}
\label{sec:LD_thermostatistics_intro}

\todo[inline]{Posible frase interesente: Large deviation theory tells us that whenever the law of large numbers holds, variational principles exist which dictate the most probable values of the observables, Oono, 1989}

{
\newcommand{\randvarparam}{N}
\newcommand{\Ncoins}{\randvarparam}
\newcommand{\probN}{\prob_{\Ncoins}}
\newcommand{\cointoss}{X}
\newcommand{\randvar}{Y_{\randvarparam}}
\newcommand{\randvarvalue}{y}
\newcommand{\randvaravg}{\randvarvalue^*}
\newcommand{\sCGFarg}{\lambda}


\begin{figure}
    \centering
    \includegraphics[width=1.\linewidth]{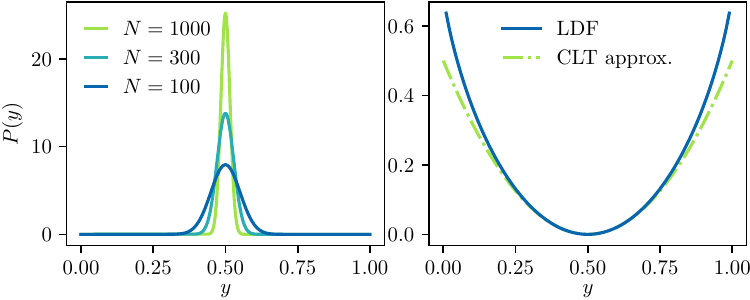}
    \caption{
        (a) Probability density of obtaining a fraction $\randvarvalue$ heads in $\Ncoins$ coin tossings.
        We see how the distribution peaks more sharply as $\Ncoins$ increases.
        (b) Large deviation function $\LDF(\randvarvalue)$ compared with the central limit theorem approximation.
        \label{fig:coin_tossing}
    }
\end{figure}

Over the past few decades, \emph{large deviation theory} has emerged as a promising framework for understanding nonequilibrium phenomena, playing a pivotal role in a plethora of breakthroughs in the field \cite{touchette09a}.
Before delving into the role it plays in nonequilibrium, let us illustrate the basic elements of this theory with a simple example: the tossing of a coin.
If we use a fair coin, the probability of each outcome---heads ($\cointoss = 1$) or tails ($\cointoss =  0$)---is equal: $\prob(\cointoss=0) = \prob(\cointoss = 1) = 1/2$.
Now, let us consider the tossing of $\Ncoins$ of such coins and ask ourselves which is the probability distribution controlling the fraction of heads, $\randvar = \frac{1}{\Ncoins}\sum_{i=1}^{\Ncoins} \cointoss_i$.
By using straightforward combinatorics, we obtain
\begin{equation}
    \probN(\randvar=\randvarvalue)
    =
    \frac{\Ncoins!}{2^{\Ncoins}(\Ncoins\randvarvalue)!(\Ncoins(1 - \randvarvalue))!}
    .
\end{equation}
In the case of a large number of coin tosses $\Ncoins$, we can apply Stirling's approximation $\ln(\Ncoins!) = \Ncoins\ln \Ncoins - \Ncoins + O(\ln \Ncoins)$ to find that the probability distribution obeys the following exponential expression
\begin{equation}
    \label{eq:LDprinciple_intro}
    \probN(\randvar=\randvarvalue)
    \asymp
    \exp(-\Ncoins\LDF(\randvarvalue))
    .
\end{equation}
with
$
    \LDF(\randvarvalue) 
    =
    \ln2  + \randvarvalue\ln\randvarvalue + (1-\randvarvalue)\ln(1 - \randvarvalue)
    ,
$
and ``$\asymp$'' standing for the asymptotic logarithmic equality, i.e.,
\begin{equation}
    \LDF(\randvarvalue)
    =
    - \lim_{\Ncoins \to \infty}\frac{1}{\Ncoins} \ln(\probN(\randvar = \randvarvalue))
    .
\end{equation}
When a random variable $\randvar$ depending on a parameter $\randvarparam$, follows an expression like Eq.~\eqref{eq:LDprinciple_intro}, we say that it satisfies a \emph{large deviation principle}.
The function $\LDF(\randvarvalue)$ is called the \emph{large deviation function} (LDF) of the variable $\randvar$, and it defines the statistic of the random variable for large values of  $\randvarparam$.
In most cases, this function has a single minimum satisfying $\LDF(\randvaravg) = 0$ that coincides with the typical or expected value $\randvaravg = \lim_{\randvarparam \to \infty} \avg{\randvar}$.
This observation allows us to understand the role played by the LDF.
On the one hand, it controls the exponential rate at which the probability decays as we move from the typical value $\randvaravg$ (for this reason, $\LDF(\randvarvalue)$ is also referred to as \emph{rate function}).
On the other hand, it defines how the probability concentrates around its average $\randvaravg$ as $\randvarparam$ increases.
This is clearly shown in Fig.~\ref{fig:coin_tossing}(a), which shows the behavior of the probability distribution in our coin-tossing example.
We see that the distribution peaks around the average $\randvaravg = 1/2$---which coincides with the zero of $\LDF(\randvarvalue)$ depicted in panel Fig.~\ref{fig:coin_tossing}(b)---, and that accumulation of probability around this value increases with the rise of $\randvarparam$.

Another key element of large deviation theory is the \emph{scaled cumulant generating function} (sCGF), defined as: 
\begin{equation}
    \label{eq:sCGF_intro}
    \sCGF(\sCGFarg)
    =
    \lim_{\randvarparam\to\infty} \frac{1}{\randvarparam} \ln \avg{e^{\sCGFarg \randvarparam \randvar}}
    ,
\end{equation}
with $\sCGFarg \in \realnumbers$.
The significance of this function stems from a fundamental result in large deviation theory, the Gärtner-Ellis theorem.
This theorem states that if $\sCGF(\sCGFarg)$ is differentiable, then the LDF can be calculated through the Legendre-Fenchel transform of the former
\begin{equation}
    \LDF(\randvarvalue)
    =
    \sup_{\sCGFarg\in\realnumbers}\left\{\sCGFarg\randvarvalue - \sCGF(\sCGFarg)\right\}
    .
\end{equation}
This is a crucial result, as the direct calculation of the LDF usually involves solving the whole probability distribution, which is unfeasible in most cases.  
The scaled cumulant generating function provides an alternative path to calculating the rate function characterizing the fluctuations of interest.
\todo{Se podría calcular theta para el coin tossing, pero dado que aquí ya tenemos la distriución de probabilidad, no sé hasta qué punto aporta}

Moreover, large deviation theory provides an extension of the central limit theorem beyond small fluctuations.
Indeed, if we expand $\LDF(\randvarvalue)$ up to the non-zero leading order around $\randvaravg$ and we substitute in Eq.~\eqref{eq:LDprinciple_intro}, we obtain 
\begin{equation}
    \probN(\randvar = \randvarvalue)
    \approx
    \exp\left(-\randvarparam\frac{\LDF''(\randvaravg)}{2} (\randvarvalue - \randvaravg)^2\right)
\end{equation}
which corresponds to the central limit theorem approximation 
\begin{equation}
    \probN(\randvar = \randvarvalue)
    \approx
    \exp\left(-\frac{(\randvarvalue - \avg{\cointoss_i})^2}{2\Var(\cointoss_i)/\Ncoins}\right)
\end{equation}
when we identify $\randvaravg = 1/2 = \avg{\cointoss_i}$ and $\LDF''(\randvaravg) = 4 = \Var(\cointoss_i)$.
Therefore, large deviation theory opens the door to exploring the behavior of stochastic processes beyond the limitations of the central limit theorem.
This is illustrated for our coin-tossing example in Fig.~\ref{fig:coin_tossing}(b), which shows a comparison between the large deviation and the Gaussian approximation, displaying how the two deviate as they get further from the average value $\randvaravg$.
These events far from the typical value, known as \emph{rare events}, often exhibit a remarkably intricate behavior and hold a key role in nonequilibrium physics, as we discuss later.

\newcommand{\tmicroconst}{\zeta} 

Large deviation theory plays a prominent role in equilibrium statistical mechanics, serving as its mathematical framework \cite{touchette09a}\todo{quizás se aun salto fuerte}.
This can be first observed in the entropy, which is intimately related to the rate function of the energy.
Indeed, from its definition in Eq.~\eqref{eq:entropy_intro},  we see that the number of microstates compatible with a given internal energy is proportional to the exponential of the entropy.
Therefore, if the probability of an energy is proportional to its number of microstates\todo{esto requeriría una aclaración}, we find that
\begin{equation}
    \label{eq:entropy_LDF_intro}
    \probN(\energyPP)
    \propto
    \exp(\Npart \entropyPP(\energyPP)/\kboltz)
    ,
\end{equation}
where $\energyPP = \internalenergy/\Npart$ and  $\entropyPP(\energyPP) = \lim_{\Npart\to\infty}\entropy_{\Npart}(\Npart \energyPP)/\Npart$ are respectively the energy and the (macroscopic) entropy per particle.
From this expression, we can readily identify the LDF 
\begin{equation}
    \LDF(\energyPP)
    =
    \tmicroconst - \entropyPP(\energyPP)/\kboltz
\end{equation}
where the constant $\tmicroconst$ comes from the normalization factor in the previous equation\footnote{
    In particular, 
    $
    \tmicroconst
    =
    \lim_{\Npart\to\infty} \frac{1}{\Npart} \ln \left(\sum_{\energy}\Nmicrostates_{\Npart}(\energy) \right)
    $
}.

The role played by large deviations is further seen in the canonical ensemble, where we can identify a close relation between the Helmholtz free energy and the scaled cumulant generating function of the energy, $\sCGF(\sCGFarg)$.
Indeed, using again the proportionality between microstates and probabilities, we have that $\avg{e^{-\invtemp\Npart\energyPP}} \propto \canonZ_{\Npart}(\invtemp)$.
Therefore from the definition of the scaled cumulant generating function and the Helmholtz free energy\todo{esto queda un poco oscuro}, we readily  obtain the expected relation
\begin{equation}
    \sCGF(-\invtemp)
    =
    -\invtemp \helmholtzFEPP(\invtemp) - \tmicroconst
    ,
\end{equation}
with $\helmholtzFEPP(\invtemp) = \lim_{\Npart \to \infty} \helmholtzFE_{\Npart}(\invtemp)/\Npart$ the Helmholtz free energy per particle in the macroscopic limit.
This allows us to use Gärtner-Ellis theorem to relate the entropy and the free energy through the following Legendre-Fenchel transformation\footnote{
    More exactly, the entropy is the Legendre-Fenchel transform of the Massieu potential or Helmholtz free entropy $\massieupot(\invtemp) = \invtemp \helmholtzFEPP(\invtemp)$ \cite{callen1985}.
}
\begin{equation}
    \entropyPP(\energyPP)
    =
    \kboltz \inf_{\invtemp \in \realnumbers} \left\{\invtemp \energyPP - \invtemp\helmholtzFEPP(\beta)\right\}
    ,
\end{equation}
which corresponds to the expected thermodynamic relation\todo{en termso es solo Legendre}.
Strikingly, this shows that the usual technique of calculating in the canonical ensemble and then using the equivalence of ensembles (thus circumventing the difficulties of the microcanonical ensemble), is just an application of large deviation theory and Gärtner-Ellis theorem.
\todo[inline]{Large deviation theory can also be found in Einstein's theory of microcanonical fluctuations, but we will stop here for the sake of brevity.}
Further analysis of the role played by large deviation theory in statistical mechanics can be found in the comprehensive review of Touchette \cite{touchette09a}.
\todo{Puede ser una buena idea comentar las diferencias entre la transformada de Legendre clásica y la de LEgendre Fenchel}


The clear link between the equilibrium statistical mechanics and large deviation theory paves the ground for the extension of the latter to nonequilibrium phenomena, providing a framework in which to derive general predictions \cite{harris05a,bertini05a,bodineau05a,lecomte07c,garrahan07a,touchette09a,derrida07a}.
In fact, LDFs are expected to serve as an analog to thermodynamic potentials in nonequilibrium.
The main difference is that, while in equilibrium we focused on configuration-dependent observables such as the energy, in nonequilibrium we have to consider the large deviations of \emph{dynamical observables} depending on the entire trajectory of the system, which capture the essential time correlations required to characterize nonequilibrium dynamics.
This choice of observable means that, instead of ensembles of configurations, we need to study \emph{ensembles of trajectories} and their associated probability distributions in order to evaluate such observables \cite{ruelle_thermodynamic_2004,lecomte07c}.

\begin{figure}
    \centering
    \includegraphics[width=0.9\linewidth]{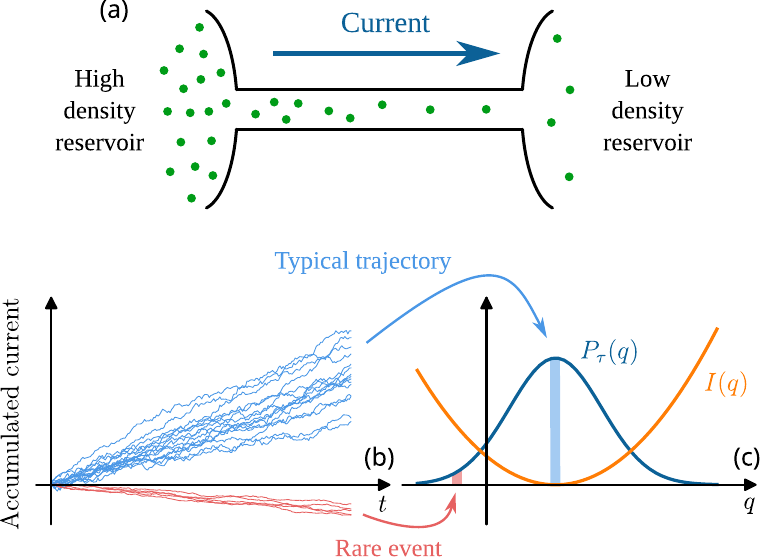}
    \caption{
        (a) Sketch of a particle-gas system between the two particle reservoirs at different densities.
        (b) Evolution of the accumulated current $\intcurrent = \int_0^{\trajduration}\dd t j(t)$ for different trajectories of the system.
        The curves in blue show the evolution during typical trajectories, while the red ones correspond to rare events with currents well below the average.
        (c) Probability distribution, $\prob_{\trajduration}(\current)$, of the time-averaged current $\current = \intcurrent/\trajduration$ and its large deviation function $\LDF(\current)$.
        The values of $\current$ corresponding to the trajectories in panel (b) are highlighted in blue and red.
    }
    \label{fig:particletransport}
\end{figure}

To illustrate the application of this approach to nonequilibrium situations, let us consider an example.
Imagine a particle gas system connected to a pair of particle reservoirs at different densities, such as the one shown in Fig.~\ref{fig:particletransport}(a).
In this system, the difference between the densities of the reservoirs generates a particle current in the direction of decreasing density, pushing it out of equilibrium.
In order to use large deviation theory to describe this system, we must first identify the relevant dynamical observable that captures its nonequilibrium behavior.
Although this choice is in general not apparent, recent developments indicate that if the system is driven out of equilibrium by a flux of some conserved quantity, the observable to choose is precisely this time-integrated flux \cite{bertini15a,derrida07a,hurtado14a}.
Therefore, we consider the large deviations of the intensive time-averaged particle current  $\current_{\trajduration} = \frac{1}{\trajduration}\int_0^{\trajduration} \dd\timev\, j(\timev)$ over trajectories of duration $\trajduration$.

To do this, we consider the ensemble of trajectories of the system $\traj$ with duration $\trajduration$.
In contrast to equilibrium, we do not have an a priori guess of the probability distribution over this ensemble of trajectories, but that does not stop us from defining a ``constrained'' ensemble, akin to the microcanonical ensemble, containing the trajectories with a particular value of the time-averaged current $\current$ \cite{evans_rules_2004,chetrite13a}.
The probability of such current, $\prob_{\trajduration}(\current)$, calculated as the sum of the probabilities on the trajectories in the constrained ensemble, is expected to follow the large deviation principle\todo{comentar que hay excepciones}
\begin{equation}
    \label{eq:LDP_current_intro}
    \prob_{\trajduration}(\current)
    \asymp
    \exp(-\trajduration \LDF(\current))
    ,
\end{equation}
which is similar to Eq.~\eqref{eq:entropy_LDF_intro} with $\LDF(\current)$ replacing the entropy and the duration of the trajectory in the role of the system size.
According to this equation, as we average over longer times, the probability distribution of the time-intensive current $\current$ sharpens around its typical value.
Therefore, the thermodynamic limit $\Npart \to \infty$ of equilibrium thermodynamics is replaced with the long-time one, $\trajduration \to \infty$.
Unfortunately, in addition to obeying similar equations, the LDF $\LDF(\current)$ and the entropy also share the difficulty in their direct calculation.
To overcome this, we can employ the same strategy as in equilibrium and we resort to the calculation of the sCGF\todo{me da la impresión de que, desde aquí, no se termina de ver la analogía con la canónica. Comentar que el lugar de fijar el observable, se controla su valor medio a través de un campo conjugado que juega el papel de la temperatura o el parámetro de ls sCGF},
\begin{equation}
    \sCGF(\sCGFarg)
    =   
    \lim_{\trajduration\to\infty}
    \frac{1}{\trajduration} \ln \avg{e^{\sCGFarg \trajduration \current}}
    ,
\end{equation}
whose computation is usually more feasible and can related to the LDF using a Legendre-Fenchel transform\todo{habría que decir que esto no es siempre posible}.
In fact, many large deviation methods in nonequilibrium systems use the \emph{biased ensemble}, which is similar to a \emph{dynamical} canonical ensemble in which the parameter $\sCGFarg$ plays the role of the temperature \cite{lecomte_chaotic_2005,lecomte07a,garrahan09a,chetrite13a}.
Indeed, we will see that $\sCGF(\sCGFarg)$ serves as a \emph{dynamical} counterpart of the free energy in nonequilibrium systems.
The different analogies drawn between equilibrium configuration statistics and nonequilibrium trajectory statistics are summarized in Table~\ref{tab:eq-noneq-analogies}.
\begin{table}
    \centering
    \begin{tabular}{c | c}
        Equilibrium & Nonequilibrium
        \\
        \hline
        configurations, $\conf$ & trajectories, $\traj$
        \\
        $\Npart \to \infty$ & $\trajduration \to \infty$
        \\
        microcanonical ensemble & constrained ensemble
        \\
        canonical ensemble & $\biasingfield$-ensemble
        \\
        entropy, $\entropyPP(\energyPP)$ & LDF, $\LDF(\current)$
        \\
        free energy, $\helmholtzFEPP(\invtemp)$ & sCGF, $\sCGF(\sCGFarg)$
    \end{tabular}
    \caption{Analogies between some features of equilibrium statistics and nonequilibrium trajectory statistics.}
    \label{tab:eq-noneq-analogies}
\end{table}

These similarities bring hope to the description of the nonequilibrium macroscopic states using large deviation theory.
However, while this approach provides a robust framework to obtain general results arbitrarily far from equilibrium, the analytical calculation of the LDFs remains a daunting task that has only been solved for a very limited number of oversimplified models \cite{derrida_exact_1992,derrida_exact_1993,schutz_phase_1993,bodineau07a,derrida07a}.
To address this, a large number of methods have been introduced in the last decades.
These include spectral methods involving the diagonalization of the \emph{tilted} generators of the Markov dynamics \cite{lecomte07c,jack2010,banuls19a}, importance sampling methods \cite{nemoto16a,ferre18,ray18a}, cloning algorithms \cite{giardina09a, lecomte07a, perez-espigares19a} and  reinforcement learning techniques \cite{das19, rose21, das21}.
Deserving a mention of its own is the Macroscopic Fluctuation Theory (MFT), which offers detailed predictions of the LDFs of diffusive models from just the knowledge of their macroscopic transport coefficients \cite{bertini01a,bertini02a,bertini15a}.
The application of these techniques has seen remarkable success in recent times, achieving important breakthroughs in diffusive and colloidal systems \cite{ciliberto_fluctuations_2010,ciliberto_experiments_2017}, glassy systems \cite{merolle_spacetime_2005,garrahan09a,hedges09a,chandler10a,banuls19a}, active matter \cite{cagnetta17a,grandpre21a,keta21a}, many-particle systems under gradients or external fields \cite{bodineau05a,bertini07a,baek17a,tizon-escamilla17b} or open quantum systems \cite{garrahan10a,manzano18a,macieszczak21a}, among many others.
\todo[inline]{These advances show that the application of LD theory to nonequilibrium is a promising enterprise, which has already shown many results.}
\todo[inline]{Habría que hablar también de teoremas de fluctuación y resultados del estilo, ¿no?}

As mentioned before, large deviation theory not only gives insights into the typical behavior of nonequilibrium systems, but it also provides a framework for the analysis of rare events, i.e., extreme fluctuations far from the typical value of the observable.
For example, in the particle gas model introduced before, current flowing against the density gradient [see the red curves in Fig.~\ref{fig:particletransport}(b)] would constitute a rare event.
This situation, while extremely unlikely, is not forbidden by any fundamental law, and its probability can be calculated from the LDF [Fig.~\ref{fig:particletransport}(c)].
Furthermore, large deviation theory also allows us to identify the \emph{optimal paths} leading to the fluctuation, that is, the particular trajectories that give rise to it.
While rare events, by definition, are extremely unlikely, they wield a significant influence in numerous processes, as their occurrence fundamentally alters their dynamics.

}

\section{Dynamical criticality at the fluctuating level}
\todo[inline]{Puedo poner algunas fotos de transicione de fase estándar y dinámicas}

Before delving into the main topic of this section, let us briefly revisit equilibrium phenomena.
As mentioned earlier, statistical mechanics aims to determine and explain emergent behaviors that are not evident from the microscopic dynamics, but which become apparent when gathering a macroscopic number of particles.
Undoubtedly, phase transitions stand as some of the most captivating of such emergent phenomena.
Loosely speaking, a phase transition can be defined as a drastic change in the arrangement of a system when one of its parameters, known as \emph{control parameter}, crosses a critical point.
These transitions are characterized by singularities in equilibrium thermodynamic potentials at this critical threshold.
Classical examples include the condensation of gasses, the freezing of liquids, the superconducting phase transition or the order-disorder transition in alloys\todo{podría extender me y hablar de lo importantes que son en la vida}.
In all these examples a new kind of order, in the form of novel structures not present in the previous phase, emerges after the transition.
A fundamental phenomenon associated with the appearance of such structures is \emph{spontaneous symmetry breaking} \cite{chaikin00a,sznajd_introduction_2003}.
This occurs during a phase transition when a symmetry of the dynamics governing the system is no longer present in the new stable states after the critical points, leading to a significant change in its structure.
An everyday example is the freezing of water into ice.
While water's properties remain invariant under any translation (i.e., it has a \emph{continuous} translational symmetry), when it freezes into ice it organizes itself into a new crystalline structure that breaks this homogeneity.
Indeed, after the phase change, only \emph{discrete} translations that coincide with the lattice spacing of the crystal leave the state invariant, so we say that it has broken the continuous symmetry present in the liquid state\todo{aquí hay un salto feísimo}.

The presence of singularities in the thermodynamic potentials during phase transitions prompts an interesting question:
given that the large deviation and scaled cumulant generating functions in nonequilibrium systems play a role akin to the thermodynamic potentials, is it possible for similar singularities to manifest in these functions? 
And if so, what implications do they carry?\todo{no estoy seguro que puedan haber singularidades en la entropía, check tochette}
Notably, research in the last decades has shown that these singularities do indeed occur and that they mark the onset of a \emph{dynamical phase transition} (DPT).
To introduce this new type of phase transition, consider a particular nonequilibrium system with such singularity and let us look at the optimal trajectories associated with the different values of the dynamical observable. 
If we observe how the behavior of the system changes as we change the considered value, a DPT manifests as a drastic change in the arrangement of the trajectories when crossing the critical point marked by the singularity.
Instead of being the consequence of a change in a physical parameter such as the temperature, DPTs arise when exploring different fluctuations of the dynamical observable. 
In this exploration, these novel structures at the level of trajectories emerge as a way to boost the probability of the corresponding fluctuation, with spontaneous symmetry breaking often playing an important role.
\todo{se puede añadir que las DPTs permiten cosas más interesantes de las estándar, nuevas fases de la materia}

The relevance of the study of DPTs is twofold.
On the one hand, DPTs often manifest in close proximity to the typical behavior of nonequilibrium systems, rendering them fundamental for a comprehensive understanding of such systems, as in the case of kinetically-constrained models \cite{garrahan07a, garrahan09a,pitard11a,abou18a} or superconducting transistors \cite{genway12a}.
On the other hand, even when DPTs appear only as very rare fluctuations, they bring precious insight into the dynamics of systems and the optimal ways they organize in order to realize particular fluctuations
\cite{bertini05a,bertini06a, derrida07a, hurtado14a, lazarescu15a, shpielberg18a,perez-espigares19a,carollo17a}.
Additionally, using tools like the generalized Doob h-transform \cite{doob57a,jack2010,chetrite15a}, these rare fluctuations can be made ``typical'', enabling the engineering of new models that exploit the particular characteristics of such fluctuations.
Thus, the study of DPTs provides not only a deeper understanding of nonequilibrium systems but also opportunities for innovative model engineering based on rare fluctuations.


\section{Breaking time-translation symmetry: time crystals}
\todo[inline]{Alargar hasta 3 páginas al menos. Comentar un poco más los ejemplos concretos por ejemplo y su desarollo histórico así como cosas bonitas}
\todo[inline]{si tengo tiempo, me haría muy bien tener aquí una figura}
\todo[inline]{Mencionar a Noether?}

As noted in the previous section, most symmetries in nature---such as rotational invariance, gauge symmetries or chiral symmetry--- can be spontaneously broken in a phase transition, with the resulting system ground state showing fewer symmetries than the associated action.
However, in contrast to the rest of the symmetries, time-translation symmetry seemed to be special and fundamentally unbreakable\todo{cita?}.
This changed with the seminal papers of Wilczek and Shapere in 2012 \cite{wilczek12a, shapere12a}.
Using the analogy with regular or \emph{space} crystals---which break continuous space-translation symmetry to give rise to a space-periodic structure---, the authors introduced the concept of \emph{time-crystals}, i.e., systems whose ground state spontaneously breaks time-translation symmetry and thus exhibit enduring periodic (or \emph{time-cristalline}) motion\todo{siento que podría decir algo más aquí} \cite{khemani_brief_2019,richerme17a}.
This seemingly natural concept stirred a vivid debate among physicists, which resulted in ruling out the possibility of time-crystals in equilibrium under rather general conditions \cite{bruno13a, nozieres13a, watanabe15a}.
However, nothing forbids their occurrence in nonequilibrium settings\todo{esto se puede reemplazar por algo que recoja la expresión comentada anterior}.

Floquet systems, i.e., systems brought out of equilibrium by a periodic driving, seemed to be the perfect candidates to observe this phenomenon.
Indeed, their study quickly led to the discovery of \emph{discrete} time crystals, i.e., systems that break the discrete translation symmetry imposed by the driving by doubling their period (or multiplying with any integer factor). 
They have been observed in both classical and quantum contexts, highlighting the versatility of the concept \cite{khemani16a,keyserlingk16a,else16a,gambetta19a,yao17a,zhang17a,choi17a, nakatsugawa17a, lazarides17a, gong18a, tucker18a, osullivan18a, lazarides19a}.
As with regular crystals, time crystals show rigidity and are robust against environmental dissipation\todo{hay que arreglarlo, pero en realidad no sé si hace falta.}.
Continuous time crystals on the other hand correspond to the original idea of the time crystal:
a ``clock'' emerging spontaneously within a time-invariant system, in Wilczek's own words.
Their discovery was more challenging, but these time-crystals have also been proposed in open quantum systems \cite{iemini18a,medenjak20a,booker_non-stationarity_2020,kongkhambut_observation_2022} and classical settings, with experimental confirmations found in pumped atom cavities \cite{kongkhambut_observation_2022} and photonic metamaterials \cite{tongjun_2023}.

The connection between time crystals and large deviations lies in that, as we show in this Thesis, time crystals can be found in the rare events of driven lattice gases.
The analysis of such rare events allows for the proposal of new models showing time-crystalline behavior \cite{hurtado-gutierrez20a}\todo{no me apasiona}.
\section{Simple models to solve complex problems}

\begin{figure}
    \centering
    \includegraphics[width=0.8\linewidth]{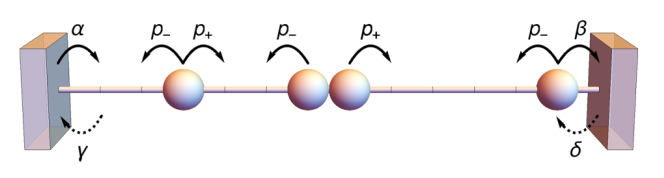}

    \includegraphics[width=0.6\linewidth]{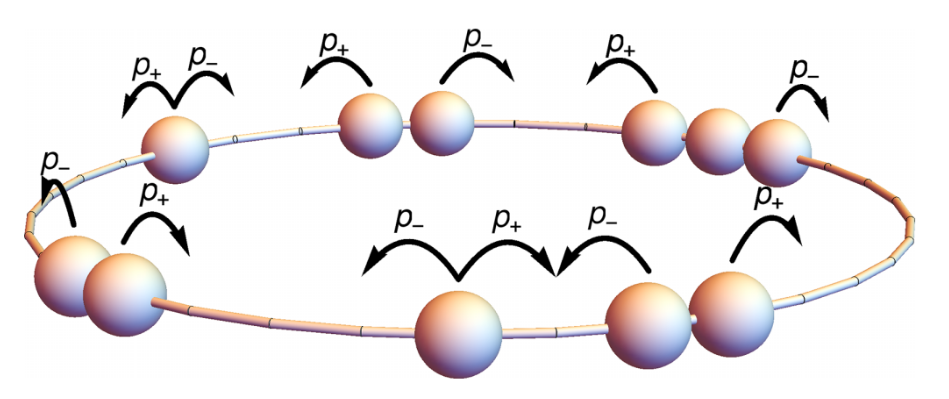}
    \caption{Sketch of a paradigmatic lattice gas model for driven diffusive transport: the Weakly Asymmetric Simple Exclusion Process (WASEP) with open (top) and closed (bottom) boundaries.}
    \label{fig:simplified_models_intro}
\end{figure}


In the preceding sections, we have introduced several methods for the analysis of systems far from equilibrium along with their associated fluctuations. 
However, the availability of these powerful analytical and computational tools does not negate the formidable challenges in their application, especially when dealing with realistic systems featuring a large number of degrees of freedom.
This underscores the relevance and necessity of employing simplified microscopic models.

While they may appear rudimentary at first glance, simplified models like cellular automata, random walk models or lattice gases, form the cornerstones in the development of nonequilibrium statistical mechanics.
Indeed, they manage to capture the fundamental ingredients of the phenomena they aim to describe while discarding the extraneous details, enabling their analysis.
Far from being overly simplistic, these models serve as invaluable stepping stones, providing profound insights that contribute to our comprehension of the intricate dynamics governing more realistic systems and allowing for the validation of underlying hypotheses within a controlled setting.

Notably, lattice gases play a fundamental role in this theory, as they serve as the perfect testing ground for transport phenomena.
Specifically, we would like to highlight three particular models.
\begin{enumerate}
    \item
    Energy transport: \emph{Kipnis-Marchioro-Pressuti model} (KMP).
    It is a one-dimensional lattice in which each site models a harmonic oscillator having a particular energy.
    The evolution of the system runs through random ``collisions'' between neighbor sites in which the energy of the pair is redistributed following a random distribution.
    This model was proven to obey Fourier's law (a fact achieved in a very reduced number of microscopic systems) \cite{kipnis82a}, and it has also served as an important workbench in the development of the study of nonequilibrium dynamics through large deviations, playing an important role in the application of macroscopic fluctuation theory to dissipative systems \cite{bertini05b,hurtado09a,bertini15a}. 
    Generalizations of the model have been proposed to include dissipation and external driving field, extending the utility of the model and setting the ground for the fundamental analysis of these processes.\todo{aquí falta una cita}
    
    \item
    Mass transport: Simple Exclusion Processes (SEP).
    These are a family of models designed to explore mass transport.
    They are defined in a (often one-dimensional) lattice in which particles perform random jumps between neighboring sites while obeying an exclusion principle.
    Several SEP models exist depending on how the transition rates depend on the spatial direction of the jump.
    Some of the most commonly used ones are the Symmetric Simple Exclusion Process (SSEP), the Asymmetric Simple Exclusion Process (ASEP) and the Weakly Asymmetric Simple Exclusion Process (WASEP) (shown in Fig.~\ref{fig:simplified_models_intro}).
    These models provide a versatile platform for the study of multiple transport phenomena, including diffusive and driven transport, and they have played a crucial role in the development and validation of methods within nonequilibrium statistical mechanics.

    \item
    Coupled mass-energy transport: Kinetic Exclusion Process (KEP). New models have been recently proposed to study more complex phenomena in which the dynamics are characterized by several conserved fields.
    A great example is the Kinetic Exclusion Process \cite{gutierrez-ariza19a}, characterized by a nonlinear diffusion and displaying the transport of both energy and mass.
    This model bears similarities with the SEP models, but with the fundamental difference that each particle is also defined by its energy,  which modulates its jump rates.
    In addition, its dynamics include random ``collisions'' in which neighboring sites may exchange energy in a way similar to the KMP model.
    Since real fluids are also characterized by multiple conserved fields, this kind of model opens the door to studying more interesting and realistic phenomena.
\end{enumerate}

\section{Outline of this Thesis}

From the previous sections, it is clear that the study of fluctuations plays a fundamental role in the modern development of statistical physics, especially in out-of-equilibrium systems.
Aligned with this objective, this Thesis delves into the critical role played by fluctuations in systems out of equilibrium, using the tools of large deviation theory.
More specifically, it will focus on unraveling the intricacies of dynamical phase transitions (DPTs), in particular those presenting the spontaneous breaking of a symmetry.
It will be demonstrated that the latter imposes a stringent structure in the spectral properties of the generator of the dynamics of the system.
This realization will pave the way for the development of a general theory regarding the behavior of its leading eigenvectors and eigenvalues in Chapter \ref{chap:SBDPTs}.
Such theory is then applied to DPTs in several paradigmatic models in Chapter~\ref{chap:DPTs_models}, validating its predictions and providing precious insights.
In Chapter \ref{chap:buildingtc}, the focus shifts to time crystals, motivated by the identification of a DPT to a time-crystal phase in the large deviations of a lattice gas. 
Using the tools developed in the aforementioned chapters, the fluctuating dynamics of the DPT are studied and its fundamental properties are distilled into a new model, which displays a similar phase transition in its typical behavior---instead of as a rare fluctuation.
Finally, in Chapter \ref{chap:genpackfield} the time-crystal mechanism observed in the previous chapter is further generalized to obtain more complex time-crystal phases in other diffusive models.
In what follows, we further detail the content of each Chapter.

First, in Chapter \ref{chap:theory} we introduce large deviation theory and its application to systems driven out of equilibrium, in particular those defined by homogeneous Markov chains.
We will define in detail the basic concepts and tools that will be used throughout the Thesis,
 such as master equations, dynamical observables and trajectory ensembles.
A fundamental concept will be the generalized Doob $h$-transform, a transformation of the generator that enables the derivation of the dynamics associated with arbitrary fluctuations of a given observable.

Chapter \ref{chap:SBDPTs} elucidates the underlying spectral mechanisms responsible for continuous DPTs in jump processes where a discrete $\Zsymm{n}$ symmetry is broken.
Through a symmetry-aided spectral analysis of the Doob-transformed dynamics, this chapter establishes the conditions for the emergence of symmetry-breaking DPTs and delineates how distinct dynamical phases arise from the specific structure of degenerate eigenvectors.
This leads us to unveil stringent restrictions on the structure of such eigenvectors, which encapsulate all the symmetry-breaking features.
Additionally, we introduce a partitioning of configuration space based on a proper order parameter, leading to a significant dimensional reduction that facilitates the quantitative characterization of the spectral fingerprints of DPTs.
Chapter \ref{chap:DPTs_models} is dedicated to the application of these results to specific cases of symmetry-breaking DPTs.
In particular, we analyze the DPTs found in the open WASEP\todo{no sé si dejarlo en siglas o no} for low current fluctuations and the three- and four-state Potts models for low magnetizations.

Chapter \ref{chap:buildingtc} is devoted to the analysis of the current fluctuations in the periodic WASEP.
When conditioned to sustain currents well below its typical value, this model displays a DPT to a periodic state characterized by the formation of a traveling particle condensate, revealing spectral features reminiscent of a time crystal phase.
Using the tools developed in the previous chapters, we successfully identify the mechanism giving rise the the phase transition.
This allows us to devise a mechanism to construct classical time-crystal generators from rare event statistics in driven diffusive systems, which is used to propose a new model, the time-crystal lattice gas (TCLG).
In this model, the phase transition to a time-crystal phase manifests in its typical behavior when the coupling to an external \emph{packing field} surpasses the critical point, instead of appearing as a rare fluctuation.
This phase transition is then analyzed in detail, confirming the expected time-crystalline behavior.
Finally, in Chapter \ref{chap:genpackfield}, we devise a generalization of this mechanism to create new time-crystal phases characterized by an arbitrary number of wave condensates.
The application of the mechanism to different driven diffusive systems is explored, with an exhaustive analysis performed in the case of the WASEP.


    }

    \chapter{A review of the statistics of trajectories in Markov processes}
    \chaptermark{Review of trajectory statistics in Markov processes}
    \label{chap:theory}
     {
         \newcommand{\exampleQmat}{A}
\newcommand{\exampleStomat}{B}
\newcommand{\permutmat}{\ophat{P}}
\newcommand{\confdepobservable}{g} 
\newcommand{\transdepobservable}[2]{\eta_{#1\to#2}} 
\newcommand{\TIobservable}{a} 
\newcommand{\confp}{\conf^{\prime}}
\newcommand{\probCa}{\probC{\TIobservable}} 
\newcommand{\observable}{O} 

\newcommand{\deltat}{\Delta\timev} 
\newcommand{\transitionprobT}[4]{G^{\biasingfield}(#1, #2;#3, #4)}
\newcommand{\transitionprobTWA}{\transitionprobT{\conf}{\confp}{\timev}{\timev+\deltat}}

This chapter aims to serve as a comprehensive introduction to the main fundamental concepts and tools that form the backbone of this thesis, building upon the ideas in Chapter~\ref{chap:intro}.
We begin by focusing on continuous-time Markov processes and the formalism that allows their study.
Following this, we delve into large deviation theory, the framework for examining rare events within these Markov processes.
This theory will offer us a structured approach to characterize the fluctuations occurring in these systems, which play a fundamental role in the understanding of their dynamics.
Finally, we will explore the biased ensemble and the Doob transform, which enable the analysis of the dynamics of fluctuations.
Together, these concepts and models will form the cornerstone of the research presented in the subsequent chapters.

\section{Markov processes and the role of stochastic models in physics}
\sectionmark{Markov processes and the role of stochastic models}
{
\label{sec:markovprocesses_theory}

Stochastic models play a pivotal role in our understanding of physical many-body systems.
Even in the case of classical physics, a deterministic approach in the determination of the evolution of the system is an unfeasible task due to the massive number of degrees of freedom.
Therefore, we often use a mesoscopic approach in which the microscopic variables are coarse-grained into new average variables subject to a noise, which carries information about the dynamics at the underlying level.
A prime example of this approach is Einstein's theory of Brownian motion,
in which the movement of the suspended particle is random and the characteristics of the noise are related to the properties of the water molecules.

An important class of stochastic systems is jump processes in a discrete state space.
In these processes, the dynamics unfolds through a sequence of transitions---or \emph{jumps}---between states in a discrete set $\conf \in \confset$.
At any given time, the system is found in a particular state $\conf$, from which it may jump to other states  $\confp$ with a transition rate or probability per unit time $\transitionrate{\conf}{\confp}$, 
defined as 
\begin{equation}
\label{eq:transrate_def_theory}
    \transitionrate{\conf}{\confp}
    =
    \left. \pdvD[\prob(\confp, t' | \conf, t)]{t'} \right|_{t'=t}
    ,
\end{equation}
where $\prob(\confp, t' | \conf, t)$ stands for the probability of being at $\confp$ at time $t'$ conditioned to being at $\conf$ at time $t$.
Specifically, our focus will be on \emph{temporally homogeneous Markov processes} in \emph{continuous time} \cite{Gillespie_1992}, in which the transition rates depend only on the pair of states $\conf$, $\confp$ involved in the transition, not on the time when it occurs (temporal homogeneity) nor the previous history of the system (Markov property).
In this way, the probability of performing a particular transition $\conf\to\confp$ in the infinitesimal interval $(\timev, \timev+\dd \timev)$ is given by $\transitionrate{\conf}{\confp}\,\dd t$, while the probability of staying in $\conf$ during the same interval is just $1 - \dd t \sum_{\confp \ne \conf} \transitionrate{\conf}{\confp}$.
The last sum, $\escaperate{\conf} = \sum_{\confp \neq \conf} \transitionrate{\conf}{\confp}$, is known as the escape rate of state $\conf$, and it corresponds to the inverse of the mean time spent at this state.
Such residence time follows an exponential distribution given by $f(t) = \escaperate{\conf}\exp(-\escaperate{\conf}t)$.
An illustrative representation of a simple Markov process is shown in Fig.~\ref{fig:markovprocess_theory}.

\begin{figure}[b!]
    \centering
    \includegraphics[width=0.9\linewidth]{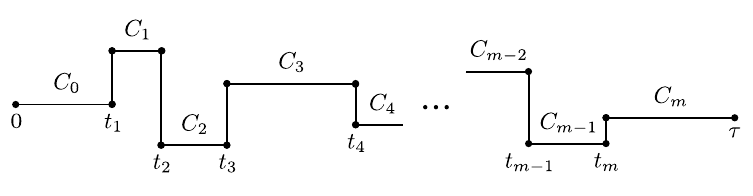} 
    \caption{
            Sketch of a trajectory in a continuous time Markov process. 
            The residence time $t_{i+1} - t_i$ at each state $\conf_i$ is distributed according to an exponential distribution with mean $\escaperate{\conf_i}$.
            Once a jump occurs, the next state $\conf_{i+1}$ is selected with probability $\transitionrate{\conf_i}{\conf_{i+1}}/\escaperate{\conf_i}$.
    }
    \label{fig:markovprocess_theory}
\end{figure}

Instead of tracking the evolution of individual trajectories, in stochastic systems it is useful to consider the evolution of the probability distribution along the different states of the system, $\prob(\conf, \timev)$.
By considering the balance of probability entering and exiting a particular configuration within an infinitesimal time interval, we can derive the following differential equation
\begin{equation}
\label{eq:probevolution}
    \pdvD[]{t}\prob(\conf, t)
    =
    \sum_{\confp \neq \conf} \transitionrate{\confp}{\conf} \prob(\confp, t)
    - \escaperate{\conf} \prob(\conf, t)
    .
\end{equation}
This is known as the \emph{master equation} of the system, and it controls the evolution of the probability distribution \cite{kampen11a,toral_stochastic_2014}.
The first term on the right-hand side accounts for the probability flow into the state $\conf$ from other states, while the second deals with the flow exiting from $\conf$ to the rest.
The conservation of the total probability is warranted by the way the escape rates are defined, as it can be readily checked by noting that $\partial_t [\sum_{\conf} \prob(\conf, t)]= 0$.

In many cases, it is convenient to rewrite the master equation in vector notation using the \emph{quantum Hamiltonian formalism} \cite{Doi_1976,schutz01a}.
In this formalism, an independent column vector $\confket$ is assigned to each state $\conf$ so that they form the basis of a $\complexnumbers^m$ vector space, with $\confsetlen$ the number of different states in the system.
Additionally, a dual basis of row vectors $\bra{\conf}$ is introduced, obeying the usual orthonormality relation $\braket{\conf}{\confp} = \delta_{\conf,\confp}$, with $\delta_{\conf,\confp}$ the Kronecker delta.
This allows us to represent the probability distribution of the system at time $\timev$ as the vector
\begin{equation}
    \probvectt = \sum_{\conf} \prob(\conf, t) \confket
    ,
\end{equation}
which in turn enables us to rewrite the master equation the following simplified form
\begin{equation}
    \label{eq:mastereqvect_theory}
    \ddt\probvectt
    =
    \genmat \probvectt
    ,
\end{equation}
where the linear operator $\genmat$, known as the \emph{generator} of the dynamics, is defined as
\begin{equation}
    \genmat
    =
    \sum_{\conf} \sum_{\confp \neq \conf} 
    \transitionrate{\conf}{\conf^{\prime}} \ket{\conf^{\prime}} \bra{\conf} - \sum_{\conf} \escaperate{\conf} \ketbra{\conf}
    .
\label{eq:genmat_def_theory}
\end{equation}
Intuitively, this operator is constructed by filling the off-diagonal terms with the corresponding transition rates $\transitionrate{\confp}{\conf}$ and the diagonal terms with minus the escape rates $\escaperate{\conf}$.
The conservation of the probability implies that the columns\footnote{
    \newcommand{\exampleop}{A}
    Unless stated otherwise, when discussing the elements of a linear operator $\ophat{\exampleop}$ we will refer to the elements of its matrix representation in the basis $\{ \ket{\conf_i}\}_{i=1}^m$ defined by the states of the system, i.e., $\exampleop_{ij} = \bra{\conf_i} \ophat{\exampleop} \ket{\conf_j}$.
}  of $\genmat$ add up to zero,
\begin{equation}
    \sum_{\confp} \confpbra \genmat \confket
    =
    \sum_{\confp \neq \conf} \transitionrate{\conf}{\confp} - \escaperate{\conf}
    = 0, \; \forall \conf
    .
\end{equation}
As a consequence, for probability-conserving generators, the so-called \emph{flat vector}, $\flatvect = \sum_{\conf} \confbra$, is a left eigenvector of the generator with eigenvalue $0$, i.e., $\flatvect\genmat = 0$.
In fact, any square matrix $\exampleQmat$ satisfying that:
\begin{enumerate}[label=(\roman*)]
    \item
    its diagonal elements are non-positive, $\exampleQmat_{ii} \le 0,\; \forall i$, 
    \item
    its non-diagonal elements are non-negative, $\exampleQmat_{ij} \ge 0,\; \forall j\neq i$,
    \item
    and each of its columns adds up to zero, $\sum_i \exampleQmat_{ij} = 0,\; \forall j$,
\end{enumerate}
defines the generator of a continuous-time Markov process.
The matrices satisfying these properties are known in general as \emph{transition-rate matrices} or \emph{Q-matrices} \cite{Norris_1997}.
Another important class of matrices is \emph{stochastic matrices}.
These are characterized for having non-negative elements, $\exampleStomat_{ij} \ge 0\; \forall i,j$, and columns adding up to one, so that $\flatvect$ is an eigenvector with eigenvalue 1, $\flatvect\ophat{\exampleStomat} = \flatvect$.
Examples of these matrices include the time evolution operator $\exp(t \genmat)$ and the transition matrices in discrete-time Markov chains\footnote{
    Although discrete-time Markov processes will not be covered in this section, we refer the interested reader to the first chapters of \cite{Freedman_1983}.
}.

One advantage of employing the quantum Hamiltonian formalism is that it allows us to easily identify the formal solution of the master equation.
Indeed, since the operational representation of the master equation in Eq.~\eqref{eq:mastereqvect_theory} is just a linear differential equation, its solution is given by
\begin{equation}
    \label{eq:mastereqsol_theory}
    \probvectt
    = 
    \exp(t \genmat) \probvecttime{0},
    \quad t \ge 0
    .
\end{equation}
That is, probability distribution at time $t$ is given by the application of the time evolution operator of the system, $\exp(t \genmat)$, to the initial probability distribution $\probvecttime{0}$.
An important property of this operator is that when $\genmat$ is irreducible\footnote{
    A matrix is \emph{irreducible} when there is no permutation matrix, $\permutmat$, that can transform it into a block upper triangular form,
    $\permutmat \genmat \transpose{\permutmat}
    =
    \Big(
    \begin{array}{cc}
        A & B
        \\
        0 & D
    \end{array}
    \Big)
    $. 
    This means that every state is accessible from every other state, indicating the ergodicity of the dynamics.
}---which is the case in most systems of interest---, all its elements can be shown to be positive, i.e., $\confbra \exp(t \genmat) \ket{\confp} > 0,\, \forall \conf, \confp$ \cite{Norris_1997}.
Therefore, the operator obeys Perron-Frobenius theorem and we know: (i) that its largest eigenvalue is positive and non-degenerate, (ii) that the corresponding left and right eigenvectors only have positive entries, (iii) that such eigenvectors are the only non-negative ones \cite{berman_nonnegative_1979}.

This time evolution operator $\exp(t\genmat)$ shares eigenvectors with the generator $\genmat$, so that if the spectral decomposition\footnote{
    In most cases of interest, the generator is either completely diagonalizable or diagonalizable when restricted to the eigenspaces associated with the largest eigenvalues, which control its long-time behavior.
} of the latter is 
\begin{equation}
    \label{eq:genmat_specdecomp_theory}
    \genmat
    =
    \sum_j \eigval{j} \eigvectR{j}\eigvectL{j}
    ,
\end{equation}
we can rewrite Eq.~\eqref{eq:mastereqsol_theory} as
\begin{equation}
    \label{eq:timeevolop_specdecomp_theory}
    \probvectt
    =
    \sum_j \expsup{t \eigval{j}} \eigvectR{j}\braket{\eigvectLnk{j}}{\probvecttimenk{0}}
    ,
\end{equation}
where $\eigvectR{j}$ and $\eigvectL{j}$ are the $j$-th right and left eigenvectors\footnote{
    In general, $\genmatT$ is not symmetric, so their left and right eigenvectors are different.
} of $\genmat$, obeying $\braket{\eigvectLnk{i}}{\eigvectRnk{j}}=\delta_{ij}$, while $\eigval{j}\in\mathbb{C}$ are the corresponding eigenvalues sorted in decreasing value of their real part.
From this equation and the previous discussion of Perron-Frobenius theorem, we can identify $\flatvect = (1, \ldots, 1)$ as the positive left eigenvector associated with the largest eigenvalue.
This implies that the largest eigenvalue is $\eigval{0} = 0$ (the one corresponding to $\flatvect$), while the rest are negative, $\eigval{j} < 0$ for $j > 0$.
Therefore, if we take the long-time limit $t\to\infty$ in Eq.~\eqref{eq:timeevolop_specdecomp_theory}, all the contributions of the eigenvectors with $j>0$ will decay, leaving $\eigvectR{0}$ as the unique steady state of the system.

Note that, except for $j=0$ where the choice $\eigvectL{0} = \flatvect$ along the orthonormality condition uniquely determines $\eigvectR{0}$, the rest of eigenvectors pairs are only specified up to a complex number.
In order to further define the eigenvectors with $j>0$, it will prove useful to impose $\max_{\conf} | \braket{\eigvectLnk{j}}{\conf}| = 1$, leaving only the freedom of a complex phase per each pair of eigenvectors.
}

\section{Capturing nonequilibrium behavior: dynamical observables}
{
\label{sec:dynamicalobs_trajensembles_theory}
\newcommand{\ntransitiontraj}{m}

As we mentioned in Chapter~\ref{chap:intro}, our focus will be set on the trajectory statistics of the stochastic models introduced in the previous section.
The study of trajectories and their fluctuations holds the key to understanding systems far from equilibrium and developing a solid framework for their study.

We will begin by introducing the probability distribution of a trajectory. 
In a continuous-time Markov process, a trajectory $\traj\equiv\{(\conf_i,\timev_i)\}_{i=0}^{\ntransitiontraj}$ of duration $\trajduration$ is completely specified by the sequence of configurations visited by the system, $\{\conf_i\}_{i=0}^{\ntransitiontraj}$, and the times when the transitions occur, $\{\timev_i\}_{i=1}^{m}$, 
\begin{equation}
    \traj :\; \conf_0 \xrightarrow{\timev_1} \conf_1 \xrightarrow{\timev_2} \conf_2 \xrightarrow{\timev_3} ( \cdots ) \xrightarrow{\timev_{m}} \conf_{m} \, ,
    \label{eq:traj_theory}
\end{equation}
with $m$ being the number of transitions throughout the particular trajectory and  $\timev_0=0$ the time origin.
From the dynamics of the Markov process, the probability of such a trajectory\footnote{
    Although this is customarily called ``probability'' of the trajectory, it is actually a probability distribution that is continuous on the transition times $\{\timev_i\}_{i=1,...,m}$ and discrete on the visited configurations $\{\conf_i\}_{i=0,1,...,m}$ and the number of transitions $m$.
}
is given by 
\begin{equation}
    \label{eq:traj_prob_theory}
    \prob[\traj]
    =
    \expsup{-(\trajduration - \timev_{\ntransitiontraj})\escaperate{\conf_{\ntransitiontraj}}}
    \left[\prod_{i=0}^{\ntransitiontraj-1} \transitionrate{\conf_i}{\conf_{i+1}} \expsup{-(\timev_{i+1} - \timev_{i})\escaperate{\conf_i}}\right]
     \prob(\conf_0, 0)
     ,
\end{equation}
where the factors in the product correspond to the probability density of remaining at the state $\conf_i$ from $\timev_i$ to $\timev_{i+1}$, $\escaperate{\conf_i}\exp(- \escaperate{\conf_i}(\timev_{i+1} - \timev_{i}) )$, and then jumping to the specific state $\conf_{i+1}$ with probability $\transitionrate{\conf_i}{\conf_{i+1}}/\escaperate{\conf_i}$.

A \emph{dynamical observable} is an observable that is evaluated on the trajectories of the system instead of just its instantaneous state.
In particular, we will be interested in time-extensive observables in the trajectories of jump processes, which may be written as
\begin{equation}
    \label{eq:TEobservable_def_theory}
    \TEobservable [\traj]
    =
    \sum_{i=0}^{\ntransitiontraj} (\timev_{i+1} - \timev_i) \confdepobservable(\conf_i) + \sum_{i=0}^{\ntransitiontraj - 1} \transdepobservable{\conf_i}{\conf_{i+1}}\,
    ,
\end{equation}
where we have defined $\timev_{\ntransitiontraj+1} = \trajduration$.
This observable may include both the time integral of configuration-dependent observables, $\confdepobservable(\conf_i)$, and the sum of observables depending on the transitions occurring along the trajectory, $\transdepobservable{\conf_i}{\conf_{i+1}}$.
To clarify this definition, let us consider some examples.
If we were interested in the large-deviation statistics of the time-integrated current in a lattice gas, we would set $\confdepobservable(\conf_i)=0$ and $\transdepobservable{\conf_i}{\conf_{i+1}}=\pm 1$ depending on the direction of the particle jump (or we would use $\transdepobservable{\conf_i}{\conf_{i+1}}=1$ if we are interested in the kinetic activity).
Conversely, to study the statistics of the time-integrated energy of the system, we would set $\transdepobservable{\conf_i}{\conf_{i+1}}=0$ and define $\confdepobservable(\conf_i)$ as the energy of configuration $\conf_i$.
These dynamical observables and their statistics play a fundamental role in the characterization of nonequilibrium systems, as they are able to encode the fundamental time- and space-correlations characterizing their dynamics.

\section{Thermodynamics of trajectories}

Following the idea of the previous section, we explore the probability distribution of the time-intensive dynamical observable $\TIobservable = \TEobservable/\trajduration$ over trajectories of duration $\trajduration$
\begin{equation}
    \label{eq:P_TEobs_typical_theory}
    \prob_\trajduration (\TIobservable)
    =
    \sum_{\traj} \prob[\traj] \delta(\TIobservable - \TEobservable[\traj]/\trajduration)
    ,
\end{equation}
where the delta $\delta(\cdot)$ is a Dirac delta.
This corresponds to a sum over those trajectories in which the dynamical observable takes the value $\TEobservable$. 
This set of trajectories, $\{\traj \mid \TEobservable[\traj] = \trajduration\TIobservable\}$, defines the \emph{constrained ensemble} \cite{evans_rules_2004,chetrite13a}, which controls the statistics of a fluctuation where the dynamical observable takes the value $\TIobservable$.
The probability of a trajectory in this ensemble is given by
\begin{equation}
\label{eq:traj_probC_theory}
    \probCa[\traj]
    =
    \left\{
    \begin{array}[]{cl}
        \prob[\traj] / \prob_{\trajduration}(\TIobservable)
        & \textrm{if}\; \TEobservable[\traj] = \trajduration\TIobservable 
        ,
        \\
        0 
        & \textrm{otherwise}
        .
    \end{array}
    \right.
\end{equation}
This ensemble plays a role analogous to the microcanonical ensemble in equilibrium, with the dynamical observable $\TEobservable$ in the place of the energy.

In general, the distribution $\prob_{\trajduration}(\TIobservable)$ is expected to obey a large deviation principle for long times \cite{hurtado14a, bertini15a, derrida07a},
\begin{equation}
    \label{eq:LDP_theory}
    \prob_{\trajduration}\left( \frac{\TEobservable}{\trajduration} = \TIobservable \right)
    \asymp
    \text{e}^{-\trajduration \LDF(\TIobservable)}\, ,
\end{equation}
where ``$\asymp$'' stands for the logarithmic asymptotic equality introduced in Section~\ref{sec:LD_thermostatistics_intro},
\begin{equation}
    \label{eq:LDF_log_asymp_theory}
    \lim_{\trajduration \to \infty} -\frac{1}{\trajduration} \ln\prob_{\trajduration}(\TIobservable)
    =
    \LDF(\TIobservable)
    \, .
\end{equation}
The function $\LDF(\TIobservable)$ is the large deviation function (LDF) of the dynamical observable, and it controls its statistics arbitrarily far from the average value for long times $\trajduration$.
It is non-negative and equal to zero only for the average or mean value of the observable, $\LDF(\langle \TIobservable \rangle) = 0$.
Therefore, it measures the exponential rate at which the probability of fluctuations away from $\langle \TIobservable\rangle$ decay as $\trajduration$ increases.
In addition, Eq.~\eqref{eq:LDF_log_asymp_theory} shows that the LDF plays a role akin to the entropy per particle in the microcanonical ensemble, with the thermodynamic limit replaced by the long-time limit $\trajduration \to \infty$.
However, as in the case of the entropy, its direct evaluation using Eq.~\eqref{eq:P_TEobs_typical_theory} poses a formidable challenge, as it requires the explicit calculation of $\prob_{\trajduration}(\TIobservable)$---a task only feasible in highly simplified models \cite{bodineau04a,bodineau06a,bodineau10a,bertini07a,gorissen12b,lazarescu15a}.
Therefore, a different approach is required to characterize the fluctuations of the dynamical observable. 

To overcome this difficulty, we employ the same strategy as in equilibrium, where we preferred to use the canonical ensemble even when dealing with isolated systems, as the lack of an energy constraint made computations significantly easier. 
Then, using the microcanonical-canonical equivalence, we brought the results back to the isolated system.
The same approach can be used when dealing with the fluctuations of dynamical observables. 
Instead of relying on the constrained ensemble of Eq.~\eqref{eq:traj_probC_theory}, it is more convenient to define a \emph{biased ensemble}---also known as \emph{s-ensemble}---in which the dynamical observable is allowed to fluctuate under the influence of a \emph{biasing field} $\biasingfield$ \cite{lecomte07c, garrahan07a, garrahan09a, jack2010}.
In this new ensemble, the probability of a trajectory is given by
\begin{equation}
    \label{eq:traj_probT_theory}
    \prob^{\biasingfield}[\traj]
    =
    \frac{\prob[\traj] \text{e}^{\biasingfield A(\traj)}} {\DPF_{\trajduration}(\biasingfield)}
    \, ,
\end{equation}
where the normalizing factor $\DPF_{\trajduration}(\biasingfield)$ is the \emph{dynamical partition function} of the biased ensemble, $Z_{\tau}(\lambda)= \avg{\expsup{\biasingfield A(\traj)}}$, with $\avg{\cdot}$ the average over the original distribution $\prob[\traj]$.
The biasing field $\biasingfield$ is conjugated to the time-integrated observable in a way similar to the inverse temperature and energy in equilibrium systems.
Positive values of $\biasingfield$ bias the dynamics towards values of $A$ larger than its average, while negative ones make the average of the observable smaller.
From Eq.~\eqref{eq:traj_probT_theory}, the average of a trajectory-dependent observable $\observable(\traj)$ in the biased ensemble  is given by
\begin{equation}
\label{eq:avgT_observable_theory}
    \avgT{\observable}
    =
    \frac{\avg{\observable~\text{e}^{\biasingfield A}}}{\DPF_{\trajduration}(\biasingfield)}
    \, .
\end{equation}

The scaled cumulant generating function (sCGF) of the original distribution is deeply related to the dynamical partition function $\DPFwa$ of the biased ensemble.
Indeed, using $\biasingfield$ as the argument of the sCGF, we see that the latter is fully determined by $\DPFwa$,
\begin{equation}
\label{eq:sCGF_from_DPF_theory}
    \sCGF(\biasingfield)
    =
    \lim_{\trajduration\to\infty} \frac{1}{\trajduration} \ln \avg{e^{\biasingfield \TEobservable}}
    =
    \lim_{\trajduration\to\infty} \frac{1}{\trajduration} \ln \DPF_{\trajduration}(\biasingfield)
    .
\end{equation}

This suggests that the sCGF plays the role of a \emph{dynamical free energy}, as its relation with $\DPF_{\trajduration}(\biasingfield)$ is analogous to the one between the Helmholtz free energy and the partition function in the canonical ensemble. 
When the sCGF is differentiable for all $\biasingfield \in \realnumbers$, we can use the Gärtner-Ellis theorem to obtain $\LDF(\TIobservable)$ as the Legendre-Fenchel transform of $\sCGF(\biasingfield)$, 
\begin{equation}
\label{eq:GE_LDFsCGF_dynamical_theory}
    \LDF(\TIobservable)
    =
    \sup_{\biasingfield\in\realnumbers}\left\{\biasingfield\TIobservable - \sCGF(\biasingfield)\right\}
    ,
\end{equation}
which is analogous to the Legendre transform between the free energy and the entropy in equilibrium.
This establishes the desired connection between the ensembles:
if we are interested in a fluctuation such that the observable's value is $\TIobservable$, we can calculate the sCGF through the dynamical partition function of the biased ensemble and then use Legendre-Fechel transform in Eq.~\eqref{eq:GE_LDFsCGF_dynamical_theory} to obtain $\LDF(\TIobservable)$.
In addition, it can be shown that the two ensembles become equivalent in the long-time limit $\trajduration\to\infty$ under the conditions of the theorem \cite{chetrite13a,chetrite15b}.
In particular, if we are interested in a particular fluctuation in which the time-intensive observable takes the value $\TIobservable$, the corresponding $\biasingfield$ in the biased ensemble is given by the relation $\sCGF'(\biasingfield) = \avgT{\TEobservable/\trajduration}$, which can be readily obtained from Eq.~\eqref{eq:avgT_observable_theory}.

A great advantage of the biased ensemble is that its statistics can be readily calculated using the so-called \emph{tilted} generator, $\genmatT$.
In order to identify this generator, we go back to Eq.~\eqref{eq:traj_probT_theory} giving the probability of a trajectory in the biased ensemble, and we use Eq.~\eqref{eq:TEobservable_def_theory} to decompose $\expsup{\biasingfield\TEobservable(\traj)}$ into the contributions of each transition.
Absorbing the different factors into the corresponding rates, we obtain
\begin{equation}
    \label{eq:traj_probT_tiltedgenmat_theory}
    \probT[\traj]    
    =
    \DPF_{\trajduration}^{-1}(\biasingfield) \expsup{-(\trajduration - \timev_{\ntransitiontraj}) \escaperateT{\conf_{\ntransitiontraj}}}
    \left[\prod_{i=0}^{\ntransitiontraj-1} \transitionrateT{\conf_i}{\conf_{i+1}} \expsup{-(\timev_{i+1} - \timev_{i})\escaperateT{\conf_i}}\right]
    \prob(\conf_0, 0)
    ,
\end{equation}
where the \emph{tilted} transition and escape rates are given by
\begin{align}
    \transitionrateT{\conf_i}{\conf_{i+1}}
    &=
    \transitionrate{\conf_i}{\conf_{i+1}}\expsup{\biasingfield\transdepobservable{\conf_i}{\conf_{i+1}}}
    ,
    \\
    \escaperateT{\conf_i}
    &=
    \escaperate{\conf_i} - \biasingfield\confdepobservable(\conf_i)
    .
\end{align} 
This expression is analogous to Eq.~\eqref{eq:traj_prob_theory} but with the normalizing factor $\DPF_{\trajduration}^{-1}(\biasingfield)$ and with the original generator replaced by the tilted generator:
\begin{equation}
\label{eq:genmatT_theory}
    \genmatT
    = 
    \sum_{\conf}\sum_{\confp\neq\conf} \expsup{\biasingfield \transdepobservable{\conf}{\confp}}\transitionrate{\conf}{\conf^{\prime}} \ket{\conf^{\prime}} \bra{\conf}
    - \sum_{\conf} (\escaperate{\conf} - \biasingfield\confdepobservable(\conf)) \ketbra{\conf}
    .
\end{equation}
Intuitively, the exponential bias in the transitions and the extra term in the diagonal push the dynamics toward transitions and states that increase or decrease (according to the sign of $\biasingfield$), the considered dynamical observable.
However, except for $\biasingfield = 0$ where the original generator is recovered, $\genmatT$ does not conserve probability, i.e., $\flatvect \genmatT \neq 0$.
This implies that it is not possible to directly retrieve the physical trajectories leading to the fluctuation from the tilted generator, since $\genmatT$ does not represent a physical dynamics. 
Still, the tilted generator contains the information about the fluctuations of parameter $\biasingfield$. 
This is first seen by noting that, summing over all possible trajectories in Eq.~\eqref{eq:traj_probT_tiltedgenmat_theory}, the dynamical partition function can be calculated from the tilted generator as \cite{lecomte07c,garrahan09a}
\begin{equation}
    \label{eq:DPF_operatordef}
    \DPF_{\trajduration}(\biasingfield)
    =
    \flatvect \expsup{\trajduration \genmatT} \probvecttime{0}
    \, .
\end{equation}

The relevance of the tilted generator can be further seen by introducing its spectral decomposition.
Let $\eigvectTR{j}$ and $\eigvectTL{j}$ be the $j$-th right and left eigenvectors of $\genmatT$ and $\eigvalT{j}\in\mathbb{C}$ the corresponding eigenvalues.
Assuming that $\genmatT$ is diagonalizable, its spectral decomposition reads
\begin{equation}
    \label{eq:genmatT_specdecomp_theory}
    \genmatT
    =
    \sum_j \eigvalT{j} \eigvectTR{j}\eigvectTL{j}
    ,
\end{equation}
with the set of left and right eigenvectors forming a complete biorthogonal basis of the vector space, such that $\braket{L^{\lambda}_i}{R^{\lambda}_j}=\delta_{ij}$.
Sorting eigenvalues in decreasing order of the real part, it is straightforward to show from Eqs.~\eqref{eq:sCGF_from_DPF_theory} and \eqref{eq:DPF_operatordef} that the sCGF or dynamical free energy is given by the largest eigenvalue of $\genmatT$,
\begin{equation}
    \DFE(\biasingfield)
    =
    \lim_{\trajduration\to\infty} \frac{1}{\trajduration} \ln \flatvect \expsup{\trajduration \genmatT} \probvecttime{0}
    =
    \eigvalT{0}
    .
\end{equation}
This confirms the importance of the tilted generator $\genmatT$ in the characterization of the fluctuations of the observable $\TEobservable$.
However, it remains to be determined how to sample trajectories in the biased ensemble, a question that will be addressed in the next section.

\section{The Doob generator: unveiling the dynamics of a fluctuation}
\sectionmark{Doob generator} 
\label{sec:Doobtransform_theory}

The question that remains is how to use $\genmatT$ to retrieve the trajectories giving rise to the fluctuation of interest.
To achieve this, a natural approach is the use of cloning Monte Carlo methods \cite{giardina06a,tailleur09a,giardina11a,hurtado14a}.
In such methods, the lack of probability conservation in the generator $\genmatT$ is interpreted as describing population dynamics.
It considers a large number of copies of the system evolving under normalized dynamics, which, after each transition, clone or erase themselves according to the tilted rates.
The trajectories replicating the most correspond to the typical ones in the considered fluctuation, and the exponential growth in the number of copies can be used to measure the sCGF $\sCGF(\biasingfield)$.
While this technique is very powerful, its implementation and interpretation is far from simple. 

A different approach is to introduce the so-called Doob-transformed generator \cite{simon2009,jack2010,popkov10a,chetrite15a}, which will be the main approach used in this Thesis.
This approach is based on a generalization of Doob's $h$-transform introduced in the context of bridge processes \cite{doob57a,doob_classical_1984}.
It defines a \emph{probability conserving} auxiliary process built on $\genmatT$ which, as we will show, describes the trajectories of a fluctuation associated with the parameter $\biasingfield$.
Given the tilted generator matrix $\genmatT$ of a process, its \emph{Doob-transformed generator} is defined as
\begin{equation}
\label{eq:genmatD_theory}
    \genmatD
    =
    \eigvectTLop{0} \genmatT (\eigvectTLop{0})^{-1} - \eigvalT{0} \identityop\, ,
\end{equation}
where $\identityop$ represents the identity matrix and $\eigvectTLop{0}$ is a diagonal matrix whose elements $(\eigvectTLop{0})_{ii}$ are the $i$-th entries of the left eigenvector $\eigvectTL{0}$.
From its definition, we can readily check that this new generator is probability-conserving.
To do so, we notice that the spectra of the generators $\genmatT$ and $\genmatD$ are simply related by a shift in their eigenvalues,
\begin{equation}
    \eigvalD{j} = \eigvalT{j} - \eigvalT{0}
    ,
\end{equation}
and a simple transformation of their left and right eigenvectors, 
\begin{align}
    \eigvectDL{j}
    &=
    \eigvectTL{j} (\eigvectTLop{0})^{-1}
    ,
    \\
    \eigvectDR{j}
    &=
    \eigvectTLop{0} \eigvectTR{j}
    .
\end{align}
As a consequence, the leading eigenvalue of $\genmatD$ becomes zero and its associated leading left eigenvector becomes $\eigvectDL{0} = \eigvectTL{0} (\eigvectTLop{0})^{-1} = \flatvect$, confirming that the Doob generator conserve probability, $\flatvect \genmatD = 0$.
In addition, the leading right eigenvector given by $\eigvectDR{0} = \eigvectTLop{0} \eigvectTR{0}$, becomes the stationary state of the Doob dynamics. 
It can be checked that the left and right eigenvectors of the Doob generator also form a complete biorthogonal basis of the Hilbert space.
As with the standard generators in Section~\ref{sec:markovprocesses_theory}, we normalize these eigenvectors by imposing $\max_{\conf} | \braket{\eigvectDLnk{j}}{\conf}| = 1$ and  $\braket{\eigvectDLnk{i}}{\eigvectDRnk{j}}=\delta_{ij}$.

While we have shown that this generator conserves probability, it remains to be proven that it generates the dynamics of the trajectories in the $\lambda$-ensemble. 
In the following, we will introduce an intuitive proof to establish this connection.
We will calculate the jump rates of the $\biasingfield$-ensemble trajectories and we will show that they correspond---in a particular time regime---to the ones of $\genmatD$.
For the sake of brevity and clarity, we will skip technical subtleties, but we refer the reader to \cite{chetrite15a} for a more throughout proof.

To find the transition rates, we will start by calculating the transition probability from $\conf$ to $\confp$ in the interval $[\timev, \timev + \deltat]$ in the biased ensemble and then we will take the limit of small $\deltat$.
This probability, denoted by $\transitionprobTWA$, is calculated as the joint probability of states $\conf$ and $\confp$ at times $\timev$ and $\timev+\deltat$ conditioned to being at state $\conf$ at time $\timev$, i.e., 
\begin{equation}
    \transitionprobTWA
    =
    \frac{\probT(\conf,\, \confp;\timev,\,\timev+\deltat)}{\probT(\conf, \timev)}
    .
\end{equation}
The numerator can be calculated from Eqs.~\eqref{eq:traj_probT_theory} and \eqref{eq:genmatT_theory} by integrating over all possible trajectories under the restriction of having $\conf$ and $\confp$ at times $\timev$ and $\timev+\deltat$.
In vector notation, this is given by
\begin{equation}
    \label{eq:probT_CtCt_theory}
    \probT(\conf,\, \confp;\timev,\,\timev+\deltat)
    =
    \frac{\flatvect \expsup{\genmatT(\trajduration-\timev-\deltat)}\ketbra{\confp}\expsup{\genmatT\deltat}\ketbra{\conf}\expsup{\genmatT\timev}\probvecttime{0}} {\DPFwa}
    .
\end{equation}
The denominator is calculated analogously,
\begin{equation}
    \label{eq:probT_Ct_theory}
    \probT(\conf,\timev)
    =
     \frac{\flatvect \expsup{\genmatT(\trajduration-\timev)}\ketbra{\conf}\expsup{\genmatT\timev}\probvecttime{0}}{\DPF_{\trajduration}(\biasingfield)}
    ,
\end{equation}
so that the transition probability is given by
\begin{equation}
    \transitionprobTWA
    =
    \frac{\flatvect \expsup{\genmatT(\trajduration-\timev)}\expsup{-\genmatT\deltat}\ketbra{\confp}\expsup{\genmatT\deltat} \ket{\conf}}
        {\flatvect \expsup{\genmatT(\trajduration-\timev)} \ket{\conf}}
    .
\end{equation}
Now we consider a small enough $\deltat$ to do a linear approximation of the exponentials to obtain
\begin{align}
    \transitionprobTWA
    \simeq
    \left\{
    \begin{array}{cc}
        1 - \deltat \left( \escaperateT{\conf}
            + \dfrac{\flatvect\expsup{\genmatT(\trajduration-\timev)}\genmatT\confket}{\flatvect\expsup{\genmatT(\trajduration-\timev)}\confket} \right)
        & \textrm{if $\confp = \conf$,}
        \\
        \deltat\, \transitionrateT{\conf}{\confp}
             \dfrac{\flatvect\expsup{\genmatT(\trajduration-\timev)}\ket{\confp}}{\flatvect\expsup{\genmatT(\trajduration-\timev)}\confket},
        & \textrm{if $\confp \neq \conf$.}
    \end{array}
    \right.
\end{align}
This equation allows us to identify the transition rates of the trajectory at time $\timev$ as
\begin{align}
    \tilde{\escaperatesymb}_{\conf}^{\biasingfield}(\timev)
    &=
    \escaperateT{\conf} + \dfrac{\flatvect\expsup{\genmatT(\trajduration-\timev)}\genmatT\confket}{\flatvect\expsup{\genmatT(\trajduration-\timev)}\confket}
    , \nonumber
    \\
    \label{eq:transratesT_timet_theory}
    \tilde{\transitionratesymb}_{\conf\to\confp}^{\biasingfield} (\timev)
    &=
    \transitionrateT{\conf}{\confp} \dfrac{\flatvect\expsup{\genmatT(\trajduration-\timev)}\ket{\confp}}{\flatvect\expsup{\genmatT(\trajduration-\timev)}\confket},
\end{align}
where we have used the tilde to distinguish the transition rates from the elements of the tilted generator---which do not represent physical rates.
These transition rates are in general time dependent.
However, if we focus on long trajectories and times $\timev$ such that $\trajduration - \timev$ is large, the exponentials in Eq.~\eqref{eq:transratesT_timet_theory} are dominated by their first eigenvectors, i.e.,
\begin{equation}
    \expsup{\genmatT(\trajduration-\timev)}
    \simeq
    \expsup{\eigvalT{0}(\trajduration-\timev)}\eigvectTR{0}\eigvectTL{0}
    ,
\end{equation}
so that the rates become 
\begin{align}
    \escaperateD{\conf}
    &=
    \escaperateT{\conf} + \eigvalT{0}
    ,
    \\
    \transitionrateD{\conf}{\confp}
    &=
    \transitionrateT{\conf}{\confp} \frac{\braket{\eigvectTLnk{0}}{\confp}}{\braket{\eigvectTLnk{0}}{\conf}}
    ,
\end{align}
which are now time-independent.
As it can be readily checked, these rates are the ones appearing in the Doob generator in Eq.~\eqref{eq:genmatD_theory}.
Therefore, we have shown that the mid-time dynamics, i.e., those corresponding to times far enough from the boundaries $t_0$ and $\trajduration$ so that the system forgets $\probvecttime{0}$ and is not affected by the time dependence of Eq.~\eqref{eq:transratesT_timet_theory}, are time-independent and follow the Doob generator $\genmatD$.
Since we are interested in the long-time limit $\trajduration\to\infty$, this mid-time dynamics will dominate the statistics of the biased ensemble, allowing us to use Doob dynamics for its analysis.

     }

    \chapter{A theory of symmetry-breaking dynamical phase transitions} 
    \chaptermark{A theory of symmetry-breaking DPTs}
    \label{chap:SBDPTs}
    {


\newcommand{\phaseweight}[1]{w_{#1,\probvecttimenk{0}}}

\section{Introduction}

\todo[inline]{Esté capítulo necesita figuras como el comer}
\todo[inline]{Revisar artículo con calma para buscar "Appendix" "paper" y demás. }
\todo[inline]{También hay que estructurarlo un poco más, los apéndices estánn simplemente pegados en medio}
\todo[inline]{Comprobar que el concepto: symmetry-eigenvale se ha introducido}
    
   
One of the most intriguing phenomena which have gained attention in the last two decades is dynamical phase transitions (DPTs) \cite{bertini05a,bodineau05a,bertini06a,jack15a,perez-espigares18b,perez-espigares19a,banuls19a}.
Unlike standard phase transitions, which occur when modifying a physical parameter, these might occur when a system sustains an atypical value, i.e., a rare fluctuation, of a trajectory-dependent observable.
These DPTs are accompanied by a drastic change in the structure of those trajectories responsible for such fluctuation, and they are revealed as non-analyticities in the associated large deviation functions, which, as we have noted in the previous chapters, play the role of thermodynamic potentials for nonequilibrium settings \cite{touchette09a}.
In this context, a myriad of emerging structures associated with DPTs have been discovered, including symmetry-breaking density profiles \cite{baek17a,baek18a,perez-espigares18a}, localization effects \cite{gutierrez21a}, condensation phenomena \cite{chleboun17a} or traveling waves \cite{hurtado11a,perez-espigares13a,tizon-escamilla17b}.
Moreover, DPTs have been also predicted and observed in active media \cite{cagnetta17a,whitelam18b,tociu19a,gradenigo19a,nemoto19a,cagnetta20a,chiarantoni20a,fodor20a,grandpre21a,keta21a,yan22a}, where individual particles can consume free energy to produce directed motion, as well as in many different open quantum systems \cite{flindt09a,garrahan10a,garrahan11a,ates12a,hickey12a,genway12a,flindt13a,lesanovsky13a,maisi14a,manzano14a,manzano18a,manzano21a}.
An interesting aspect of these DPTs is that, as in equilibrium phase transitions, many involve the spontaneous breaking of a symmetry.
\todo[inline]{
    Often, this corresponds to a $\mathbb{Z}_n$ symmetry, which can be understood as an invariance under particular discrete rotations of angles $2\pi/m$ with $m=1,2,...,n$ in the order parameter space.
}
In these phase transitions, the trajectories adopted by the system to give rise to the fluctuation stop obeying the underlying symmetry, in an apparent violation of the rule of the dynamics.
However, the symmetry is still displayed in the dynamics by the appearance of a number of new equiprobable trajectories realizing the same fluctuation.
\todo[inline]{In addition, such trajectories have the properties of being connected by the action of the symmetry.
As we will see, these simple realizations have enormous consequences on the special properties of the generators governing the dynamics of the systems undergoing the transition.
}
     

From a spectral perspective, the hallmark of a symmetry-breaking DPT is the emergence of a collection of Doob eigenvectors with a vanishing spectral gap \cite{gaveau98a,hanggi82a,ledermann50a,gaveau06a,minganti18a}.
The degenerate subspace spanned by these vectors
defines the stationary subspace of the Doob stochastic generator, so that the typical states responsible for a given fluctuation in the original system can be retrieved from these degenerate Doob eigenvectors.
This is similar to what is found in standard phase transitions, where the equivalence between the emergent degeneracy of the leading eigenspace of the generator and the appearance of a phase transition has been demonstrated \cite{gaveau98a,gaveau06a}.
\todo[inline]{También sería buena idea resaltat que existe un problema y resolverlo}

\todo[inline]{
    a common underlying spectral mechanism giving rise to the different dynamical phases when was still lacking.

    a discrete $\mathbb{Z}_n$ symmetry is broken is k

    the general structure of the Doob eigenvectors, their relation to the underlying symmetry of the stochastic generator, and the particular mechanism of symmetry breaking have been elusive so far.
    Previous works have focused on particular models, but the common underlying spectral mechanism giving rise to the different dynamical phases when a discrete $\mathbb{Z}_n$ symmetry is broken is still lacking.
}
\todo[inline]{Hay que cambiar a partir de aquí abajo para hacerlo menos terrible}

\todo[inline]{Hay que reescribir esto establenciendo la estructura del capítulo}

\todo[inline]{Remarkably, the above-described symmetry-breaking spectral mechanism demonstrated here, for DPTs in the large deviation statistics of time-averaged observables, is completely general for $\mathbb{Z}_n$-invariant systems and expected to hold valid also in standard (non anomalous) critical phenomena \cite{gaveau98a,gaveau06a,binney92a}.}
   
\todo[inline]{check aand update/rewrite}
\todo[inline]{The chapter is organized as follows.}

\section{$\mathbb{Z}_n$ symmetries on Markov processes}
\todo[inline]{Pensar título alternativo}
\label{ssecSym}

As previously noted, our interest in this chapter is focused on DPTs involving the spontaneous breaking of a cyclic $\mathbb{Z}_n$ symmetry.
Therefore, some general remarks about such symmetries in stochastic processes are in order.
The symmetry group of a stochastic process
is the set of transformations $\symmop$ that leave the process invariant (in a sense to be described below).
In particular, the \emph{cyclic group} of order $n$, $\Zsymm{n}$, is the symmetry group built from the repeated application of a single operator $\symmop\in \mathbb{Z}_n$, satisfying $\symmop^n=\identityop$.
When dealing with discrete state spaces, this operator is unitary, invertible and stochastic (i.e. with non-negative elements and $\bra{-}\hat{S}=\bra{-}$), acting as a permutation between the different states of the system \cite{hanggi82a}, 
\begin{equation}
    \symmop\ket{C} = \ket{SC}
    .
\end{equation}

We say a stochastic process, as defined by a Markov generator $\genmat$, is invariant under a symmetry transformation $\symmop$ when $\commutator{\genmat}{\symmop} = 0$, or equivalently 
\begin{equation}
\label{symcond}
    \genmat
    =
    \symmop \genmat \symmop^{-1}\, .
\end{equation}
This means that $\genmatD$ and $\symmop$ share a common eigenbasis, i.e. they can be diagonalized at the same time, so that the eigenvectors $\eigvectL{j}$ and $\eigvectR{j}$ are also eigenvectors of $\symmop$. 
In addition, due to the unitarity and cyclic character of $\symmop$, its eigenvalues simply correspond to the $n$ roots of unity $\text{e}^{i 2\pi k/n}$ with $k=0,1,...,n-1$.



In addition, the symmetry operator $\symmop$ induces a map $\cal{S}$ in trajectory space
\begin{eqnarray}
\nonumber
&\traj :\; \conf_0 \xrightarrow{\timev_1} \conf_1 \xrightarrow{\timev_2} \conf_2 \xrightarrow{\timev_3} ( \cdots ) \xrightarrow{\timev_{m}} \conf_{m} \\ \nonumber
&\mathcal{S}\big\downarrow \\ 
&\mathcal{S}\traj :\; S\conf_0 \xrightarrow{\timev_1} S\conf_1 \xrightarrow{\timev_2} S\conf_2 \xrightarrow{\timev_3} ( \cdots ) \xrightarrow{\timev_{m}} S\conf_{m} \, ,
\label{Rconf}
\end{eqnarray}
which transforms the configurations visited along the path but leaves unchanged the transition times $\{t_i\}_{i=0,1,\ldots m}$ between configurations.
As it can be readily checked from Eq.~\eqref{eq:traj_prob_theory}, if $\symmop$ is a symmetry of $\genmat$, then the probabilities of the trajectories $\traj$ and $\mathcal{S}\traj$ remain the same.

This discussion illustrates some of the implications that symmetries have on the dynamics of $\genmat$.
However, we are interested in the dynamics associated with the fluctuations of a dynamical observable $\TEobservable[\traj]$, as given by Doob's generator $\genmatD$.
Therefore, in the section we will consider under which conditions a  symmetry of $\genmat$ is inherited by $\genmatD$.

\todo[inline]{Lo que hay abajo habría que borrarlo o pasarlo a la siguiente sección}
\todo[inline, size=\normalsize]{
    For the symmetry to be inherited by the Doob auxiliary process $\genmatD$ associated with the fluctuations of an observable $A$, see Eq.~\eqref{eq:genmatD_theory}, it is necessary that this trajectory-dependent observable remains invariant under the trajectory transformation, i.e. $A({\cal{S}}\omega_{\tau})=A(\omega_{\tau})$. 
    This condition, together with the invariance of the original dynamics under $\symmop$, see Eq.~\eqref{symcond}, crucially implies that both the tilted and the Doob generators are also invariant under $\symmop$, as demonstrated in Section~\ref{sec:inheritance_symmetry_SDPTs},
    \begin{equation}\genmatT = \symmop \genmatT \symmop^{-1}\, ,~~~~\genmatD = \symmop \genmatD \symmop^{-1}.
    \end{equation}As a consequence, both $\genmatD$ and $\symmop$ share a common eigenbasis, i.e. they can be diagonalized at the same time, so $\ket{R_{j,\text{D}}^{\lambda}}$ and $\bra{L_{j,\text{D}}^{\lambda}}$ are also eigenvectors of $\symmop$ with eigenvalues $\phi_j$, i.e. $\hat{S}\ket{R_{j,\text{D}}^{\lambda}}=\phi_j\ket{R_{j,\text{D}}^{\lambda}}$ and $\bra{L_{j,\text{D}}^{\lambda}}\hat{S}=\phi_j\bra{L_{j,\text{D}}^{\lambda}}$, designated as symmetry eigenvalues. Due to the unitarity and cyclic character of $\symmop$, the eigenvalues $\phi_j$ simply correspond to the $n$ roots of unity, i.e. $\phi_j=\text{e}^{i 2\pi k_j/n}$ with $k_j=0,1,...,n-1$.
}

\section{Symmetry inheritance in Doob's generator}
\todo[inline]{Cambiar el título, o al menos hacer una versión corta para la cabecera}
\label{sec:inheritance_symmetry_SDPTs}


We will start by analyzing the conditions under which a symmetry of $\genmat$ is inherited by the tilted generator $\genmatT$.
As explained in the previous section, the time-extensive observables $\TEobservable[\traj]$ whose fluctuations we want to consider might depend on the state of the process and its transitions over time. For jump processes as the ones considered here, such trajectory-dependent observables can be written in general as 
\begin{equation*}
    \label{eq:current_defA}
    A [\omega_{\tau}]
    =
    \sum_{i=0}^{m } (t_{i+1} - t_i)g(C_i) + \sum_{i=0}^{m - 1} \eta_{C_i,C_{i+1}}
    \, ,
\end{equation*}
see Eq.~\eqref{eq:TEobservable_def_theory} in Section~\ref{sec:dynamicalobs_trajensembles_theory}.
The first sum above corresponds to the time integral of configuration-dependent observables, $g(C_i)$, while the second one stands for observables that increase by $\eta_{C_i,C_{i+1}}$ in the transitions from $C_i$ to $C_{i+1}$ (we have defined $t_0=0$ and $t_{m+1}=\trajduration$).

Since the biased ensemble modifies the original ensemble with an exponential bias on $\TEobservable[\traj]$ and we already have that $\prob[\mathcal{S} \traj] = \prob[\traj]$, it seems reasonable to expect that the tilted generator $\genmatT$ will be symmetric under $\symmop$ if $\TEobservable[\mathcal{S}\traj] = \TEobservable[\traj]$.
Demanding $A[\omega_{\tau}]$ to remain invariant under the symmetry transformation for any trajectory implies that both the configuration-dependent $g(C)$ and the transition-dependent $\eta_{C,C'}$ functions are invariant under such transformation, i.e., that $g(C)=g(C_S)$ and $\eta_{C,C'}=\eta_{C_S,C'_S}$, with the definitions $\ket{C_S}=\hat{S}\ket{C}$ and $\ket{C'_S}=\hat{S}\ket{C'}$.
From this and the definition of $\genmatT$ in Eq.~\eqref{eq:genmatT_theory} we obtain the expected result,
\begin{equation*}
\begin{split}
    \symmop \genmatT \symmop^{-1}
    = 
    \sum_{\conf, \conf^{\prime} \neq \conf} &\text{e}^{\lambda \eta_{C,C'}}\transitionrate{\conf}{\conf^{\prime}} \symmop\ket{\conf^{\prime}} \bra{\conf}\symmop^{-1}
    \\ 
    - \sum_{\conf} &\escaperate{\conf} \symmop\ketbra{\conf}\symmop^{-1} + \lambda \sum_{\conf} g(C)\symmop\ketbra{\conf}\symmop^{-1}
    \\ 
    =
    \sum_{\conf_S, \conf^{\prime}_S \neq \conf_S} &\text{e}^{\lambda \eta_{C_S,C'_S}}\transitionrate{\conf_S}{\conf^{\prime}_S} \ket{\conf^{\prime}_S} \bra{\conf_S}
    \\ 
    - \sum_{\conf_S} &\escaperate{\conf_S} \ketbra{\conf_S} + \lambda \sum_{\conf_S} g(C_S)\ketbra{\conf_S} = \genmatT \, .
\end{split}
\label{eq:genmatTA}
\end{equation*}
Therefore we have that $[\genmatT,\hat{S}]=0$, provided the above conditions on observable $A$ hold.

Finally, in order to prove that the generator, $\genmatD = \eigvectTLop{0} \genmatT (\eigvectTLop{0})^{-1} - \eigvalT{0} \identityop$, also inherits this symmetry, 
we have to show that $\eigvectTLop{0}$ commutes with $\symmop$. 
To do so, we notice that since $\genmatT$ entries are non-negative, by virtue of Perron-Frobenius theorem the eigenvector $\eigvectTL{0}$ associated with the largest eigenvalue $\eigvalT{0}$ must be non-degenerate and with positive entries.
Moreover, as $\genmatT$ commutes with $\symmop$, $\eigvectTL{0}$ must also be an eigenvector of $\symmop$, $\eigvectTL{0} \symmop = \symmeigval{0} \eigvectTL{0}$.
Finally, because all components of both $\eigvectTL{0}$ and $\symmop$ are real and positive, we must have the same for $\symmeigval{0} \eigvectTL{0}$, which is only possible if $\symmeigval{0} = 1$.
In this way $\eigvectTL{0} \symmop = \eigvectTL{0}$, and the operator $\eigvectTLop{0}$ used in the Doob transform commutes with $\symmop$,
\begin{equation}
\begin{split}
\symmop \eigvectTLop{0} \symmop^{-1} &= \sum_{\conf} \symmop\ket{C} \braket{\eigvectTLnk{0}}{\conf} \bra{\conf}\symmop^{-1} \\
&= \sum_{C_S} \ket{C_S} \braket{\eigvectTLnk{0}}{C_S} \bra{C_S} = \eigvectTLop{0} \, , 
\end{split}
\end{equation}
where we have used in the second equality that $\braket{\eigvectTLnk{0}}{\conf}=\bra{\eigvectTLnk{0}}\symmop\ket{C}$.

Thus, we have shown that, as long as $\TEobservable[\mathcal{S}\traj] = \TEobservable[\traj]$, the invariance of the original generator $\genmat$ with respect to the symmetry operator $\symmop$, also implies the invariance of the Doob generator, i.e, $[\genmatD,\hat{S}]=0$.
As a consequence, both $\genmatD$ and $\symmop$ will also share a common eigenbasis, so that $\ket{R_{j,\text{D}}^{\lambda}}$ and $\bra{L_{j,\text{D}}^{\lambda}}$ are also eigenvectors of $\symmop$ with eigenvalues $\phi_j$, i.e. $\hat{S}\ket{R_{j,\text{D}}^{\lambda}}=\phi_j\ket{R_{j,\text{D}}^{\lambda}}$ and $\bra{L_{j,\text{D}}^{\lambda}}\hat{S}=\phi_j\bra{L_{j,\text{D}}^{\lambda}}$. 
In what follows we will assume that the Doob generator is symmetric under $\symmop$.

\todo[inline]{Creo que las tres siguintes secciones se podrían agrupar en una sola con subsecciones. Ahora, eso implicaría escribir un pequeño párrafo introductorio}
\section{Stationary state degeneracy in dynamical phase transitions}
\label{ssecDeg}

As it was shown in Section~\ref{sec:Doobtransform_theory}, the steady state associated with the Doob stochastic generator $\genmatD$ describes the statistics of trajectories during a large deviation event of parameter $\currenttilt$ of the original dynamics.
The formal solution of the Doob master equation for any time $\timev$ and starting from an initial probability vector $\ket{P_0}$ can be written as $\probTvecttime{\timev} = \exp(+ \timev \genmatD) \ket{P_0}$.
Using the spectral decomposition of this formal solution, we have
\begin{equation}
    \probTvecttime{t} = \eigvectDR{0} + \sum_{j>0} \text{e}^{\timev \eigvalD{j}} \eigvectDR{j} \braket{\eigvectDLnk{j}}{P_0}  \, ,
    \label{specdec}
\end{equation}
where we have used that the leading eigenvalue of Doob's generator is $\eigvalD{0} = 0$.
Furthermore, since $\braket{-}{\eigvectDRnk{j}} = \braket{\eigvectDLnk{0}}{\eigvectDRnk{j}} = \delta_{0j}$, all the probability of $\probTvecttime{t}$ is contained in $\ket{\eigvectDRnk{0}}$, i.e. $\braket{-}{\probTvecttimenk{t}}=\braket{-}{\eigvectDRnk{0}}=1$.
Thus, each term with $j>0$ 
in the r.h.s of Eq.~\eqref{specdec} corresponds to a different redistribution of the probability. 
Moreover, as the symmetry operator $\hat{S}$ conserves probability, we get $1 = \flatvect \symmop \eigvectDR{0} = \flatvect \symmeigval{0} \eigvectDR{0}$, i.e.
\begin{equation}
\phi_0=1 \, ,
\end{equation}
for the symmetry eigenvalue of the leading eigenvector, so that $\eigvectDR{0}$ is invariant under $\symmop$.

In order to study the steady state of the Doob dynamics, $\probTvectst\equiv \lim_{\timev\to\infty} \probTvecttime{\timev}$, we now define the spectral gaps as $\Delta^{\lambda}_j=\Re(\theta_0^{\lambda}-\theta_j^{\lambda})=-\Re(\eigvalD{j})\ge 0$, which control the exponential decay of the corresponding eigenvectors, cf. Eq.~\eqref{specdec}.
Remember that $0\le \Delta^{\lambda}_1\le \Delta^{\lambda}_2\le \ldots$ due to way we sorted the eigenvalues (in decreasing value of their real part) in Chapter~\ref{chap:theory}.
When $\Delta^{\lambda}_1$ is strictly positive, $\Delta^{\lambda}_1>0$, so that the spectrum is gapped (usually $\Delta^{\lambda}_1$ is referred to as \emph{the} spectral gap), all subleading eigenvectors decay exponentially fast for timescales $\timev \gg 1/\Delta^{\lambda}_1$ and the resulting Doob steady state is unique,
\begin{equation}\probTvectst =  \eigvectDR{0}.
\label{ssnopt}
\end{equation}
This steady state preserves the symmetry of the generator, $\symmop \probTvectst = \probTvectst$, so no symmetry-breaking phenomenon at the fluctuating level is possible whenever the spectrum of the Doob generator $\genmatD$ is gapped.
This is hence the spectral scenario before the emergence of any DPT.

Conversely, any symmetry-breaking phase transition at the trajectory level will demand an emergent degeneracy in the leading eigenspace of the associated Doob generator. This is equivalent to the spectral fingerprints of standard symmetry-breaking phase transitions in stochastic systems \cite{gaveau98a,hanggi82a,ledermann50a,gaveau06a,minganti18a}. As the Doob auxiliary process $\genmatD$ 
is indeed stochastic, these spectral fingerprints \cite{gaveau98a} will characterize also DPTs at the fluctuating level. In particular, for a many-body stochastic system undergoing a $\mathbb{Z}_n$ symmetry-breaking DPT, we expect that the difference between the real part of the first and the subsequent $n-1$ eigenvalues $\eigvalD{j}$ goes to zero in the thermodynamic limit once the DPT kicks in. In this case the Doob \emph{stationary} probability vector is determined by the first $n$ eigenvectors defining the degenerate subspace. Note that, in virtue of the Perron-Frobenius theorem, for any finite system size the steady state is non-degenerate, highlighting the relevance of the thermodynamic limit.

In general, the gap-closing eigenvalues associated with these eigenvectors may exhibit non-zero imaginary parts, $\Im(\theta_{j,\text{D}}^\lambda)\ne 0$, thus leading to a time-dependent Doob \emph{stationary} vector in the thermodynamic limit
\begin{equation}
\probTvectst (t) = \eigvectDR{0} + \sum_{j=1}^{n-1} \text{e}^{+i t \Im(\theta_{j,\text{D}}^\lambda)}\eigvectDR{j} \braket{\eigvectDLnk{j}}{\probvecttimenk{0}}\, .
\label{Psstime}
\end{equation}
Moreover, if these imaginary parts display band structure, the resulting Doob \emph{stationary} state will exhibit a periodic motion characteristic of a time crystal phase \cite{hurtado-gutierrez20a}, as we will analyze later in the particular example of Chapter~\ref{chap:buildingtc}.
However, in many cases the gap-closing eigenvalues of the Doob eigenvectors in the degenerate subspace are purely real, so $\Im(\theta_{j,\text{D}}^\lambda)= 0$ and the resulting Doob steady state is truly stationary,
\begin{equation}
\probTvectst = \eigvectDR{0} + \sum^{n-1}_{j=1}  \eigvectDR{j} \braket{\eigvectDLnk{j}}{\probvecttimenk{0}}\, .
\label{eq:DPT_probvect}
\end{equation}
The number $n$ of vectors that contribute to the Doob steady state corresponds to the different number of phases that appear once the $\Zsymm{n}$ symmetry is broken. Indeed, a $n$th-order degeneracy of the leading eigenspace implies the appearance of $n$ different, linearly independent stationary distributions \cite{hanggi82a,ledermann50a}, as we shall show below. 
As in the general time-dependent solution, Eq.~\eqref{specdec}, all the probability is concentrated on the first eigenvector $\eigvectDR{0}$, which preserves the symmetry, $\symmop\eigvectDR{0}=\eigvectDR{0}$, while the subsequent eigenvectors in the degenerate subspace describe the \emph{redistribution} of this probability according to their projection on the initial state, containing at the same time all the information on the symmetry-breaking process.
Notice that, even if the degeneration of the $n$ first eigenvalues is complete, we can still single out $\eigvectDR{0}$ as the only eigenvector with eigevalue $\phi_0=1$ under $\symmop$ (all the gap-closing eigenvectors have different eigenvalues under $\symmop$, as it is shown in Section~\ref{ssecPhase})\todo{adaptar esto a la nueva estructura (antes era un apéndice)}.
This means that the steady state Eq.~\eqref{eq:DPT_probvect} does not preserve in general the symmetry of the generator, i.e. $\symmop \probTvectst \ne \probTvectst$, and hence the symmetry is broken in the degenerate phase. The same happens for the time-dependent Doob asymptotic state Eq.~\eqref{Psstime}.

\section{Calculating the phase probability vectors}
\todo[inline]{Aquí he mezclado una sección y un apéndice del artículo original. Revisar que todo está coherente}
\label{ssecPhase}

Our next task consists in finding the $n$ different and linearly independent stationary distributions $\probDphasevect{l}\in {\cal H}$, with $l=0,1,\ldots n-1$, that emerge at the DPT once the degeneracy kicks in \cite{gaveau98a,hanggi82a,ledermann50a,gaveau06a,minganti18a}.
Each one of these phase probability vectors $\probDphasevect{l}$ corresponds to the probability disitribution of a single symmetry-broken phase $l\in[0 \isep n-1]$, such as the two ferromagetic phases in the Ising model.
The set spanned by these vectors 
defines a new basis of the degenerate subspace. In this way, a phase probability vector $\probDphasevect{l}$ can be always written as a linear combination of the Doob eigenvectors in the degenerate subspace,
\begin{equation}\probDphasevect{l} = \sum_{j=0}^{n-1} b_{l,j} \eigvectDR{j}\, ,
\label{ppvlin}
\end{equation}with complex coefficients $b_{l,j}\in \mathbb{C}$. Moreover, the phase probability vectors must be normalized, $\braket{-}{\Pi_l^\lambda}=1$ $\forall l\in[0\isep n-1]$, and crucially they must be related by the action of the symmetry operator, 
\begin{equation}
    \probDphasevect{l+1} = \symmop \probDphasevect{l} \, ,
\label{ppvsym}
\end{equation}which implies that $\probDphasevect{l} = \symmop^l \probDphasevect{0}$ and therefore $b_{l,j} = b_{0,j} (\phi_j)^l$,  
with $\phi_j$ the eigenvalues of the symmetry operator.



In order to obtain the coefficients $b_{0,j}$, we impose now that the Doob stationary distribution can be written as a statistical mixture (or convex sum)
of the different phase probability vectors
\begin{equation}
\probTvectst = \sum_{l=0}^{n-1} w_l \probDphasevect{l} = \sum_{j=0}^{n-1} \sum_{l=0}^{n-1} w_l b_{0,j} (\phi_j)^l \eigvectDR{j} \, , 
\end{equation}
with $\sum_{l=0}^{n-1} w_l = 1$ and $0\le w_l\le 1$, where we have used Eq.~\eqref{ppvlin} in the second equality. Comparing this expression with the spectral decomposition of the Doob steady state, $\probTvectst = \sum^{n-1}_{j=0}  \eigvectDR{j} \braket{\eigvectDLnk{j}}{\probvecttimenk{0}}$, we find
\begin{equation}\braket{\eigvectDLnk{j}}{\probvecttimenk{0}} = \sum_{l=0}^{n-1} w_l b_{0,j} (\symmeigval{j})^l  \, .
\label{leftA}
\end{equation}Taking now the modulus on both sides of the equation, using the triangular inequality 
and noticing that eigenvalues $\phi_j$ lie in the complex unit circle so $|\phi_j|=1$, we obtain $\abs{\braket{\eigvectDLnk{j}}{\probvecttimenk{0}}} \le \abs{b_{0,j}}$. This inequality is saturated whenever the initial vector $\probvecttime{0}$ is chosen so that the Doob stationary vector coincides with one phase, i.e., $w_l = \delta_{l,l'}$ for some $l'$, see Eq.~\eqref{leftA}, so that $\probTvectst=\probDphasevect{l'}$ for this particular initial $\ket{\probvecttimenk{0}}$. In this way, we have found that in general
\begin{equation}
    \abs{b_{0,j}}
    =
    \max_{\ket{P_0}} \abs{\braket{\eigvectDLnk{j}}{\probvecttimenk{0}}} \, . 
\end{equation}
We can now write
\begin{equation}
    \max_{\ket{P_0}} \left| \braket{\eigvectDLnk{j}}{\probvecttimenk{0}}\right|
    =
    \max_{\ket{P_0}} \left| \sum_C \braket{\eigvectDLnk{j}}{C}\braket{C}{\probvecttimenk{0}}\right| 
    \, ,
\end{equation}
and since we have chosen to normalize the left eigenvectors such that $\max_{C} \abs{\braket{\eigvectDLnk{j}}{\conf}} = 1$ (see Section~\ref{sec:markovprocesses_theory}), and noting that $\braket{C}{\probvecttimenk{0}}\le 1$ $\forall\ket{C}$, it is clear that the maximum over $\ket{P_0}$ is reached when $\ket{P_0}=\ket{C^*}$, the configuration where $\abs{\braket{\eigvectDLnk{j}}{\conf^*}}$ takes it maximum value 1. Therefore we find that $\abs{b_{0,j}} = 1$. Note also that the normalization condition $\max_{C} \abs{\braket{\eigvectDLnk{j}}{\conf}} = 1$ specifies left eigenvectors up to an arbitrary complex phase, which can be now chosen so that $b_{0,j} = 1$ $\forall j < n$. This hence implies that the coefficients in the expansion of Eq.~\eqref{ppvlin} are just $b_{l, j} = (\symmeigval{j})^l$, and we can obtain the final form of the probability vector of the phases in terms of the degenerate right eigenvectors,
\begin{equation}
\probDphasevect{l} =   \sum_{j=0}^{n-1} (\phi_j)^l\eigvectDR{j} = \eigvectDR{0} +  \sum_{j=1}^{n-1} (\phi_j)^l\eigvectDR{j} \, .
\label{eq:probphasevect}
\end{equation}

In addition, the structure in the phase vectors $\probDphasevect{l}$ has implications on the eigenvalues of the symmetry operator.
In particular, the fact that the different phases must be linearly independent implies that the first $n$ eigenvalues $\symmeigval{j}$ must be different. 
If there were two eigenvalues such that $\symmeigval{j'} = \symmeigval{j''}$, then all the vectors $\probDphasevecttime{l}$ would live in the hyperplane given by the constraint $(\eigvectDL{j'} - \eigvectDL{j''})\ket{v} = 0$, in contradiction to our initial assumption.
Therefore, for the symmetry-breaking DPT to occur, the first $n$ eigenvalues $\symmeigval{j}$ must be different, which in turn implies that they must correspond to all the different $n$-th roots of unity.

It will prove useful to introduce the left duals $\bra{\pi_l^\lambda}$ of the phase probability vectors, i.e., the row vectors satisfying the biorthogonality relation $\braket{\pi_{l'}^\lambda}{\Pi_{l}^\lambda}=\delta_{l',l}$. These must be linear combinations of the left eigenvectors associated with the degenerate subspace,
\begin{equation}\bra{\pi_l^\lambda} = \sum_{j=0}^{n-1} \eigvectDL{j} D_{l,j}  \, ,
\label{ppvlin2}
\end{equation}with complex coefficients $D_{l,j}\in \mathbb{C}$. Imposing biorthogonality, using the spectral expansion in Eq.~\eqref{ppvlin}, and recalling that
the first $n$ eigenvalues $\phi_j$ correspond exactly to the $n$-th roots of unity,
we thus find $D_{l,j} = \frac{1}{n}(\phi_j)^{-l}$, so that
\begin{equation}\bra{\pi_l^\lambda} = \sum_{j=0}^{n-1} \frac{1}{n}(\phi_j)^{-l} \eigvectDL{j} \, .
\label{ppvlin3}
\end{equation}
With these left duals, we can now easily write the right eigenvectors $\eigvectDR{j}$ in the degenerate subspace, $\forall j\in[0\isep n-1]$, in terms of the different phase probability vectors,
\begin{equation}
    \eigvectDR{j} = \sum_{l=0}^{n-1} \probDphasevect{l} \braket{\pi_l^\lambda}{R_{j,D}^\lambda} = \frac{1}{n} \sum_{l=0}^{n-1} (\symmeigval{j})^{-l} \probDphasevect{l}\, ,
\label{rinv}
\end{equation}where we have used that $\braket{\pi_l^\lambda}{R_{j,D}^\lambda} = \frac{1}{n} (\symmeigval{j})^{-l}$. Using this decomposition in Eq.~\eqref{Psstime}, we can thus reconstruct the (degenerate) Doob \emph{steady} state as a weighted sum of the phase probability vectors $\probDphasevect{l}$ associated with each of the $n$ symmetry-broken phases,
\begin{equation}\probTvectst (t)= \sum_{l=0}^{n-1} w_l(t) \probDphasevect{l} \, ,
\label{Pssdecomp}
\end{equation}with the weights $w_l(t)=\braket{\pi_l^\lambda}{P_{\text{ss},P_0}^\lambda}(t)$ given by
\begin{equation}w_l(t)= \frac{1}{n} + \frac{1}{n}\sum_{j=1}^{n-1}  \text{e}^{+i t \Im(\theta_{j,\text{D}}^\lambda)} (\phi_j)^{-l} \braket{\eigvectDLnk{j}}{\probvecttimenk{0}}\, .
\label{coeft}
\end{equation}These weights are time-dependent if the imaginary part of the gap-closing eigenvalues are non-zero, though in many applications the relevant eigenvalues are purely real. In such cases
\begin{equation}\begin{split}
&\probTvectst = \sum_{l=0}^{n-1} w_l \probDphasevect{l}\, , \\
&w_l = \frac{1}{n} + \frac{1}{n}\sum_{j=1}^{n-1}  (\phi_j)^{-l} \braket{\eigvectDLnk{j}}{\probvecttimenk{0}}\, .
\end{split}
\label{coef}
\ee 
This shows that the Doob steady state can be described as a statistical mixture of the different phases (as described by their unique phase probability vectors $\probDphasevect{l}$), i.e. $\sum_{l=0}^{n-1} w_l=1$, with $0\le w_l\le 1$ $\forall l\in[0\isep n-1]$. These statistical weights $w_l$ are determined by the projection of the initial state on the different phases, which is in turn governed by the overlaps of the degenerate left eigenvectors with the initial state and their associated symmetry eigenvalues.

Equation~\eqref{coef} shows how to prepare the system initial state $\ket{\probvecttimenk{0}}$ to single out a given symmetry-broken phase $\probDphasevect{l'}$ in the long-time limit.
Indeed, by comparing Eqs.~\eqref{eq:DPT_probvect} and~\eqref{eq:probphasevect}, it becomes evident that choosing $\ket{\probvecttimenk{0}}$ such that $\braket{\eigvectDLnk{j}}{\probvecttimenk{0}}=(\phi_j)^{l'}$ $\forall j \in [1\isep n-1]$ leads to a \emph{pure} steady state $\probTvectst = \probDphasevect{l'}$, i.e. such that $w_{l}=\delta_{l,l'}$. 
This strategy provides a simple phase-selection mechanisms by initial state preparation somewhat similar to those already described in open quantum systems with strong symmetries \cite{manzano14a,manzano18a,manzano21a}.

It is important to note that, in general, degeneracy of the leading eigenspace of the stochastic generator is only possible in the thermodynamic limit. This means that for finite-size systems one should always expect small but non-zero spectral gaps $\Delta^{\lambda}_j$, $j\in[1\isep n-1]$,  and hence the long-time Doob steady state is $\probTvectst = \eigvectDR{0}$, see Eq.~\eqref{ssnopt}. This steady state, which can be written as $\eigvectDR{0} = \frac{1}{n}\sum_{l=0}^{n-1} \probDphasevect{l}$, preserves the symmetry of the generator, so that no symmetry-breaking DPT is possible for finite-size systems. 
Rather, for large but finite system sizes, one should expect an emerging quasi-degeneracy \cite{gaveau98a,gaveau06a} in the parameter regime where the DPT emerges,  i.e. with $\Delta^{\lambda}_j / \Delta^{\lambda}_{n} \ll 1$ $\forall j\in[1\isep n-1]$. In this case, and for time scales $t\ll 1/\Delta^{\lambda}_{n-1}$ but $t\gg 1/\Delta^{\lambda}_{n}$, we expect to observe a sort of metastable symmetry breaking captured by the physical phase probability vectors $\probDphasevect{l}$, with punctuated jumps between different symmetry sectors at the individual trajectory level. 
This leads in the long-time limit to an effective restitution of the original symmetry as far as the system size is finite.

It is worth mentioning that the symmetry group which is broken in the DPT may be larger than $\mathbb{Z}_n$. In such case, as long as it contains a $\mathbb{Z}_n$ group whose symmetry is also broken, the results derived in this section are still valid. Actually, this occurs for the $r$-state Potts model discussed in Section \ref{secPotts}, which despite breaking the symmetry of the dihedral group $\mathbb{D}_r$, it breaks as well the symmetry of the $\mathbb{Z}_r$ subgroup.

\section{Structure of the eigenvectors in the degenerate subspace}
\label{ssecSpec}
\todo[inline]{Esto hay que cambiarlo por algo mejor}

A key observation is that, once a symmetry-breaking phase transition kicks in, be it either configurational (i.e. standard) or dynamical, the associated \emph{typical} configurations fall into well-defined symmetry classes, i.e. the symmetry is broken already at the individual configurational level. As an example, consider the paradigmatic 2D Ising model and its (standard) $\mathbb{Z}_2$ symmetry-breaking phase transition at the Onsager temperature $T_c$, separating a disordered paramagnetic phase for $T>T_c$ from an ordered, symmetry-broken ferromagnetic phase for $T<T_c$ \cite{binney92a}. For temperatures well below the critical one, the stationary probability of e.g. completely random (symmetry-preserving) spin configurations is extremely low, while high-probability configurations exhibit a net non-zero magnetization typical of symmetry breaking. This means that statistically-relevant configurations do belong to a specific symmetry phase, in the sense that they can be assigned to the \emph{basin of attraction} of a given symmetry sector \cite{gaveau06a}.

Something similar happens in $\mathbb{Z}_n$ symmetry-breaking DPTs. In particular, once the DPT kicks in and the symmetry is broken, statistically-relevant configurations $\ket{C}$ (i.e. such that $\braket{C}{P_{\text{ss},P_0}^\lambda}=P_{\text{ss},P_0}^\lambda(C)$ is significantly different from zero) 
belong to a well-defined symmetry class with index $\ell_C\in[0\isep n-1]$. In terms of phase probability vectors, this means that
\begin{equation}
\frac{\braket{C}{\probDphasevectnk{{l}}}}{\braket{C}{\probDphasevectnk{{\ell_{\conf}}}}} \approx 0, \qquad \forall l\ne \ell_C \, .
\label{condC}
\end{equation}This property arises from the large deviation scaling of $\braket{C}{\probDphasevectnk{{l}}}$.
In other words, statistically-relevant configurations in the symmetry-broken Doob steady state can be partitioned into disjoint symmetry classes. This simple but crucial observation can be used now to unveil a hidden spectral structure in the degenerate subspace, associated with such configurations. In this way, if $\ket{C}$ is  one of these configurations belonging to phase $\ell_C$, from Eq.~\eqref{rinv} we deduce that
\begin{equation}\braket{C}{\eigvectDRnk{j}}= \frac{1}{n} \sum_{l=0}^{n-1} (\symmeigval{j})^{-l} \braket{C}{\probDphasevectnk{{l}}}\approx \frac{1}{n} (\symmeigval{j})^{-\ell_C} \braket{C}{\probDphasevectnk{{\ell_C}}} \, .
\ee    
In particular, for $j=0$ we have that $\braket{C}{\eigvectDRnk{0}} \approx \frac{1}{n} \braket{C}{\probDphasevectnk{{\ell_C}}}$ since $\phi_0=1$, and therefore
\begin{equation}
    \braket{C}{\eigvectDRnk{j}} \approx (\symmeigval{j})^{-\ell_C} \braket{C}{\eigvectDRnk{0}} 
\label{eq:eigvectsfromeigvect0}
\end{equation}for $j\in[1\isep n-1]$. In this way, the components $\braket{C}{\eigvectDRnk{j}}$ of the subleading eigenvectors in the degenerate subspace associated with the statistically-relevant configurations are (almost) equal to those of the leading eigenvector $\eigvectDR{0}$ except for a complex argument given by $(\phi_j)^{-\ell_C}$. This highlights how the $\mathbb{Z}_n$ symmetry-breaking phenomenon imposes a specific structure on the degenerate eigenvectors involved in a continuous DPT. Of course, this result is based on the (empirically sound) assumption that statistically-relevant configurations can be partitioned into disjoint symmetry classes. We will confirm \emph{a posteriori} 
these results in the three examples considered in the next chapter.

\todo[inline]{Cambiar aalgo mejor}
\section{Reduction to the order parameter vector space}
\label{ssecOparam}

The direct analysis of the eigenvectors in many-body stochastic systems is typically an unfeasible task, as the dimension of the configuration Hilbert space usually grows exponentially with the system size. Moreover, extracting useful information from this analysis is also difficult as configurations are not naturally categorized according to their symmetry properties. This suggests to introduce a partition of the configuration Hilbert space ${\cal H}$ into equivalence classes according to a proper order parameter for the DPT under study, grouping \emph{similar} configurations together (in terms of their symmetry properties) so as to reduce the effective dimension of the problem, while introducing at the same time a natural parameter to analyze the spectral properties.

We define an order parameter $\mu$ for the DPT of interest as a map $\mu:{\cal H} \to \mathbb{C}$ that gives for each configuration $\ket{C}\in {\cal H}$ a complex number $\mu(C)$ whose modulus measures the \emph{amount of order}, i.e. how deep the system is into the symmetry-broken regime, and whose argument determines in which phase it is. Of course, other types of DPTs may have their own natural order parameters, but for $\mathbb{Z}_n$ symmetry-breaking DPTs a simple complex-valued number suffices, as we shall show below. Associated with this order parameter, we now introduce a reduced Hilbert space ${\cal H}_\mu=\{\kket{\nu}\}$ representing the possible values of the order parameter as vectors $\kket{\nu}$ of a biorthogonal basis satisfying $\bbraket{\nu'}{\nu} = \delta_{\nu,\nu'}$. In general, the dimension of ${\cal H}_\mu$ will be significantly smaller than that of ${\cal H}$ since the possible values of the order parameter typically scale linearly with the system size.

In order to transform probability vectors from the original Hilbert space to the reduced one,
we define a surjective application $\widetilde{\cal T}:{\cal H} \to {\cal H}_\mu$ that maps all configurations $\ket{C}\in {\cal H}$ with order parameter $\nu$ onto a single vector $\kket{\nu}\in {\cal H}_\mu$ of the reduced Hilbert space.
Crucially, this mapping $\widetilde{\cal T}$ from configurations to order parameter equivalence classes must conserve probability, i.e. the accumulated probability of all configurations with a given value of the order parameter in the original Hilbert space ${\cal H}$ must be the same as the probability of the equivalent vector component in the reduced space. In particular, let $\ket{P}\in {\cal H}$ be a probability vector in configuration space, and $\kket{P}=\widetilde{\cal T} \ket{P}$ its corresponding reduced vector in ${\cal H}_\mu$. Conservation of probability then means that
\begin{equation}P(\nu)
= \sum_{\substack{\ket{C}\in{\cal H}: \\ \mu(C)=\nu}} \braket{C}{P} = \bbraket{\nu}{P} \, , \qquad \forall \nu \, .
\label{Pconserv}
\end{equation}This probability-conserving condition thus constraints the particular form of the map $\widetilde{\cal T}:{\cal H} \to {\cal H}_\mu$. In general, if $\ket{\psi}\in{\cal H}$ is a vector in the original configuration Hilbert space, the reduced vector $\kket{\psi}=\widetilde{\cal T}\ket{\psi}\in {\cal H}_\mu$ is hence defined as
\begin{equation}\kket{\psi} = \sum_\nu \bbraket{\nu}{\psi}~\kket{\nu} = \sum_\nu \Bigg[\sum_{\substack{\ket{C}\in{\cal H}: \\ \mu(C)=\nu}} \braket{C}{\psi}\Bigg] \kket{\nu} \, .
\label{Pconserv3}
\end{equation}
But what makes a \emph{good} order parameter $\mu$? In short, a good order parameter must be sensitive to the different symmetry-broken phases and to how the symmetry operator moves one phase to another. More in detail, let $\{\ket{C}\}_\nu=\{\ket{C}\in{\cal H}: \mu(C)=\nu\}$ be the set of all configurations $\ket{C}\in{\cal H}$ with order parameter $\mu(C)=\nu$, i.e. the set of all configurations defining the equivalence class represented by the reduced vector $\kket{\nu}\in{\cal H}_\mu$. Applying the symmetry operator $\hat{S}$ to all configurations in $\{\ket{C}\}_\nu$ defines a new set $\hat{S}(\{\ket{C}\}_\nu)$. We say that $\mu$ is a good order parameter iff: (i) The new set $\hat{S}(\{\ket{C}\}_\nu)$ 
corresponds to the equivalence class
$\{\ket{C}\}_{\nu'}$
associated with another order parameter vector $\kket{\nu'}\in{\cal H}_\mu$, and (ii) it can distinguish a symmetry-broken configuration from its symmetry-transformed configuration. This introduces a bijective mapping $\hat{S}_\mu \kket{\nu}=\kket{\nu'}$ between equivalence classes that defines a reduced symmetry operator $\hat{S}_\mu$ acting on the reduced order-parameter space. Mathematically, this mapping can be defined from the relation $\widetilde{\cal T} \symmop \ket{C} = \hat{S}_\mu \widetilde{\cal T}\ket{C}$ $\forall \ket{C}\in{\cal H}$, where condition (i) ensures that $\hat{S}_\mu$ is a valid symmetry operator.

As an example, consider again the 2D Ising spin model for the paramagnetic-ferromagetic phase transition mentioned above \cite{binney92a}. Below the critical temperature, this model breaks spontaneously a $\mathbb{Z}_2$ spin up-spin down symmetry, a phase transition well captured by the total magnetization $m$, the natural order parameter. The symmetry operation consists in this case in flipping the sign of all spins in a configuration, and this operation induces a one to one, bijective mapping between opposite magnetizations. An alternative, plausible order parameter could be $m^2$. This parameter can certainly distinguish the ordered phase ($m^2\ne 0$) from the disordered one ($m^2\approx 0$), but still cannot discern between the two symmetry-broken phases, and hence it is not a good order parameter in the sense defined above.

As we shall illustrate below, the reduced eigenvectors $\kket{R_{j,\text{D}}^\lambda} = \widetilde{\cal T} \ket{R_{j,\text{D}}^\lambda}$ associated with the 
spectrum of the Doob generator in the original configuration space encode the most relevant information regarding the DPT, and can be readily analyzed. Indeed, it can be easily checked that the results obtained in the previous subsections also apply in the reduced order parameter space. In particular, before the DPT happens, the reduced Doob steady state is unique, see Eq.~\eqref{ssnopt},
\begin{equation}\kket{P_{\text{ss},P_0}^\lambda} = \kket{R_{0,\text{D}}^\lambda} \, ,
\label{redeq1}
\end{equation}while once the DPT kicks in and the symmetry is broken, degeneracy appears and
\begin{equation}\kket{P_{\text{ss},P_0}^\lambda} = \kket{R_{0,\text{D}}^\lambda} + \sum^{n-1}_{j=1}  \kket{R_{j,\text{D}}^\lambda}~\braket{\eigvectDLnk{j}}{\probvecttimenk{0}}\, ,
\label{redeq2}
\end{equation}see Eq.~\eqref{eq:DPT_probvect} for purely real eigenvalues [and similarly for eigenvalues with non-zero imaginary parts, see Eq.~\eqref{Psstime}]. Notice that since $\widetilde{\cal T}$ is a linear transformation, the brackets $\braket{\eigvectDLnk{j}}{P_0}$ do not change under $\widetilde{\cal T}$ as they are scalars. Reduced phase probability vectors can be defined in terms of the reduced eigenvectors in the degenerate subspace, see Eq.~\eqref{eq:probphasevect},
\begin{equation}\kket{\Pi_l^\lambda} = \kket{R_{0,\text{D}}^\lambda} + \sum_{j=1}^{n-1} (\phi_j)^l \kket{R_{j,\text{D}}^\lambda} \, ,
\label{redeq3}
\end{equation}and the reduced Doob steady state can be written in terms of these reduced phase probability vectors, $\kket{P_{\text{ss},P_0}^\lambda}  = \sum_{l=0}^{n-1} w_l \kket{\Pi_l^\lambda}$, see Eq.~\eqref{coef}. Finally, the structural relation between the Doob eigenvectors spanning the degenerate subspace, Eq.~\eqref{eq:eigvectsfromeigvect0}, is also reflected in the order-parameter space. In particular, for a statistically-relevant value of $\mu$
\begin{equation}
\label{redeq4}
    \bbraket{\mu}{R_{j,\text{D}}^\lambda}
    \approx
    \phi_j^{-\ell_\mu} \bbraket{\mu}{R_{0,\text{D}}^\lambda} \, ,
\end{equation}
for $j\in[1\isep n-1]$, where $\ell_\mu=[0\isep n-1]$ 
is an indicator function which maps the different possible values of the order parameter $\mu$ with their corresponding phase index $\ell_\mu$.
It is worth mentioning that this implies that, if the steady-state distribution of $\mu$ follows a large-deviation principle, $\bbraket{\mu}{R_{0,\text{D}}^\lambda} \asymp \text{e}^{-L F(\mu)}$, with $L$ being the system size, then the rest of gap closing reduced eigenvectors obey the following property for the statistically-relevant values of $\mu$
\begin{equation*}
    \bbraket{\mu}{R_{j,\text{D}}^\lambda}
    \asymp
    \phi_j^{-\ell_\mu} \text{e}^{-L F(\mu)}
    .
\end{equation*}
In the following sections we shall illustrate our main results by projecting the spectral information in the order-parameter reduced Hilbert space for three paradigmatic many-body systems exhibiting continuous DPTs.

\section{Conclusion}
\label{sec:conclu_SBDPT}

In this chapter, we have unveiled the spectral signatures of symmetry-breaking DPTs. Such DPTs appear in the fluctuating behavior of many-body systems as non-analyticities in the large deviation functions describing the fluctuations of time-averaged observables, and are accompanied by singular changes in the trajectories responsible for such rare events. The main tools used in this chapter include the quantum Hamiltonian formalism for the master equation, describing the dynamics of stochastic many-body systems, together with large deviation theory whereby the symmetry of the microscopic dynamics has been fully exploited. A cornerstone in our analysis has been the Doob transform to build a driven stochastic process that makes typical a rare fluctuation of the original dynamics. Crucially, the steady state of the resulting Doob dynamics contains all the information of the most likely path leading to such rare fluctuation of the original process.
In this way, the spectral hallmark of a symmetry-breaking DPT is the emergence of a degeneracy in the stationary subspace of Doob eigenvectors. The degenerate eigenvectors exhibit different behavior under the symmetry transformation, and we show how symmetry and degeneracy cooperate to yield different, coexisting steady states once the DPT has kicked in. Such steady states are characterized by physical phase probability vectors, connected via the symmetry transformation, that we explicitly build from the gapless Doob eigenvectors in the degenerate subspace. Moreover, a generic steady state can be then written as a weighted sum of these phase probability vectors, with the different weights controlled by the initial state. This mechanism explains how the system breaks the symmetry by singling out a particular dynamical phase out of the multiple possible phases present in the first Doob eigenvector. By conjecturing that statistically relevant configurations in the symmetry-broken regime can be partitioned into different symmetry classes, we further derive an expression for the components of the subleading Doob eigenvectors in the degenerate subspace in terms of the leading eigenvector and the symmetry eigenvalues, showcasing the stringent spectral structure imposed by symmetry on DPTs. Finally, we introduce a reduced Hilbert space based on a suitable order parameter for the DPT, with appropriate transformation properties under the symmetry operator. All the spectral signatures of the DPT are reflected in this reduced order-parameter space, which hence allows for the empirical verification of our results while providing a natural classification scheme for configurations in terms of their symmetry properties.


The spectral symmetry-breaking mechanism described in this chapter is completely general for $\mathbb{Z}_n$-invariant systems, so we expect these results to hold valid also in standard (steady-state) critical phenomena, where the dimensional reduction introduced by the projection on the order-parameter, reduced Hilbert space can offer new perspectives on well-known phase transitions \cite{gaveau98a,gaveau06a,binney92a}.

It would be also interesting to extend the current analysis to more complex DPTs. For instance, it would be desirable to investigate the spectral signatures of DPTs in realistic high-dimensional driven diffusive systems, as e.g. the DPTs discovered in the current vector statistics of the 2D closed WASEP \cite{tizon-escamilla17a}. In this case, the complex interplay among the external field, lattice anisotropy, and vector currents in 2D leads to a rich phase diagram, with different symmetry-broken dynamical phases separated by lines of first- and second-order DPTs, and competing time-crystal phases. The spectral fingerprints of this complex competition between DPTs would further illuminate future developments. It would be also interesting to explore the spectral signatures of possible DPTs in driven dissipative systems \cite{prados11a,prados12a,hurtado13a} , or for diffusive systems characterized by multiple local conservation laws, as e.g. the recently introduced kinetic exclusion process \cite{gutierrez-ariza19a}. Finally, though the interplay between symmetry and DPTs in open quantum systems has been investigated in recent years \cite{manzano14a,manzano18a}, the range of possibilities offered by the order-parameter reduced Hilbert space calls for further investigation.

    }

    \chapter{Unveiling symmetry-breaking DPTs in microscopic models}
    \chaptermark{Symmetry-breaking DPTs in microscopic models}
    {
        \label{chap:DPTs_models}
        In the introduction, we mentioned the crucial role played by simplified models in the development of new theories.
Building on this premise, this chapter will focus on the application of the concepts introduced in Chapter~\ref{chap:SBDPTs} to a set of paradigmatic simple microscopic models exhibiting DPTs.
This will not only validate the results of the previous chapter but will also enhance our understanding of the phase transitions in such models.

We will focus on two specific models: the Weakly Asymmetric Simple Exclusion Process (WASEP) and the Potts model.
The WASEP is a paradigmatic one-dimensional lattice gas used for the study of driven diffusive transport.
We will explore its behavior when its ends are connected to two distinct particle reservoirs, a configuration commonly known as \emph{open} WASEP. 
In particular, we will analyze the DPT displayed by this lattice gas when observing fluctuations of the particle current well below the typical value\todo{under particular circumstances}.
We will see that, under some conditions, this model presents a $\Zsymm{2}$ particle-hole symmetry that is spontaneously broken by the DPT.
We will also study the one-dimensional Potts model of ferromagnetism, defined as a lattice of spins in which each one can take a discrete number $\nstatesP$ of orientations.
This model exhibits a DPT from a paramagnetic to a ferromagnetic phase when observing low energy fluctuations, which breaks the $\Zsymm{\nstatesP}$ rotational symmetry present in the original state.
This makes the model a perfect example for the analysis of spontaneous symmetry breaking of a $\Zsymm{\nstatesP}$ for arbitrary values of $\nstatesP$.
In particular, we will analyze the three- and four-state Potts models with periodic boundaries.

Another model of special interest will be the WASEP with periodic boundaries.
However, as this lattice gas exhibits some distinctive and intriguing features not present in the aforementioned models, we will postpone its analysis to Chapter~\ref{chap:buildingtc}.

\section{Dynamical criticality in the boundary-driven WASEP}
\label{sec:openWASEP_DPTsmodels}

\subsection{Model and dynamical phase transition}

\begin{figure}
    \centering
    \includegraphics[width=0.9\linewidth]{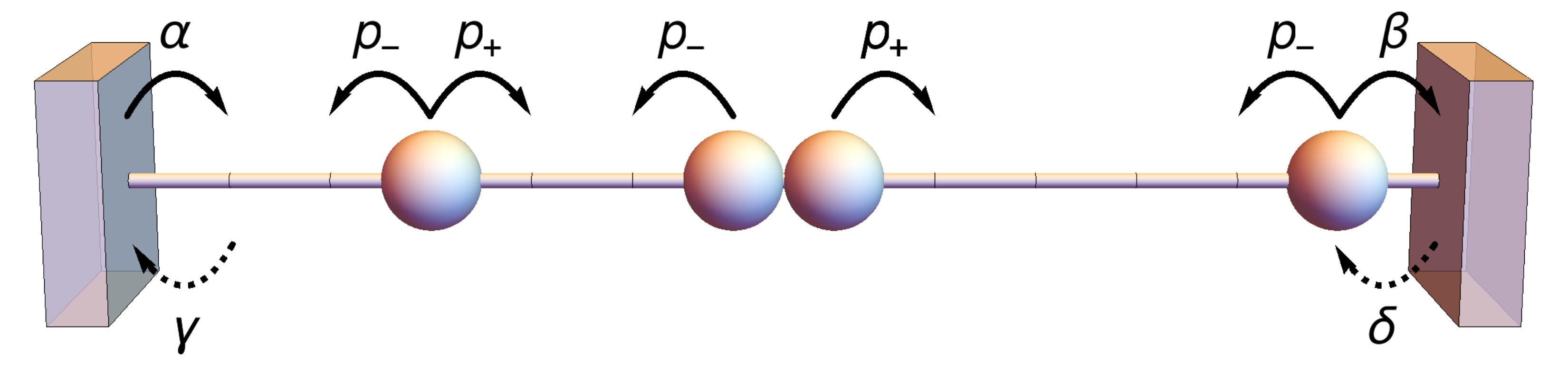}
    \caption{{\bf Sketch of the boundary-driven WASEP.} $N$ particles in a lattice of $L$ sites which randomly jump to empty neighboring sites with asymmetric rates  $p_\pm$, and coupled to boundary reservoirs which inject and remove particles at different rates.
    \label{fig:WASEP_def_open}}
\end{figure}

We shall first illustrate the ideas introduced above using the boundary-driven (or open) weakly asymmetric simple exclusion process (WASEP), which is an archetypical stochastic lattice gas modeling a driven diffusive system \cite{gartner87a, de-masi89a, bodineau05a}.
It consists of $N$ particles in a one-dimensional lattice of $\lattsize$ sites which can be either empty or occupied by one particle at most, so that the state $\conf$ of the system at any time is defined by the set of occupation numbers of all sites, $\conf = \{\occup_k\}_{k=1, ..., \lattsize}$, with $n_k=0,1$. Such state is encoded in a column vector $\ket{\conf} = \bigotimes^{\lattsize}_{k=1} (\occup_k, 1 - \occup_k)^{\transposesymb}$, where $\transposesymb$ denotes transposition, in a Hilbert space ${\cal H}$ of dimension $2^L$. Particles hop randomly to empty adjacent sites with asymmetric rates $\jumprateRL = \frac{1}{2} \text{e}^{\pm \field / \lattsize}$ to the right and left respectively, due to the action of an external field $\field$ applied to the particles of the system, see Fig.~\ref{fig:WASEP_def_open}. The ends of the lattice are connected to particle reservoirs which inject and remove particles with rates $\rateinL$, $\rateoutL$ respectively in the leftmost site, and rates $\rateinR$, $\rateoutR$ in the rightmost one. These rates are related to the densities of the boundary particle reservoirs as $\densL = \rateinL/(\rateinL + \rateoutL)$ and $\densR = \rateinR/(\rateinR + \rateoutR)$ \cite{derrida07a}. Overall, the combined action of the external field and the reservoirs drive the system into a nonequilibrium steady state with a net particle current.

At the macroscopic level, namely after a diffusive scaling of the spatiotemporal coordinates, the WASEP is described by the following diffusion equation for the particle density field $\rho(x,t)$ \cite{spohn12a},
\begin{equation}
\pdv{\rho}{t} = - \pdv{\left[-\diffusivity(\dens)\pdv{\dens}{x} + \mobility(\dens) \field \right]}{x}\, ,
\label{eq:WASEP_hydro}
\end{equation}
where $\diffusivity(\dens) = \frac{1}{2}$ is the diffusion transport coefficient and $\mobility(\dens) =  \dens (1 - \dens)$ is the mobility.

At the microscopic level, the stochastic generator of the dynamics reads
\begin{equation}
\begin{split}
\label{eq:WASEP_gen}
\genmat = \sum_{k=1}^{\lattsize-1} \Big[\;
          &\jumprateR \big( \creationop_{k+1} \destructop_{k} 
        - \occupop_{k  } (\identityop - \occupop_{k+1} )\big)
    \\ 
        + &\jumprateL (\creationop_{k  } \destructop_{k+1}
        - \occupop_{k+1} (\identityop - \occupop_{k  } )\big)\Big]
    \\
    + &\rateinL [ \creationop_{1} - (\identityop - \occupop_{1}) ]
        + \rateoutL [ \destructop_{1} - \occupop_{1} ]
    \\
    + &\rateinR [ \creationop_{\lattsize} - (\identityop - \occupop_{\lattsize}) ]
        + \rateoutR [ \destructop_{\lattsize} - \occupop_{\lattsize} ]
    ,
\end{split}
\end{equation}
where $\creationop_{k}$, $\destructop_{k}$ are respectively the creation and annihilation operators given by $\hat{\sigma}^{\pm}_k= (\hat{\sigma}^x_k \pm i \hat{\sigma}^y_k)/2  $, with $\hat{\sigma}^{x,y}_k$ the standard $x,y$-Pauli matrices, while $\occupop_{k}=\hat{\sigma}^+_k \hat{\sigma}^-_k$ and $\identityop_k$ are the occupation and identity operators acting on site $k$.
The first row in the r.h.s of the above equation corresponds to transitions where a particle on site $k$ jumps to the right with rate $p_+$, whereas the second one corresponds to the jumps from site $k+1$ to the left with rate $p_-$. The last two rows correspond to the injection and removal of particles at the left and right boundaries.

Interestingly, if the boundary rates are such that $\rateinL =  \rateoutR$ and $\rateoutL = \rateinR$, implying that $\densR = 1 - \densL$, the dynamics becomes invariant under a particle-hole (PH) transformation \cite{perez-espigares18a}, $\symmop_{\textrm{PH}}$, which thus commutes with the generator of the dynamics, $[\symmop_{\textrm{PH}},  \genmat] = 0$. This transformation simply amounts to changing the occupation of each site, $\occup_k \to 1 - \occup_k$, and inverting the spatial order, $k \to \lattsize -k +1$, and it is represented by the microscopic operator
\begin{equation}
\begin{split}
\symmop_{\textrm{PH}} = \prod_{k'=1}^{\lattsize} \hat{\sigma}^x_{k'} \prod_{k=1}^{\floor{\lattsize/2}} \Big[\creationop_{k}\destructop_{\lattsize-k+1} + \creationop_{\lattsize-k+1}\destructop_{k} \\
+ \frac{1}{2}(\hat{\sigma}^z_k \hat{\sigma}^z_{\lattsize-k+1} + \identityop)\Big] \, ,
\end{split}
\end{equation}
where $\floor{\cdot}$ is the floor function and $\hat{\sigma}^z_k = \identityop - 2\occupop_{k}$ is the $z$-Pauli matrix.
The operator in brackets exchanges the occupancy of sites $k$ and $\lattsize-k+1$. 
In particular, the first two terms act on particle-hole pairs, while the last one affects pairs with the same occupancy.
Notice that $\symmop_{\textrm{PH}}$ is a $\Zsymm{2}$ symmetry, since $(\symmop_{\textrm{PH}})^2 = \identityop$. Macroscopically the PH transformation means to change $\rho\to 1-\rho$ and $x\to 1-x$, which leaves invariant Eq. (\refeq{eq:WASEP_hydro}) due to the symmetry of the mobility $\sigma(\rho)$ around $\rho=1/2$ and the constant diffusion coefficient \cite{baek17a,baek18a,perez-espigares18a}.

\todo[inline]{Creo que habría que decir que también está space averaged}
A key observable in this model is the time-integrated and space-averaged current $\intcurrent$, and the corresponding time-intensive observable $\current=Q/\tau$. The current $Q$ is defined as the number of jumps to the right minus those to the left per bond (in the bulk) during a trajectory of duration $\tau$. For any given trajectory $\omega_\tau$, this observable remains invariant under the PH transformation, $Q(\mathcal{S}_{\textrm{PH}}\omega_\tau) = Q(\omega_\tau)$, since the change in the occupation and the inversion of the spatial order gives rise to a double change in the flux sign, leaving the total current invariant. In this way, the symmetry under the transformation $\symmop_{\textrm{PH}}$ will be inherited by the Doob process associated with the fluctuations of the current, see Section~\ref{ssecSym}. Indeed, the current statistics and the corresponding driven process can be obtained by biasing or \emph{tilting} the original generator according to Eq.~\eqref{eq:genmatT_theory},
\begin{equation}
\begin{split}
\genmatT = \sum_{k=1}^{\lattsize-1} \Big[\; &\jumprateR \big( \text{e}^{\currenttilt/(\lattsize - 1)}\creationop_{k+1} \destructop_{k} - \occupop_{k  } (\identityop_{k+1} - \occupop_{k+1} )\big) \\ 
+ &\jumprateL (\text{e}^{-\currenttilt/(\lattsize - 1)} \creationop_{k  } \destructop_{k+1} - \occupop_{k+1} (\identityop_{k} - \occupop_{k  } )\big)\Big] \\
+ &\rateinL [ \creationop_{1} - (\identityop_{1} - \occupop_{1}) ] + \rateoutL [ \destructop_{1} - \occupop_{1} ] \\
+ &\rateinR [ \creationop_{\lattsize} - (\identityop_{\lattsize} - \occupop_{\lattsize}) ] + \rateoutR [ \destructop_{\lattsize} - \occupop_{\lattsize} ] \, ,
\end{split}
\label{eq:WASEPopen_tilted_gen}
\end{equation}
where we have used that the contribution to the time-integrated current $\intcurrent$ of a particle jump in the transition $C\to C'$ is $\eta_{C\to C'}=\pm 1/(\lattsize - 1)$, depending on the direction of the jump. 

\begin{figure}
    \centering
    \includegraphics[width=1.0\linewidth]{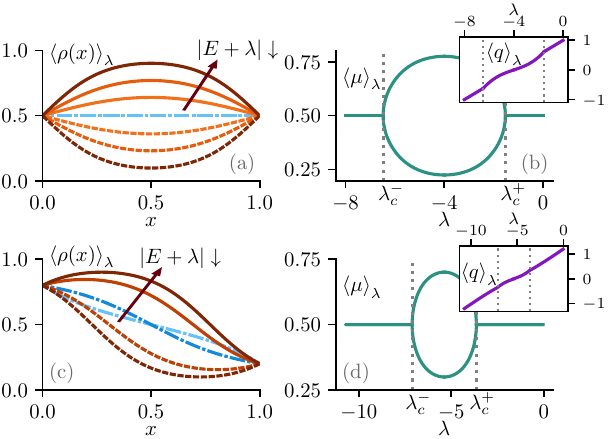}
    \caption{{\bf Particle-hole dynamical symmetry breaking in the open WASEP.} Macroscopic Fluctuation Theory results for the DPT in current fluctuations observed in the boundary-driven WASEP for $\field = 4>E_c$ \cite{perez-espigares18a}. (a) Density profiles of the system for $\rho_L=\rho_R=0.5$ and different values of the biasing field $\currenttilt$. The flat, PH-symmetry-preserving profile (blue, dash-dotted line) is the optimal solution for moderate current fluctuations, corresponding to values above $\crit{\currenttilt}^+ = -1.52$, while the PH-symmetry-breaking density profiles (reddish, full and dashed lines) correspond to $\currenttilt = -1.6, -1.75, -2, -2.5$ and $\lambda=-4 =-E$. (b) Mean lattice occupation $\langle \mu\rangle_\lambda$ in terms of the biasing field $\lambda$ for $\rho_L=\rho_R=0.5$. The observed bifurcations at $\lambda_c^\pm$ signal the symmetry-breaking DPT. The inset shows the average current as a function of $\lambda$, which becomes nonlinear in the symmetry-breaking regime. Panels (c) and (d) are equivalent to (a) and (b), respectively, but for $\rho_L=0.8$ and $\rho_R=1-\rho_L=0.2$.}
\label{fig:DPT_open}
\end{figure}

Remarkably, when $\densR = 1 - \densL$ and the stochastic dynamics is hence invariant under $\symmop_{\textrm{PH}}$, the boundary-driven WASEP displays, for a sufficiently strong external field $E$, a second-order DPT for fluctuations of the particle current below a critical threshold. Such DPT, illustrated in Fig.~\ref{fig:DPT_open}, was predicted in \cite{baek17a,baek18a} and further explored in \cite{perez-espigares18a} from a macroscopic perspective. In order to sustain a long-time current fluctuation, the system adapts its density field so as to maximize the probability of such an event, according to the MFT action functional \cite{bertini15a,baek17a,baek18a,perez-espigares18a}. For moderate current fluctuations, this optimal density profile might change, but retains the PH symmetry of the original action. However, for 
current fluctuations below a critical threshold in absolute value, the PH symmetry of the original action breaks down and two different but equally probable optimal density fields appear, both connected via the symmetry operator [see full and dashed reddish lines in Fig.~\ref{fig:DPT_open}(a) and Fig.~\ref{fig:DPT_open}(c)]. The emergence of these two action minimizers can be understood by noticing that, when $\densR = 1 - \densL$, either crowding the lattice with particles or depleting the particle population define two equally-optimal strategies to hinder particle motion, thus reducing the total current through the system \cite{baek17a,baek18a,perez-espigares18a}. More precisely, the DPT appears for an external field $|E|>E_c=\pi$ and current fluctuations such that $\abs{\current} \leq \currentcrit = \sqrt{\field^2 - \pi^2}/4$, which correspond to biasing fields in the range $\currenttiltcritmin < \currenttilt < \currenttiltcritmax$, with $\currenttiltcritminmax = -\field \pm \sqrt{\field^2 - \pi^2}$, see insets to Fig.~\ref{fig:DPT_open}(b) and Fig.~\ref{fig:DPT_open}(d). To characterize the phase transition, a suitable choice of the order parameter is the mean occupation of the lattice, defined as $\mass = \lattsize^{-1} \sum_{k=1}^{\lattsize} \occup_k$, with $\mu\in [0,1]$ since $n_k=0,1$ due to the exclusion rule. The behavior of this order parameter as a function of the biasing field is displayed in Fig.~\ref{fig:DPT_open}(b) and Fig.~\ref{fig:DPT_open}(d). 

\begin{figure}
    \centering
    \includegraphics[width=1\linewidth]{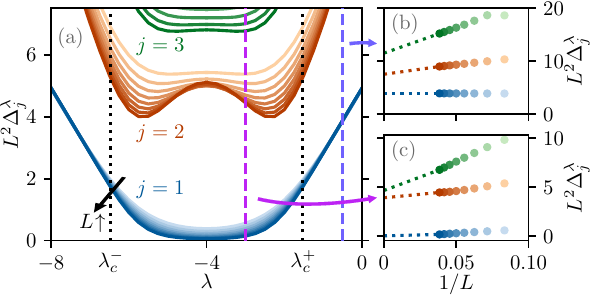}
    \caption{ {\bf DPT and quasi-degeneracy in the open WASEP.} (a) Scaled spectral gaps, $L^2\Delta_j^{\lambda}$, with $j=1,2,3$, as a function of the biasing field $\currenttilt$ for different lattice sizes ($L= 12, 14, 16, 18, 20, 22, 24, 26$) in the open WASEP.  Each set of lines is associated with a different value of $j$, with darker colors corresponding to larger system sizes. The small panels show the scaled spectral gaps as a function of the inverse lattice size for (b) $\lambda=-0.5>\lambda_c^+$ and (c) $\lambda=-3.5<\lambda_c^+$. The spectral gap $L^2\Delta_1^{\lambda}$ vanishes as $L$ increases for $\lambda_c^-<\lambda<\lambda_c^+$, but remains finite outside this region.}
    \label{fig:eigvalues_open}
\end{figure}

\subsection{Spectral fingerprints of a $\Zsymm{2}$ dynamical phase transition}
   
We now proceed to explore the spectral fingerprints of the phase transition for current fluctuations in the open WASEP, a $\mathbb{Z}_2$ symmetry-breaking DPT. In the following, we set $E=4$ and $\alpha=\beta=\gamma=\delta=0.5$, corresponding to equal densities $\rho_L=\rho_R=0.5$, though all our results apply also to arbitrary strong drivings as far as $\rho_L=1-\rho_R$ and $|E|>\pi$. As discussed in previous sections, the hallmark of any symmetry-breaking DPT is the emergence of a degenerate subspace for the leading eigenvectors of the Doob driven process. Our first goal is hence to analyze the scaled spectral gaps $L^2 \Delta_j^{\lambda}$ associated with the first eigenvalues of the Doob generator $\genmatD$ for the boundary-driven WASEP. 
Note that the $L^2$ scaling in the spectral gaps is required because the system dynamics is diffusive \cite{derrida07a,perez-espigares18a}\todo{esto no estaría mal eplicarlo mejor}. These spectral gaps, obtained from the numerical diagonalization of the tilted generator in Eq.~\eqref{eq:WASEPopen_tilted_gen}, are displayed in Fig.~\ref{fig:eigvalues_open}(a) as a function of $\lambda$ for different system sizes. Recall that by definition $\Delta_0^{\lambda}(\lattsize)=0$ $\forall \lambda$ since $\genmatD$ is a probability-conserving stochastic generator. Moreover, we observe that $L^2 \Delta_{2,3}^{\lambda}(L)>0$ for all $\lambda$ and $L$, so their associated eigenvectors do not contribute to the Doob stationary subspace. On the other hand, $L^2 \Delta_1^{\lambda}(L)$ exhibits a more intricate behavior.
In particular, for $\lambda_c^-< \lambda < \lambda_c^+$ this spectral gap vanishes as $L$ increases, 
signaling that the DPT has already kicked in, while outside of this range $L^2 \Delta_1^{\lambda}(L)$ converges to a non-zero value as $L$ increases.
Note however that the change in the spectral gap behavior across $\lambda_c^{\pm}$ is not so apparent due to the moderate system sizes at reach with numerical diagonalization. In any case, these two markedly different behaviors are more clearly appreciated in Figs.~\ref{fig:eigvalues_open}(b)-\ref{fig:eigvalues_open}(c), which display $L^2\Delta_j^{\lambda}$ for $j=1,2,3$ as a function of $1/L$ for $\lambda=-0.5>\lambda_c^+$ (top panel) and $\lambda=-3.5<\lambda_c^+$ (bottom panel). In particular, a clear decay of $L^2\Delta_1^{\lambda}(L)$ to zero as $1/L\to 0$ is apparent in Fig~\ref{fig:eigvalues_open}(c), while $L^2 \Delta_{2,3}^{\lambda}(L)$ remain non-zero in this limit. In this way, outside the critical region, i.e. for $\lambda > \lambda_c^+$ or $\lambda < \lambda_c^-$, the Doob steady state is unique and given by the first Doob eigenvector, $\probTvectst = \eigvectDR{0}$, as shown in Eq.~\eqref{ssnopt}. This Doob steady state preserves the PH symmetry of the original dynamics. On the other hand, for $\lambda_c^-< \lambda < \lambda_c^+$, the spectral gap vanishes in the asymptotic $L\to\infty$ limit. As a consequence, the second eigenvector of $\genmatD$, $\eigvectDR{1}$, enters the degenerate subspace so that the Doob steady state is now doubly degenerated in the thermodynamic limit, and given by $\probTvectst = \eigvectDR{0} + \eigvectDR{1} \braket{\eigvectDLnk{1}}{\probvecttimenk{0}}$, see Eq.~\eqref{eq:DPT_probvect}, thus breaking spontaneously the PH symmetry of the original dynamics. Note that this is exact in the macroscopic limit, while for finite sizes it is just an approximation valid on timescales $1/\Delta_2^\lambda(L) \ll t \ll 1/\Delta_1^\lambda(L)$, the long time limit being always $\probTvectst \approx  \eigvectDR{0}$ for finite system sizes.

\begin{figure}
    \centering
    \includegraphics[width=1\linewidth]{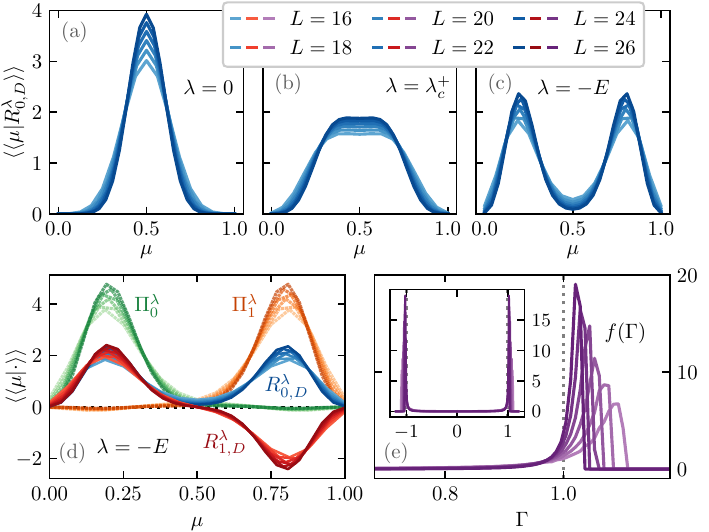}
    \caption{{\bf Structure of the degenerate subspace in the open WASEP.}
    The top panels show the structure of the leading reduced Doob eigenvector $\bbraket{\mu}{R_{0,\text{D}}^\lambda}$ for different values of $\currenttilt$ across the DPT and varying $L$. From left to right: (a) $\currenttilt = -0.5$ (symmetry-preserving phase, before the DPT), (b) $\currenttilt = \lambda_c^+=-1.52$ (critical), (c) $\currenttilt = -4$ (symmetry-broken phase, after the DPT). (d) Comparative structure of the first and second reduced Doob eigenvectors in the degenerate subspace,  $\bbraket{\mu}{R_{j,\text{D}}^\lambda}$ with $j=0, 1$, in the symmetry-broken phase ($\currenttilt = -4$, full lines). The structure of the resulting reduced phase probability vectors $\bbraket{\mu}{\Pi_l^\lambda}$, $l=0,1$, is also shown (dotted lines). (e) Histogram for $\Gamma=\braket{C}{\eigvectDRnk{1}}/\braket{C}{\eigvectDRnk{0}}$ obtained for a large set of configurations $\ket{C}$ sampled from the Doob steady-state distribution for $\lambda = -4$.}
    \label{fig:eigvect_open}
\end{figure}
       
In order to analyze the structure of eigenvectors contributing to the Doob steady state as predicted in Section~\ref{chap:SBDPTs}, we now turn to the order parameter space and the reduced vectors defined in Eq.~\eqref{Pconserv3}, using as order parameter the mean occupation of the lattice $\mass = \lattsize^{-1} \sum_{k=1}^{\lattsize} \occup_k$. This is a good order parameter as defined by its behavior under the symmetry operator $\symmop_{\textrm{PH}}$, see discussion in Section~\ref{ssecOparam}. In this way we extract the relevant macroscopic information contained in the leading Doob eigenvectors, which is displayed in Fig.~\ref{fig:eigvect_open} for different values of the biasing field $\lambda$. In particular, Figs.~\ref{fig:eigvect_open}(a)-\ref{fig:eigvect_open}(c) show the order parameter structure of the leading reduced Doob eigenvector $\kket{R_{0,\text{D}}^\lambda}$ before the DPT [$\lambda=0$, Fig.~\ref{fig:eigvect_open}(a)], at the critical point [$\lambda=\lambda_c^+$, Fig.~\ref{fig:eigvect_open}(b)], and after the DPT [$\lambda=-E$, Fig.~\ref{fig:eigvect_open}(c)]. These panels fully confirm the predictions of Chapter~\ref{chap:SBDPTs}. In particular, before the DPT happens there is only a single phase contributing to the Doob steady state, $\kket{R_{0,\text{D}}^\lambda} = \kket{\Pi_0^\lambda}$, which preserves the $\mathbb{Z}_2$ symmetry of the original dynamics. This is reflected in the unimodality of the distribution $\bbraket{\mu}{R_{0,\text{D}}^\lambda}$ shown in Fig.~\ref{fig:eigvect_open}(a) for different system sizes. Indeed, in this phase $\bbraket{\mu}{R_{0,\text{D}}^\lambda}$ is nothing but the steady state probability distribution of the order parameter $\mu$, see Eq.~\eqref{Pconserv}. Upon approaching the critical point $\lambda=\lambda_c^+$, the distribution $\bbraket{\mu}{R_{0,\text{D}}^\lambda}$ is still unimodal but becomes flat around the peak (i.e. with zero second derivative), see Fig.~\ref{fig:eigvect_open}(b), a feature very much reminiscent of second-order, $\mathbb{Z}_2$ symmetry-breaking phase transitions \cite{binney92a}. In fact, once $\currenttilt$ enters the symmetry-broken region, the distribution $\bbraket{\mu}{R_{0,\text{D}}^\lambda}$ becomes bimodal as shown in Fig.~\ref{fig:eigvect_open}(c), where two different but symmetric peaks around $\mu \approx 0.25$ and $\mu \approx 0.75$, respectively, can be distinguished. The leading reduced Doob eigenvector is still invariant under the symmetry operation, i.e. $\bbraket{\mu}{R_{0,\text{D}}^\lambda}=\bbra{\mu}\hat{S}_\mu\kket{R_{0,\text{D}}^\lambda}$ (recall that $\mass \to 1 -\mass$ under such transformation), but the degenerate subspace also includes now (in the $L\to\infty$ limit) the second reduced Doob eigenvector $\kket{R_{1,\text{D}}^\lambda}$, whose order parameter structure is compared to that of $\kket{R_{0,\text{D}}^\lambda}$ in Fig.~\ref{fig:eigvect_open}(d). Clearly, while $\kket{R_{0,\text{D}}^\lambda}$ is invariant under $\hat{S}_\mu$ as stated, i.e. it has a symmetry eigenvalue $\phi_0=1$, $\kket{R_{1,\text{D}}^\lambda}$ is on the other hand antisymmetric, $\hat{S}_\mu\kket{R_{1,\text{D}}^\lambda}=-\kket{R_{1,\text{D}}^\lambda}$, i.e. $\phi_{1} = \text{e}^{i\pi}=-1$. This result can be also confirmed numerically for the unreduced eigenvectors $\ket{R_{0,\text{D}}^\lambda}$ and $\ket{R_{1,\text{D}}^\lambda}$ in the space of configurations. The reduced phase probability vectors can be written according to Eq.~\eqref{redeq3} as $\kket{\Pi_l^\lambda} = \kket{R_{0,\text{D}}^\lambda} + (-1)^l \kket{R_{1,\text{D}}^\lambda}$, with $l=0,1$, and they simply correspond to the degenerate reduced Doob steady states in each of the symmetry branches, see Fig.~\ref{fig:eigvect_open}(d). Such distributions correspond to each of the symmetry-broken profiles previously shown in Fig.~\ref{fig:DPT_open}(a) for $\lambda=-E$. In general, the reduced Doob steady state in the $L\to\infty$ limit will be a weighted superposition of these two degenerate branches, 
\be
\kket{P_{\text{ss},P_0}^\lambda}  = w_0 \kket{\Pi_0^\lambda} + w_1 \kket{\Pi_1^\lambda} \, , 
\ee
see Eq.~\eqref{coef} and Section~\ref{ssecOparam}, with weights $w_l=(1 + (-1)^l \braket{\eigvectDLnk{1}}{\probvecttimenk{0}})/2$ depending on the initial state distribution $\ket{P_0}$. This illustrates as well the phase selection mechanism via initial state preparation discussed in Section~\ref{ssecPhase}: choosing $\ket{\probvecttimenk{0}}$ such that $\braket{\eigvectDLnk{1}}{\probvecttimenk{0}}=(-1)^l$ leads to a \emph{pure} steady state $\probTvectst = \probDphasevect{l}$.

To end this section, we want to test the detailed structure imposed by symmetry on the Doob degenerate subspace. In particular we showed in Section~\ref{ssecSpec} that, in the symmetry-broken regime, the components $\braket{C}{\eigvectDRnk{j}}$ of the subleading Doob eigenvectors $j\in[1,n-1]$ associated with statistically-relevant configurations $\ket{C}$ are almost equal to those of the leading eigenvector, $\braket{C}{\eigvectDRnk{0}}$, except for a complex phase, see Eq.~\eqref{eq:eigvectsfromeigvect0}. This will be true provided that $\ket{C}$ belongs to a given symmetry class $\ell_C$. For the open WASEP, we have that
\be
\braket{C}{\eigvectDRnk{1}} \approx (-1)^{-\ell_C} \braket{C}{\eigvectDRnk{0}} \, , 
\label{eq:eigvectsfromeigvect0v2}
\ee
where $\ell_C=0,1$ depending whether configuration $\ket{C}$ belongs to the high-$\mu$ or low-$\mu$ symmetry sector, respectively. To test this prediction, we sample a large number of statistically-relevant configurations in the Doob steady state, and study the histogram for the quotient $\Gamma(\conf) = \braket{\conf}{\eigvectDRnk{1}} / \braket{C}{\eigvectDRnk{0}}$, see Fig.~\ref{fig:eigvect_open}(e). As expected from the previous equation, the frequencies $f(\Gamma)$ peak sharply around $(\phi_1)^0=1$ and $\phi_1=-1$, see also inset to Fig.~\ref{fig:eigvect_open}(e), and concentrate around these values as $\lattsize$ increases. This confirms that the structure of the subleading Doob eigenvector $\ket{R_{1,\text{D}}}$ is enslaved to that of $\ket{R_{0,\text{D}}}$ depending on the symmetry basin of each configuration. Moreover, this observation also supports \emph{a posteriori} 
that statistically-relevant configurations can be partitioned into disjoint symmetry classes. In the reduced order-parameter space, the relation between eigenvectors in the degenerate subspace implies that $\bbraket{\mu}{R_{1,\text{D}}^\lambda} \approx (-1)^{-\ell_\mu} \bbraket{\mu}{R_{0,\text{D}}^\lambda}$, where $\ell_\mu=\Theta(\mu-0.5)$ is an indicator function identifying each phase in $\mu$-line, with $\Theta(x)$ the Heaviside step function. In this way, the magnitude and shape of the peaks of $\bbraket{\mu}{R_{1,\text{D}}^\lambda}$ are directly related to those of $\bbraket{\mu}{R_{0,\text{D}}^\lambda}$, despite their antisymmetric (resp. symmetric) behavior under $\hat{S}_\mu$, as corroborated in Fig.~\ref{fig:eigvect_open}(d).

We end by noting that, despite the moderate lattice sizes at reach with numerical diagonalization, the results presented above for the boundary-driven WASEP show an outstanding agreement with the macroscopic predictions of Chapter~\ref{chap:SBDPTs}

\section{Energy fluctuations in spin systems: the $r$-state Potts model}
\label{secPotts}

\subsection{Definition of the model}

The next example is the one-dimensional Potts model of ferromagnets \cite{potts52a}, a generalization of the Ising model. The system consists of a 1D periodic lattice with $\lattsize$ spins $\{s_k\}_{k=1,...,L}$, which can be in any of $\nstatesP$ different states ${s}_k \in \{0, 1, ..., \nstatesP-1\}$ distributed in a unit circle with angles $\spinangle_k = 2\pi s_k /\nstatesP$, as sketched in Fig.~\ref{fig:sketch_potts} for the particular case $r=3$. Nearest-neighbor spins interact according to the Hamiltonian
\begin{equation}
\hamilt = -J \sum_{k=1}^{\lattsize} \cos(\spinangle_{k+1} - \spinangle_k) \, ,
\label{HPott}
\end{equation}    
with $J>0$ a coupling constant favouring the parallel orientation of neighboring spins. Configurations $\{C\}=\{s_k\}_{k=1,...,L}$ can be represented as vectors in a Hilbert space ${\cal H}$ of dimension $r^L$,    
\be
\ket{C}=\bigotimes_{k=1}^L \left( \delta_{s_k,0},\delta_{s_k,1},...,\delta_{s_k,r-1} \right)^T\, , 
\ee
such that $s_k=0$, $s_k=1$, $\ldots$ , $s_k=r-1$ correspond to $\ket{0}_k=(1,0,\ldots,0)$, $\ket{1}_k=(0,1,\ldots,0)$, $\ldots$, $\ket{r-1}_k=(0,0,\ldots,1)$, respectively. Spins evolve stochastically in time according to the single spin-flip Glauber dynamics at inverse temperature $\beta$ \cite{glauber63a}. The stochastic generator $\genmat$ for this model can be hence obtained from the Hamiltonian as $\bra{C'}\genmat\ket{C}=\transitionrate{\conf}{\conf^{\prime}} = 1/\left(1 + \text{e}^{\invtemp \Delta E_{\conf', \conf}}\right)$, where $\Delta E_{\conf', \conf}$ is the energy change in the transition $C\to C'$, which involves a spin rotation. The explicit operator form for $\genmat$ can be then easily obtained from these considerations, but is somewhat cumbersome (see e.g. Appendix B of \cite{marcantoni20a} for an explicit expression of $\genmat$ in the case $\nstatesP=3$). 

\begin{figure}
    \centering
    \includegraphics[width=0.8\linewidth]{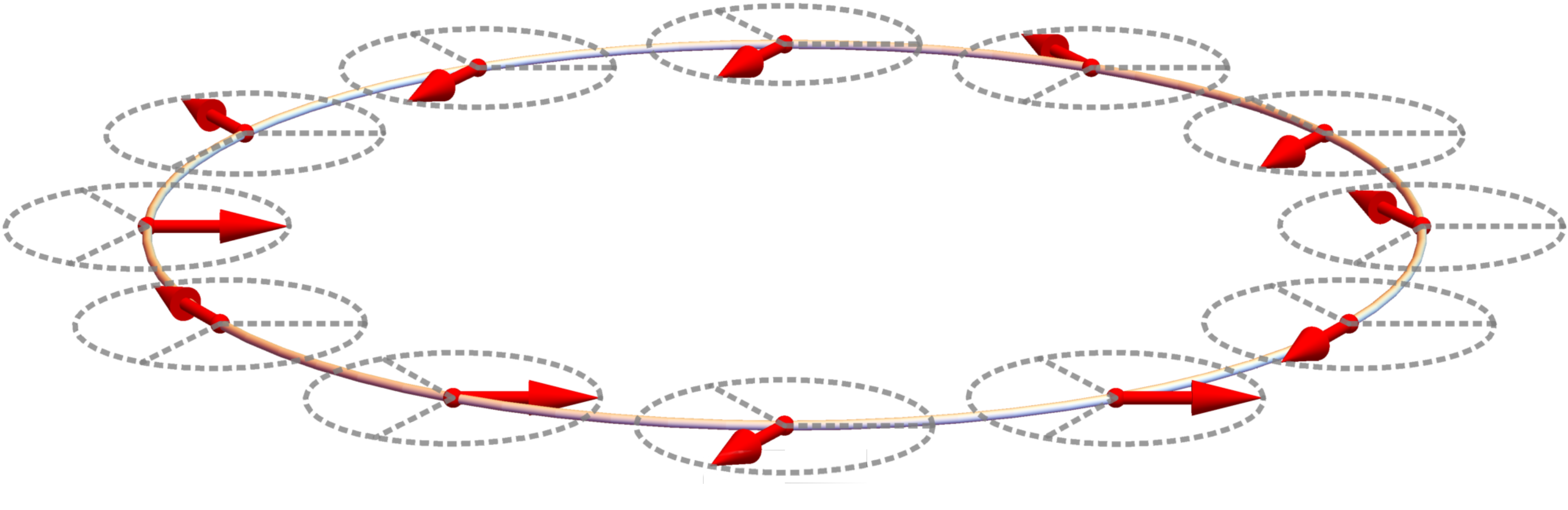}
    \caption{{\bf Sketch of the three-state Potts model.} Each lattice site contains a spin variable with 3 possible in-plane orientations, and neighboring spins interact depending on their relative orientations as described by the Hamiltonian~\eqref{HPott}.}
    \label{fig:sketch_potts}
\end{figure}

Interestingly, the Hamiltonian~\eqref{HPott} is invariant under any global rotation of all the spins for angles multiple of $2\pi/r$. For convenience we define an elementary rotation of angle $2\pi/r$, which transforms $\ket{s}_k$ into $\ket{s+1}_k$ for every spin $k$, with the operator
\begin{equation}
\symmopR = \bigotimes_{k=1}^L \sum_{s=0}^{\nstatesP-1} \left(\ket{s+1}\bra{s}\right)_k \, ,
\label{symmopR}
\end{equation}
where we identify $\ket{\nstatesP}_k = \ket{0}_k$. Note that 
\be
(\symmopR)^r=\hat{S}_{2\pi}=\identityop \, .
\label{symmopR2}
\ee
Since the Glauber dynamics inherits the symmetries of the Hamiltonian, the generator is also invariant under the action of the rotation operator $\symmopR$, i.e.~$[\genmat, \symmopR]=0$. This hence implies that $\genmat$ has a $\Zsymm{r}$ symmetry in the language of Section~\ref{ssecSym}, see Eq.~\eqref{symmopR2}, and makes the $r$-state Potts model a suitable candidate for illustrating our results beyond the $\Zsymm{2}$ symmetry-breaking phenomenon of the previous section, provided that this model exhibits a DPT in its fluctuation behavior.

It is well known that the 1D Potts model does not present any standard phase transition for finite values of $\invtemp$ \cite{baxter16a}. However, we shall show below that it does exhibit a paramagnetic-ferromagnetic DPT for $r=3$ and $r=4$ when trajectories are conditioned to sustain a fluctuation of the time-averaged energy per spin well below its typical value. Indeed, in order to sustain such a low energy fluctuation, the $r$-spin system eventually develops ferromagnetic order so as to maximize the probability of such event, aligning a macroscopic fraction of spins in the same (arbitrary) direction and thus breaking spontaneously the underlying $\Zsymm{\nstatesP}$ symmetry. This symmetry breaking process in energy fluctuations leads to $\nstatesP$ different ferromagnetic phases, each one corresponding to one of the $\nstatesP$ possible spin orientations.
This DPT is well captured by the average magnetization per spin $\meanmag = L^{-1}\sum_{k=1}^{\lattsize} \text{e}^{i\spinangle_k}$, a complex number which plays the role of the order parameter in this case.  
Note that a similar DPT has been reported for the 1D Ising model \cite{jack2010}, which can be seen as a  $\nstatesP = 2$ Potts model.

Notice that, apart from the higher order $\Zsymm{\nstatesP}$ symmetry-breaking process involved in this DPT, a crucial difference with the one observed in the open WASEP is that here the observable whose fluctuations we are interested in (i.e. the energy) is \emph{configuration-dependent}, as opposed to the particle current in WASEP, which depends on state transitions [see Eq.~\eqref{eq:TEobservable_def_theory} and related discussion]. Note also that, as in \cite{jack2010}, temperature does not play a crucial role in this DPT. In particular, a change in the temperature just amounts to a modification of the critical bias $\lambda_c$ at which the DPT occurs, which becomes more negative as the temperature increases. Since the aim of this section is not to characterize in detail the DPT but to analyze the spectral fingerprints of the symmetry-breaking process, we consider $\invtemp = 1$ in what follows without loss of generality.

\begin{figure}
    \centering
    \includegraphics[width=1\linewidth]{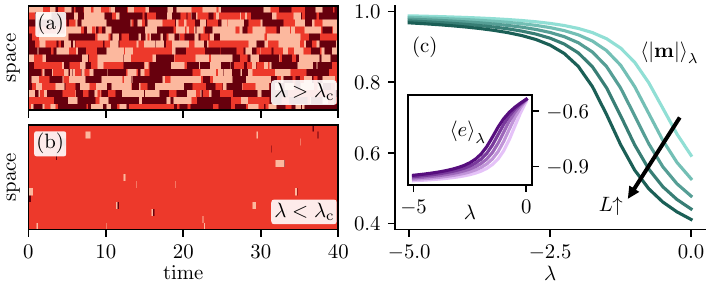}
    \caption{{\bf $\mathbb{Z}_3$ dynamical symmetry breaking in the three-state Potts model.} Left panels: Spatio-temporal raster plots of typical trajectories of the spin system (a) before the DPT ($\lambda>\lambda_c\approx -1$), and (b) once the DPT kicks in ($\lambda<\lambda_c$) for $\lattsize=16$ and $\beta=1$. Each color corresponds to one of the three posible spin states. (c) Magnitude of magnetization  as a function of the biasing field $\energytilt$ for increasing $L$ in the Doob stationary regime. The inset shows the average energy per spin vs $\lambda$. Each color represents a different $\lattsize = 8, 10, 12, 14, 16$, with darker colors corresponding to larger sizes.}
    \label{fig:eigvals_Pottsq3}
\end{figure}

The statistics of the biased trajectories can be obtained from the tilted generator, see Eq.~\eqref{eq:genmatT_theory}, which now reads 
\begin{equation}
\genmatT = \genmat + \lambda \sum_{\conf}e(C) \ketbra{\conf} \, ,
\label{genmatT_theory}
\end{equation}
where $e(C)=\hamilt(\conf)/\lattsize$ is the energy per spin in configuration $\conf$, see Eq.~\eqref{HPott}.

\subsection{Symmetry breaking in the three-state Potts model}

We start by analyzing the three-state model.
The main features of the Potts DPT are illustrated in Fig.~\ref{fig:eigvals_Pottsq3}. In particular, Figs.~\ref{fig:eigvals_Pottsq3}(a)-\ref{fig:eigvals_Pottsq3}(b) show two typical trajectories for different values of the biasing field $\lambda$. These trajectories are obtained using the Doob stochastic generator for each $\lambda$, see Eq.~\eqref{eq:genmatD_theory}. Interestingly, typical trajectories for moderate energy fluctuations [Fig.~\ref{fig:eigvals_Pottsq3}(a), $\lambda>\lambda_c\approx -1$] are disordered, paramagnetic while, for energy fluctuations well below the average, trajectories exhibit clear ferromagnetic order, breaking spontaneously the $\mathbb{Z}_3$ symmetry [Fig.~\ref{fig:eigvals_Pottsq3}(b), $\lambda<\lambda_c$]. The emergence of this ferromagnetic dynamical phase is well captured by the magnetization order parameter. Fig.~\ref{fig:eigvals_Pottsq3}(c) shows the average magnitude of the magnetization $\avgT{|\meanmag|}$ as a function of $\lambda$, while the inset shows the behavior of the average energy per spin, $\avgT{e}$. These observables are calculated from the Doob stationary distribution. As expected, the energy decreases as $\energytilt$ becomes more negative, while the magnitude of the magnetization order parameter increases sharply, the more the larger $L$ is. The presence of a DPT around $\energytilt =\lambda_c \approx -1$ is apparent, although the system sizes at reach via numerical diagonalization (recall that the generator is a $3^L\times 3^L$ matrix) do not allow for a more precise determination of the critical threshold.

\begin{figure}
    \centering
    \includegraphics[width=1\linewidth]{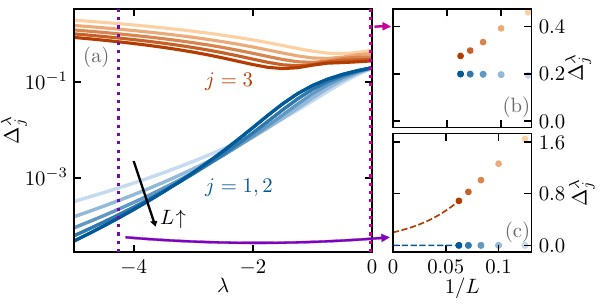}
    \caption{{\bf DPT and quasi-degeneracy in the three-state Potts model.} (a) Evolution of the three leading spectral gaps $\Delta_j^\lambda$, $j=1,2,3$, with the biasing field $\energytilt$ for different lattice sizes $\lattsize$. Note that $\Delta_1^{\lambda}=\Delta_2^{\lambda}$ $\forall \lambda,L$. The right panels show the spectral gaps as a function of the inverse lattice size for (b) $\lambda=0>\lambda_c$ and (c) $\lambda=-4.25<\lambda_c$. For $\lambda>\lambda_c$ the system remains gapped $\forall L$ so the Doob steady state is unique and no symmetry breaking happens. On the other hand, for $\lambda<\lambda_c$ the (equal) spectral gaps $\Delta_{1,2}^{\lambda}$ vanish as $L\to \infty$, leading to a $\mathbb{Z}_3$ dynamical symmetry breaking. Blue symbols correspond to eigenvectors $j=1,2$ and orange symbols to $j=3$, while the dashed lines display the expected behavior. The lattices sizes used are $\lattsize = 8, 10, 12, 14, 16$ (ordered by increasing color intensity).}
    \label{fig:gap_Pottsq3}
\end{figure}

As explained before, the alignment of a macroscopic fraction of spins along a preferential (but arbitrary) direction breaks spontaneously the underlying $\mathbb{Z}_3$ symmetry. We hence expect this DPT to be accompanied by the appearance of a degenerate subspace spanned by the three leading Doob eigenvectors, and the corresponding decay of the second and third spectral gaps $\Delta_j^\lambda$ in the thermodynamic limit ($j=1, 2$, recall that $\Delta_0^\lambda=0$ $\forall \lambda$). This is confirmed in Fig.~\ref{fig:gap_Pottsq3}, which explores the spectral signatures of the Potts DPT. In particular, Fig.~\ref{fig:gap_Pottsq3}(a) shows the evolution with $\lambda$ of the three leading spectral gaps $\Delta_j^\lambda$, $j=1,2,3$, for different system sizes. As expected, while the system remains gapped for $\lambda>\lambda_c$ $\forall L$ [see Fig.~\ref{fig:gap_Pottsq3}(b)], once the DPT kicks in ($\lambda<\lambda_c$) the spectral gaps $\Delta_{1,2}^{\lambda}$ vanish as $L\to \infty$, as confirmed in Fig.~\ref{fig:gap_Pottsq3}(c).
On the other hand, $\Delta_3^{\lambda}$ is expected to remain non-zero, although this is not evident in Fig.~\ref{fig:gap_Pottsq3}(c) due to the limited system sizes at our reach.
Note that  $\Delta_1^{\lambda}=\Delta_2^{\lambda}$ $\forall \lambda$ since $\Delta_j^\lambda = -\Re(\eigvalD{j})$ and $\eigvalD{1}=(\eigvalD{2})^*$ (and similarly for eigenvectors, $\eigvectDR{1} = \eigvectDR{2}^{*}$), as complex eigenvalues and eigenvectors of real matrices such as $\genmatD$ and $\hat{S}_{\frac{2\pi}{3}}$ come in complex-conjugate pairs.
In this case only the eigenvectors are complex, the second and third eigenvalues of $\genmatD$ are real and therefore equal, $\eigvalD{1}=\eigvalD{2}$.
 In fact the corresponding eigenvalues of the symmetry operator $\hat{S}_{\frac{2\pi}{3}}$ are $\symmeigval{0}=1$, $\symmeigval{1}=\text{e}^{i2\pi/3}$, and $\symmeigval{2}=\text{e}^{-i2\pi/3}$.
 Therefore, for $\lambda>\lambda_c$, where the spectrum is gapped, the resulting Doob steady state will be unique as given by the leading Doob eigenvector, $\probTvectst = \eigvectDR{0}$, which remains invariant under $\hat{S}_{\frac{2\pi}{3}}$. 
On the other hand, for $\lambda<\lambda_c$ the two subleading spectral gaps $\Delta_{1,2}^{\lambda}$ vanish as $L\to\infty$, so the Doob stationary subspace is three-fold degenerated in the thermodynamic limit, and the Doob steady state depends on the projection of the initial state along the eigen-directions of the degenerate subspace,
\begin{equation}
\begin{split}
\probTvectst = \eigvectDR{0} &+ \eigvectDR{1} \braket{\eigvectDLnk{1}}{\probvecttimenk{0}} \\
&+ \eigvectDR{2} \braket{\eigvectDLnk{2}}{\probvecttimenk{0}} \, .
\end{split}
\end{equation}
Since we also have that $\eigvectDL{2} = \eigvectDL{1}^{*}$, the above expression can be rewritten as 
\be
\probTvectst = \eigvectDR{0} + 2\Re\left[\eigvectDR{1} \braket{\eigvectDLnk{1}}{\probvecttimenk{0}}\right] \, ,
\label{probPotts}
\ee    
that is, the Doob steady state in the thermodynamic limit is completely specified by the magnitude and complex argument of $\braket{\eigvectDLnk{1}}{\probvecttimenk{0}}$. This steady state for $\lambda<\lambda_c$ breaks the $\mathbb{Z}_3$ symmetry of the spin dynamics since $\hat{S}_{\frac{2\pi}{3}} \probTvectst \ne \probTvectst$. 

\begin{figure}
    \centering
    \vspace{-7mm}
    \includegraphics[width=0.72\linewidth]{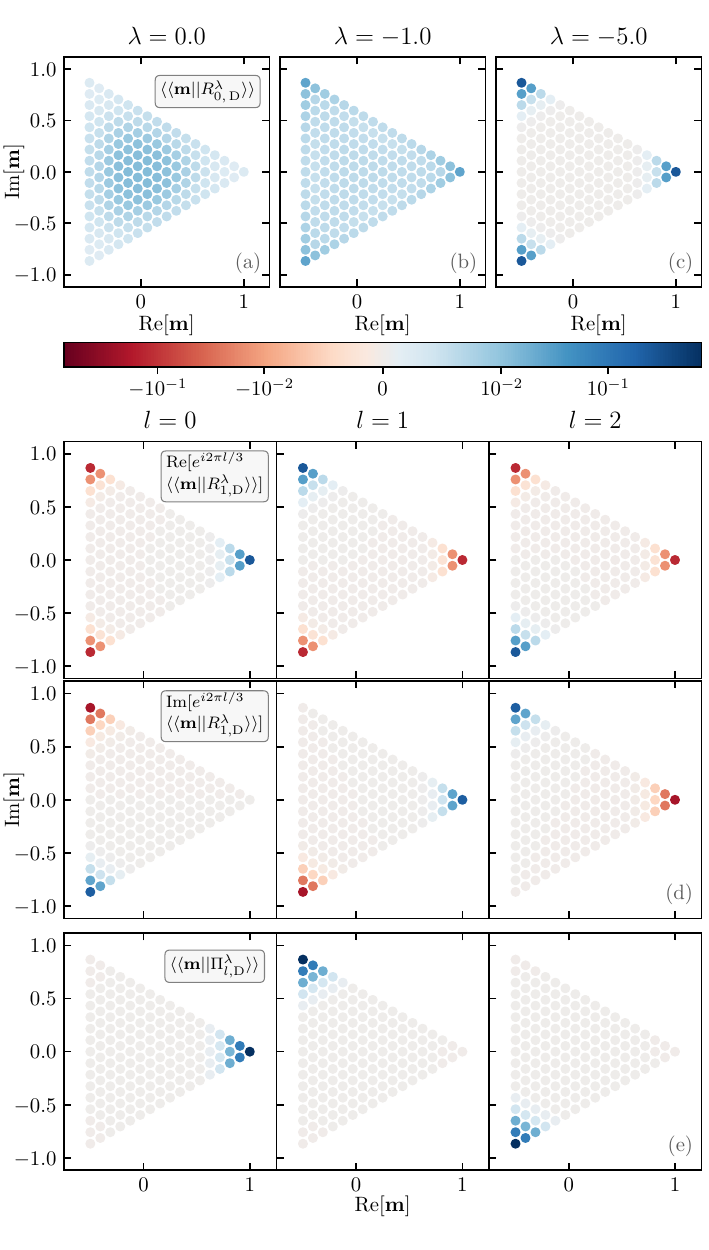}
    \vspace{-5mm}
    \caption{{\bf Quasi-degenerate reduced eigenvectors in the three-state Potts model.} (a)-(c) Structure of $\bbraket{\meanmag}{R_{0,\text{D}}^\lambda}$ in the complex $\meanmag$-plane for different values of $\currenttilt$ across the DPT and $L=16$. From left to right: (a) $\currenttilt = 0$ (symmetry-preserving phase, before the DPT), (b) $\currenttilt = -1 \approx \lambda_c$, (c) $\currenttilt = -5$ (symmetry-broken phase, after the DPT). 
    (d) Structure of $\bbraket{\meanmag}{R_{1,\text{D}}^\lambda}$. The top panels display $\Re[\text{e}^{i 2\pi l/3}\bbraket{\meanmag}{R_{1,\text{D}}^\lambda}]$, while the mid panels show $\Im[\text{e}^{i 2\pi l/3}\bbraket{\meanmag}{R_{1,\text{D}}^\lambda}]$ for $l=0,1,2$. This enables to illustrate the phase selection mechanism of Eq.~\eqref{redeq3}. (e) Structure of the resulting reduced phase probability vectors $\bbraket{\meanmag}{\Pi_l^\lambda}$.
    Note the symmetric logarithmic scale in the color bar.
    \todo[inline]{Hay que dividir esta figura para que entre bien}
    }
    \label{fig:Potts3_eigvect}
\end{figure}

Again, as in the example for the boundary-driven WASEP, we now turn to the reduced magnetization Hilbert space to analyze the structure of the eigenvectors spanning the Doob stationary subspace. In particular, Fig.~\ref{fig:Potts3_eigvect} shows the structure of the leading reduced Doob eigenvector $\kket{R_{0,\text{D}}^\lambda}$ in terms of the (complex) magnetization order parameter $\meanmag = L^{-1}\sum_{k=1}^{\lattsize} \text{e}^{i\spinangle_k}$ before the DPT [$\lambda=0>\lambda_c$, Fig.~\ref{fig:Potts3_eigvect}(a)], around the critical point [$\lambda=-1\approx \lambda_c$, Fig.~\ref{fig:Potts3_eigvect}(b)], and once the DPT is triggered [$\lambda=-5< \lambda_c$, Fig.~\ref{fig:Potts3_eigvect}(c)]. We recall that
\be
\bbraket{\meanmag}{R_{0,\text{D}}^\lambda} = \sum_{\substack{\ket{C}\in{\cal H}: \\ \meanmag(C)=\meanmag}} \braket{C}{R_{0,\text{D}}^\lambda} \, , 
\ee
see Eq.~\eqref{Pconserv}, and note that $\ket{R_{0,\text{D}}^\lambda}$ is always real, and so is the projection $\bbraket{\meanmag}{R_{0,\text{D}}^\lambda}$ which can be hence considered as a probability distribution in the complex $\meanmag$-plane. For $\lambda>\lambda_c$, before the DPT happens, $\bbraket{\meanmag}{R_{0,\text{D}}^\lambda}$ is peaked around $|\meanmag|=0$, see Fig.~\ref{fig:Potts3_eigvect}(a). In this case the spectrum is gapped and there exists a unique, symmetry-preserving reduced Doob steady state $\kket{P_{\text{ss},P_0}^\lambda} = \kket{R_{0,\text{D}}^\lambda}$. When $\lambda\approx\lambda_c$ the distribution $\bbraket{\meanmag}{R_{0,\text{D}}^\lambda}$ flattens and spreads out, see Fig.~\ref{fig:Potts3_eigvect}(b), hinting at the emerging DPT which becomes apparent once $\lambda<\lambda_c$, when $\bbraket{\meanmag}{R_{0,\text{D}}^\lambda}$ develops well-defined peaks in the complex $\meanmag$-plane around regions with $|\meanmag|\approx 1$ and complex phases $\varphi=0,$ $2\pi/3$ and $4\pi/3$, see Fig.~\ref{fig:Potts3_eigvect}(c). In all cases $\kket{R_{0,\text{D}}^\lambda}$ is invariant under the reduced symmetry transformation, $\bbra{\meanmag}\hat{S}_\meanmag\kket{R_{0,\text{D}}^\lambda}=\bbraket{\meanmag}{R_{0,\text{D}}^\lambda}$, where $\hat{S}_\meanmag$ amounts now to a rotation of angle $2\pi/3$ in the complex $\meanmag$-plane that keeps constant $|\meanmag|$. However, for $\lambda<\lambda_c$ the Doob stationary subspace is three-fold degenerate in the $L\to\infty$ limit, and includes now the complex-conjugate pair of eigenvectors $\kket{R_{1,\text{D}}^\lambda}$ and $\kket{R_{2,\text{D}}^\lambda}$, which transform under the reduced symmetry operator as $\hat{S}_\meanmag\kket{R_{j,\text{D}}^\lambda} = \phi_j \kket{R_{j,\text{D}}^\lambda}$ with $\phi_j=\text{e}^{i2\pi j/3}$ and $j=1,2$. The reduced phase probability vectors now follow from Eq.~\eqref{redeq3},
\be
\begin{split}
\kket{\Pi_l^\lambda} &= \kket{R_{0,\text{D}}^\lambda} +  \text{e}^{i 2\pi l/3}\kket{R_{1,\text{D}}^\lambda} + \text{e}^{-i2\pi l/3}\kket{R_{2,\text{D}}^\lambda} \\
& = \kket{R_{0,\text{D}}^\lambda} +  2\Re[\text{e}^{i 2\pi l/3}\kket{R_{1,\text{D}}^\lambda}] \, , 
\end{split}
\ee
with $l=0,1,2$, and define the 3 degenerate Doob steady states, one for each symmetry branch, once the DPT appears. The order parameter structure of these reduced phase probability vectors, as well as that of the real and imaginary parts of the reduced eigenvector $\kket{R_{2,\text{D}}^\lambda}=\kket{R_{1,\text{D}}^\lambda}^*$, are shown in Figs.~\ref{fig:Potts3_eigvect}(d)-\ref{fig:Potts3_eigvect}(e). In particular we display $\Re[\text{e}^{i 2\pi l/3}\bbraket{\meanmag}{R_{1,\text{D}}^\lambda}]$ and $\Im[\text{e}^{i 2\pi l/3}\bbraket{\meanmag}{R_{1,\text{D}}^\lambda}]$, instead of $\Re[\bbraket{\meanmag}{R_{1,\text{D}}^\lambda}]$ and $\Im[\bbraket{\meanmag}{R_{1,\text{D}}^\lambda}]$, respectively, to illustrate the phase selection mechanism of Eq.~\eqref{eq:probphasevect} while conveying the full complex structure of this eigenvector. For instance, the phase vector $\kket{\Pi_0^\lambda}$ can be selected by just adding $2\Re[\kket{R_{1,\text{D}}^\lambda}]$ to $\kket{R_{0,\text{D}}^\lambda}$, see the $l=0$ in Figs.~\ref{fig:Potts3_eigvect}(d)-\ref{fig:Potts3_eigvect}(e), while for the other two $\kket{\Pi_l^\lambda}$ a complex phase is required to \emph{rotate} $\kket{R_{1,\text{D}}^\lambda}$ and cancel out the undesired peaks in $\kket{R_{0,\text{D}}^\lambda}$. A generic steady state will correspond to a weighted superposition of these phase probability vectors, $\ket{\probTvectst} = \sum_{l=0}^{2} w_l \ket{\probDphasevect{l}}$, with the statistical weights depending on the initial state, see Eq.~\eqref{coef}.

\begin{figure}
    \centering
    \includegraphics[width=0.8\linewidth]{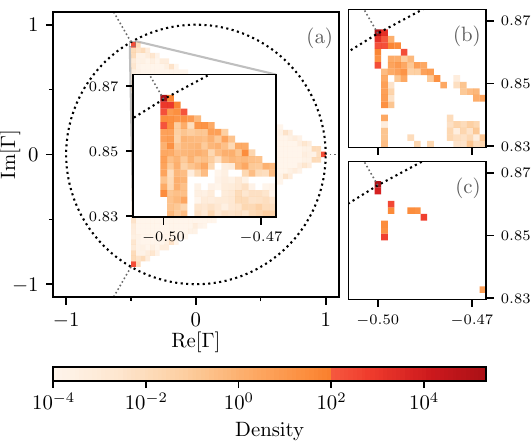}
    \caption{{\bf Structure of the degenerate subspace in the three-state Potts model.}
    (a) Density plot for $\Gamma=\braket{C}{\eigvectDRnk{1}}/\braket{C}{\eigvectDRnk{0}}$ in the complex $\Gamma$-plane obtained for a large set of configurations $\ket{C}$ sampled from the Doob steady-state distribution for $L=16$ and $\lambda = -5$. The inset is a zoom on one of the compact regions around the complex unit circle where points converge. The small panels to the right show the same density plot, as obtained for different system sizes (b) $L=12$ and (c) $L=8$. 
    A different color is used for the highest values of the density to highlight the sharp concentration around the cube roots of unity.
    }
    \label{fig:Potts3_deg}
\end{figure}

The intimate structural relation among the leading eigenvectors defining the degenerate subspace in the symmetry-broken regime ($\lambda<\lambda_c$) can be studied now by plotting $\Gamma(\conf) = \braket{\conf}{\eigvectDRnk{1}} / \braket{C}{\eigvectDRnk{0}}$ in the complex plane for a large sample of statistically-relevant configurations $\ket{C}$. As predicted in Eq.~\eqref{eq:eigvectsfromeigvect0}, this quotient should converge as $L$ increases to
\be
\Gamma(C) \approx (\text{e}^{i 2\pi/3})^{-\ell_C} \, ,
\label{cregions}
\ee
with $\ell_C=0,1,2$ identifying the symmetry sector to which configuration $\ket{C}$ belongs in. Fig.~\ref{fig:Potts3_deg} shows the density histogram in the complex $\Gamma$-plane obtained in this way for $\lambda=-5<\lambda_c$ and a large sample of important configurations. As predicted, all points concentrate sharply around three compact regions around the complex unit circle at phases $\varphi=0,~2\pi/3$ and $4\pi/3$, see Eq.~\eqref{cregions} and the inset in Fig.~\ref{fig:Potts3_deg}.
Notice that, even though a log scale is used to appreciate the global structure, practically all density is contained in a very small region around the cube roots of unity.
Moreover, the convergence to the predicted values as $L$ increases is illustrated in Figs.~\ref{fig:Potts3_deg}(b)-\ref{fig:Potts3_deg}(c). Equivalently, in the reduced order parameter space we expect that
\be
\bbraket{\meanmag}{R_{j,\text{D}}^\lambda} \approx (\text{e}^{i 2\pi j/3})^{-\ell_\meanmag} \bbraket{\meanmag}{R_{0,\text{D}}^\lambda} \, ,
\ee
for $j=1,2$, where $\ell_\meanmag=0,1,2$ is a characteristic function identifying each phase in the $\meanmag$-plane. This relation implies that the size and shape (and not only the positions) of the different peaks in $\bbraket{\meanmag}{R_{j,\text{D}}^\lambda}$, $j=0,1,2$, in the complex plane are the same, see Eq.~\eqref{redeq4}, a general relation also confirmed in the boundary-driven WASEP.

\begin{figure}
    \vspace{-4mm}
    \hspace{0.03\linewidth}\includegraphics[width=0.9\linewidth]{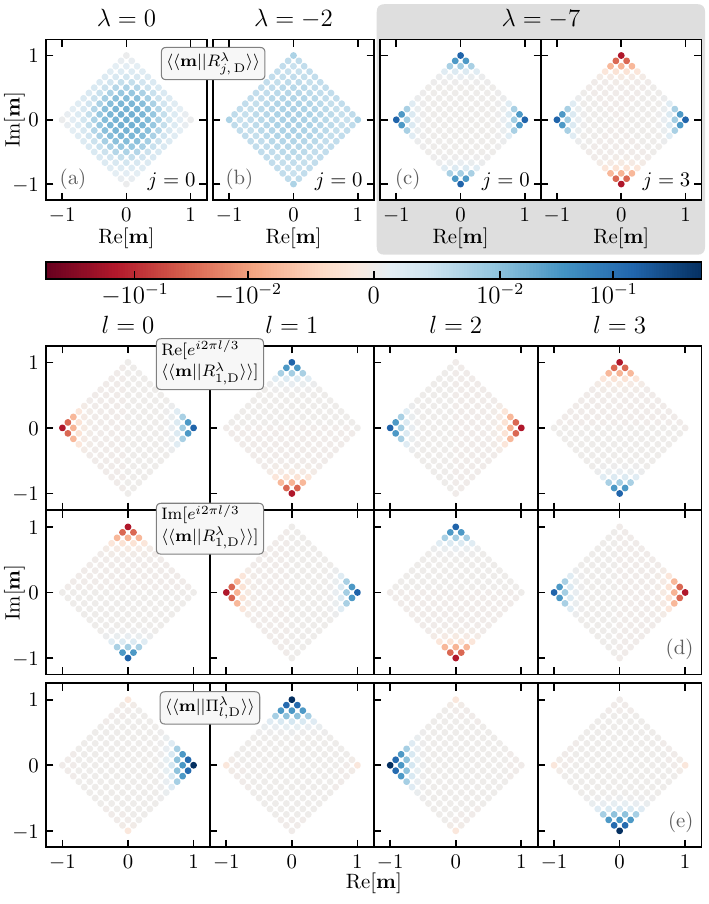}
    \vspace{-3mm}
    \caption{{\bf Quasi-degenerate reduced eigenvectors in the four-state Potts model.} 
    Panels (a)-(c) show the structure of $\bbraket{\meanmag}{R_{0,\text{D}}^\lambda}$ in the complex $\meanmag$-plane for different values of $\currenttilt$ across the DPT and $L=12$. From left to right: (a) $\currenttilt = 0$ (symmetry-preserving phase, before the DPT), (b) $\currenttilt = -2 \approx \lambda_c$, (c) $\currenttilt = -7$ (symmetry-broken phase, after the DPT). The panel to the right of (c)
    shows $\bbraket{\meanmag}{R_{3,\text{D}}^\lambda}$, which is also real. (d) Real and imaginary structure of $\bbraket{\meanmag}{R_{1,\text{D}}^\lambda}$. The top panels display $\Re[\text{e}^{i 2\pi l/4}\bbraket{\meanmag}{R_{1,\text{D}}^\lambda}]$, while the mid panels show $\Im[\text{e}^{i 2\pi l/4}\bbraket{\meanmag}{R_{1,\text{D}}^\lambda}]$ for $l=0,1,2,3$. This illustrates the phase selection mechanism of Eq.~\eqref{redeq3}. Panels (e) show the structure of the resulting reduced phase probability vectors $\bbraket{\meanmag}{\Pi_l^\lambda}$. 
    Note the symmetric logarithmic scale in the color bar.
    \todo[inline]{Me gustaría centrar perfectamente la figura}
    }
    \label{fig:Potts4_eigvect}
\end{figure}

\begin{figure}
    \centering
    \includegraphics[width=0.8\linewidth]{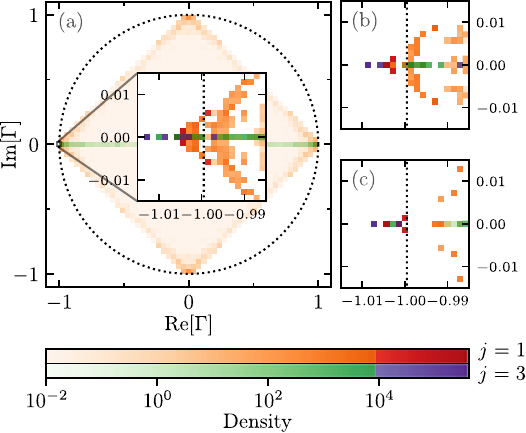}
    \caption{{\bf Structure of the degenerate subspace for the four-state Potts model.} 
    (a) Density plot for $\Gamma_j=\braket{C}{\eigvectDRnk{j}}/\braket{C}{\eigvectDRnk{0}}$ in the complex $\Gamma$-plane for $j=1,3$ obtained for a large set of configurations $\ket{C}$ sampled from the Doob steady-state distribution with $\lattsize = 12$ and $\lambda = -7$. The inset zooms on one of the regions, given by Eq.~\eqref{eq:eigvectsfromeigvect0} where the density peaks. The small panels to the right show the same density plot, as obtained for different system sizes: (b) $L = 10$  and (c) $L = 8$. A comparison of the different panels allows us to appreciate the convergence to the predicted values as $L$ increases, even though in this case the difference is subtle due to the similar lattice sizes. Notice the log scale in the color bar, which shows that almost all the density is contained in a small region around $\pm 1$ and $\pm i$, as predicted.
    }
    \label{fig:Potts4_eigvect_complete}
\end{figure}

\subsection{Phase transition in the four-state Potts model}
\todo[inline]{Me parece un título terrible}

Equivalent ideas hold valid for the ferromagnetic dynamical phase found in the four-state Potts model. In this case, when the system is conditioned to sustain a large time-averaged energy fluctuation well below its average, a similar DPT to a dynamical ferromagnetic phase appears, breaking spontaneously the $\Zsymm{4}$ discrete rotational symmetry of this model. Hence we expect a degenerate Doob stationary subspace spanned by the first four leading eigenvectors, with eigenvalues under the symmetry operator given by $\symmeigval{0} = 1$, $\symmeigval{1} = \text{e}^{i 2\pi/4}$ and $\symmeigval{2} = \text{e}^{-i 2\pi/4}$ and $\symmeigval{3} = -1$. The generic Doob steady state can be then written as

\begin{equation}
\begin{split}
\probTvectst = \eigvectDR{0} + 2\Re[\eigvectDR{1} \braket{\eigvectDLnk{1}}{\probvecttimenk{0}}]  \\
+ \eigvectDR{3} \braket{\eigvectDLnk{3}}{\probvecttimenk{0}} \, , 
\end{split}
\end{equation}
where $\eigvectDR{3}$ is purely real and we have used that $\eigvectDR{2} = \eigvectDR{1}^{*}$. Fig.~\ref{fig:Potts4_eigvect} summarizes the spectral signatures of the DPT in the reduced order parameter space for the four-state Potts model. In particular, Figs.~\ref{fig:Potts4_eigvect}(a)-\ref{fig:Potts4_eigvect}(c) show $\bbraket{\meanmag}{R_{0,\text{D}}^\lambda}$ for different values of $\lambda$ across the DPT. This distribution, which exhibits now four-fold symmetry, goes from unimodal around $|\meanmag|=0$ for $\lambda> \lambda_c$ to multimodal, with 4 clear peaks, for $\lambda< \lambda_c$, as expected. Figure~\ref{fig:Potts4_eigvect}(c) also includes $\bbraket{\meanmag}{R_{3,\text{D}}^\lambda}$ for this $\lambda$, which is purely real. Figures~\ref{fig:Potts4_eigvect}(d) capture the real and imaginary structure of $\bbraket{\meanmag}{R_{1,\text{D}}^\lambda}$ for $\lambda< \lambda_c$, in a manner equivalent to Fig.~\ref{fig:Potts3_eigvect}(e) (recall that $\bbraket{\meanmag}{R_{2,\text{D}}^\lambda}$ is simply its complex conjugate). Interestingly, the presence of a fourth eigenvector in the degenerate subspace for $\lambda<\lambda_c$ makes for a richer phase selection mechanism. Indeed, the reduced phase probability vectors are now
\begin{equation}
\kket{\Pi_l^\lambda} = \kket{R_{0,\text{D}}^\lambda} +  2\Re[\text{e}^{i \pi l/2}\kket{R_{1,\text{D}}^\lambda}] + (-1)^l  \kket{R_{3,\text{D}}^\lambda} \, . 
\end{equation}
Their order parameter structure $\bbraket{\meanmag}{\Pi_l^\lambda}$ is displayed in Figs.~\ref{fig:Potts4_eigvect}(e). The $j=3$ eigenvector transfers probability from the configurations with magnetizations either in the horizontal or vertical orientation to the other one, see Fig.~\ref{fig:Potts4_eigvect}(c), while the combination of the second and third eigenvectors transfers probability between the two directions as dictated by the complex argument of $\braket{\eigvectDLnk{1}}{\probvecttimenk{0}}$, see Eq.~\eqref{coef}. Finally, Fig.~\ref{fig:Potts4_eigvect_complete} confirms the tight symmetry-induced structure in the degenerate subspace for $\lambda<\lambda_c$ by plotting $\Gamma_j(\conf) = \braket{\conf}{\eigvectDRnk{j}} / \braket{C}{\eigvectDRnk{0}}$ in the complex plane for $j=1,3$ and a large sample of statistically-relevant configurations $\ket{C}$. 
As expected, the point density associated with each eigenvector peaks around $(\symmeigval{j})^{-\ell_C}$, which in this case corresponds to $\pm 1$, $\pm i$ 
for $\eigvectDR{1}$, and $\pm 1$ for $\eigvectDR{3}$. Also, the density concentrates more and more around these points as $L$ increases. 


\section{Conclusion}

In this chapter, we have illustrated the general results in Chapter~\ref{chap:SBDPTs} by analyzing three distinct DPTs in several paradigmatic many-body systems.
In particular, we have analyzed the one-dimensional boundary-driven WASEP, which exhibits a particle-hole symmetry-breaking DPT for current fluctuations and the Potts model for spin dynamics (with $r=3$ and $4$ states), which displays discrete rotational symmetry-breaking DPTs for energy fluctuations.
 Our results on the spectral fingerprints of symmetry-breaking DPTs are fully confirmed in these examples, offering a fresh view of spontaneous symmetry-breaking phenomena at the fluctuating level and allowing a deeper understanding of their DPTS.
    }

    \chapter{Building continuous time crystals from rare events: the periodic WASEP}
    \chaptermark{Building continuous time crystals from rare events}
    \label{chap:buildingtc}
    \todo[inline]{qué es un coherent rare event?}
    {

\newcommand{\macrofraction}{\mathcal{O}(\lattsize)}

As suggested in the previous chapter, here we will delve into the analysis of the DPT displayed by the WASEP with periodic boundaries when sustaining currents well below the typical value.
As we will see, this dynamical phase transition exhibits the characteristics of a time crystal, spontaneously breaking the continuous symmetry under rotations and time translations.
This phenomenon will lead to more complex and interesting behaviors than those observed in the previous examples, which will be analyzed in detail.

\section{Introduction}

The concept of time crystal, first introduced by Wilczek and Shapere in 2012 \cite{wilczek12a, shapere12a}, describes many-body systems that spontaneously break time-translation symmetry, a phenomenon that leads to persistent oscillatory behavior and fundamental periodicity in time  \cite{zakrzewski12a,richerme17a,yao18a, sacha18a,sacha20a}.
The fact that a continuous symmetry might appear broken comes as no surprise in general.
Indeed spontaneous symmetry-breaking phenomena, where a system ground or lowest-energy state can display fewer symmetries than the associated action, are common in nature.
However, time translation symmetry had resisted this picture for a long time, as it seemed fundamentally unbreakable. 

The progress made over the last decade has challenged this scenario showing that both continuous and discrete time-translation symmetry can be spontaneously broken, giving rise to the so-called continuous and discrete time crystals, respectively.
In quantum settings, the formers are prohibited in isolated systems by virtue of a series of no-go theorems \cite{bruno13a,nozieres13a,watanabe15a}, which are, however, circumvented in a dissipative context allowing for continuous time crystals \cite{iemini18a,buca19b,kessler19a,carollo20a1,carollo22a}.
An experimental realization of such dissipative continuous time crystals has been recently reported in a dissipative atom-cavity system \cite{kongkhambut_observation_2022}.
On the other hand, quantum discrete time crystals can emerge as a subharmonic response to a periodic driving by exhibiting an oscillating behavior at a slower frequency than the one of the driving, and have been theoretically proposed \cite{else16a,moessner17a,yao17a,gong18a,gambetta19a,khemani_brief_2019,Lazarides20a,Else20a} and experimentally observed in isolated \cite{zhang17a,choi17a,Rovny18a,smits18a,Autti18a,Sullivan20a,Kyprianidis21a,Randall21a,Mi22a} and dissipative settings \cite{kessler21a,Kongkhambut21a}.
As for classical systems \cite{shapere12a}, there has been as well intensive research leading to the finding of discrete time crystals in a periodically-driven two-dimensional Ising model \cite{gambetta19b} and in a one-dimensional system of coupled, non-linear pendula at finite temperature \cite{yao20a}; as well as to their experimental demonstration in a classical network of dissipative parametric resonators \cite{Heugel19a}. Further, continuous time crystals have been lately reported in a two-dimensional array of plasmonic metamolecules displaying a superradiant-like state of transmissivity oscillations \cite{Liu23a}. However, despite such recent examples, a general approach to devise classical systems displaying continuous time-crystalline phases has been elusive so far.   

\begin{figure}
    \centering
    \includegraphics[width=0.7\linewidth]{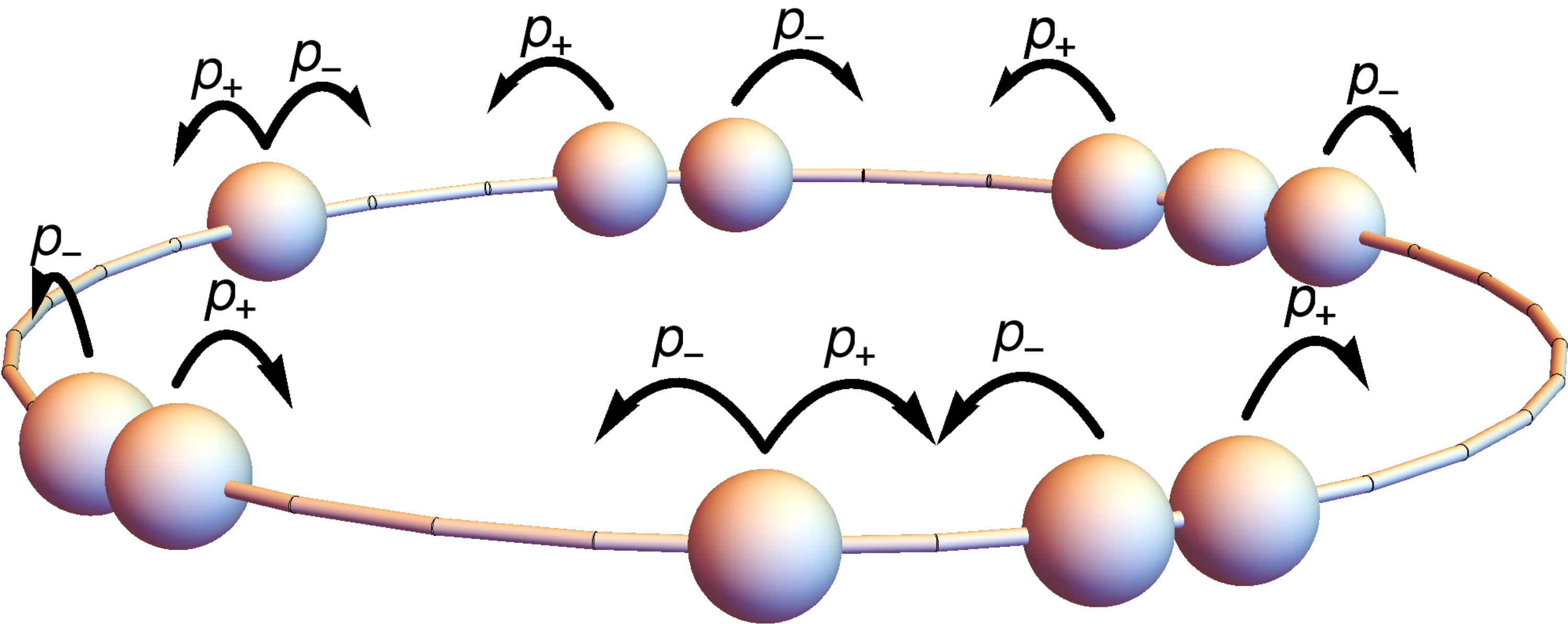}

    \vspace{0.5cm}
    \caption{Sketch of the WASEP with periodic boundary conditions. 
    The stochastic particle jumps occur now in a periodic lattice, so the total number of particles is conserved during the evolution.}
    \label{fig:WASEP_def_closed}
\end{figure}

In this chapter, we propose an alternative route to search for time-crystalline order in classical settings, based on the recent observation of spontaneous symmetry breaking in the dynamical fluctuations of many-body systems \cite{bertini05a, bodineau05a, lecomte07c, garrahan07a, 
hurtado11a, 
jack15a, baek15a, 
}. 
As mentioned earlier, such fluctuations or rare events concern time-integrated observables and are highly unlikely to occur, since their probability decays exponentially with time, thus following a large deviation principle \cite{touchette09a}.
However, when these fluctuations come about, they may lead to dynamical phase transitions (DPTs), which manifest as drastic changes in the trajectories of the system.
This is the case of the one-dimensional periodic WASEP \cite{spitzer70a, derrida98a, derrida98b, golinelli06a, perez-espigares13a, hurtado14a, tizon-escamilla17b}, depicted in Fig.~\ref{fig:WASEP_def_closed}.
When conditioned to sustain a time-integrated current fluctuation well below its average, this model develops a jammed density wave or rotating condensate that hinders particle transport and thus facilitates the fluctuation \cite{bodineau05a,perez-espigares13a}.
As we will see, such a DPT breaks the continuous time-translational symmetry of the original model and gives rise to the emergence of a periodic condensate, thus opening the door to its use as a resource to build continuous time crystals.
We will first demonstrate that the rotating condensate that emerges after the DPT indeed corresponds to a time-crystal phase at the fluctuating level. 
This will done by throughoutly characterizing the spectral fingerprints of the DPT, using the tools of Chapter~\ref{chap:SBDPTs}.
Then, the particular dynamics of the time-crystal phase will be studied by means of Doob's transform.
This will allow us to discover that this dynamics can be interpreted in terms of the original dynamics supplemented with a \emph{packing field},
This field acts by pushing particles that lag behind the condensate's center of mass while restraining those moving ahead, which amplifies naturally occurring fluctuations of the 
density and eventually leads to the formation of a traveling condensate and the onset of the time-crystal phase.
Finally, the key properties of the mechanism found in the periodic WASEP will be distilled in order to introduce a new model, the time-crystal lattice gas (tcLG).
This model will be shown to display a phase transition to a time crystal in its typical behavior, instead of as a rare event as in the WASEP.

\section{Model and dynamical phase transition}

We start by defining the dynamics of the periodic---or \emph{closed}---WASEP.
The dynamics of this model, depicted in Fig.~\ref{fig:WASEP_def_closed}, are similar to the one of the open WASEP of Section~\ref{sec:openWASEP_DPTsmodels} but with the lattice arranged in a ring instead of being connected to reservoirs at its ends.
\todo{se entiende?}
The generator of the microscopic dynamics is thus
\begin{equation}
\begin{split}
    \genmat
    =
    \sum_{k=1}^{\lattsize} \Big[\; &\jumprateR \big( \creationop_{k+1} \destructop_{k} - \occupop_{k  } (\identityop_{k+1} - \occupop_{k+1} )\big) \\ 
    + &\jumprateL ( \creationop_{k  } \destructop_{k+1} - \occupop_{k+1} (\identityop_{k} - \occupop_{k  } )\big)\Big]
    \, ,
\end{split}
\label{eq:WASEPclosed__gen}
\end{equation}
where jump rates are $\jumprate_{\pm} = \expsup{\pm\E/\lattsize}$, with $\lattsize$ the number of lattice sites, and where we identify the site $k=L+1$ with the first site $k=1$ to implement the periodic boundaries.
As in the open WASEP, $\creationop_{k}$, $\destructop_{k}$ are the creation and annihilation operators respectively, given by $\hat{\sigma}^{\pm}_k= (\hat{\sigma}^x_k \pm i \hat{\sigma}^y_k)/2  $, with $\hat{\sigma}^{x,y}_k$ the standard $x,y$-Pauli matrices, while $\occupop_{k}=\hat{\sigma}^+_k \hat{\sigma}^-_k$ and $\identityop_k$ are the occupation and identity operators acting on site $k$.

Unlike the boundary-driven case, where the reservoirs allowed the exchange of particles, in this model the total number of particles $\nprtcls$ remains constant during its evolution, so that the mean density $\dens_0 = \nprtcls/\lattsize$ becomes a parameter.
As a consequence, the particle-hole (PH) symmetry that appeared in the boundary-driven WASEP for reservoir densities $\rho_\text{R}=1-\rho_\text{L}$, is only present in the closed WASEP when the global density is $\dens_0 = 1/2$, since it is the only case in which the PH transformation keeps the $\meandens$ constant. 
Instead, the closed WASEP is invariant under the translation operator, $\symmop_{T}$, which shifts all the particles one site to the right, $k \to k+1$.
Such operator reads,
\begin{equation}
    \label{eq:trans_operator_buildingtc}
    \symmop_{\textrm{T}}
    =
    \prod_{k=1}^{\lattsize-1} \Big[\creationop_{k}\destructop_{k+1} + \creationop_{k+1}\destructop_{k} + \frac{1}{2}(\hat{\sigma}^z_k \hat{\sigma}^z_{k+1} + \identityop) \Big] \, , 
\end{equation}
so that the invariance under this symmetry transformation means that $[\genmat,\symmop_{\textrm{T}}]=0$.
This corresponds to a $\Zsymm{\lattsize}$ symmetry, with $\lattsize$ the number of sites in the lattice, as it can be readily checked that $(\symmop_{\textrm{T}})^L=\identityop$, since a whole turn leaves the ring lattice invariant.
This symmetry is different from the examples in Chapter~\ref{chap:DPTs_models}, where the order of the symmetry was constant.
In this case, the order of the symmetry increases with the system size, approaching a continuous symmetry in the thermodynamic limit.
While this will introduce some changes from the original assumption in Chapter~\ref{chap:SBDPTs}, we will see that the main results still hold after some adaptations.
Regarding its description at the macroscopic level, the WASEP is characterized by a density field $\rho(x,t)$ which obeys a hydrodynamic equation \cite{spohn12a}
\begin{equation}
    \label{eq:hydroWASEP_buildingtc}
    \partial_t \dens
    =
    -\partial_x \Big(-\diffcoef(\dens)\partial_x \dens + \mobility(\dens) \E \Big) \, ,
\end{equation}
with $\diffcoef(\dens)$ and $\mobility(\dens)$ the diffusivity and mobility coefficients, which for WASEP are $\diffcoef(\dens)=1/2$ and $\mobility (\dens)=\dens (1-\dens)$.

\begin{figure}
    \centering
    \includegraphics[width=0.9\linewidth]{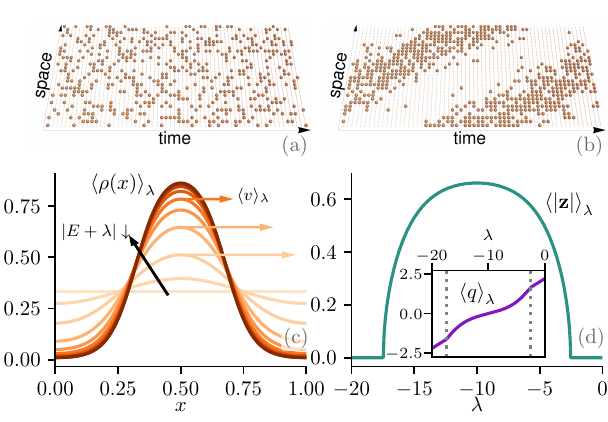}
    \caption{
        $\mathbb{Z}_L$ symmetry-breaking DPT in the closed WASEP.
        Top panels: Typical spacetime trajectories of the closed WASEP for current fluctuations above (a) and below (b) the critical current.
        Note the periodic boundary conditions and the emergence of a jammed matter wave below the critical current.
        (c) Density profile of the rotating condensate for different values of $\currenttilt$.
        (d) Average magnitude of the packing order parameter as a function of $\lambda$.
        The inset shows the average current vs $\lambda$, which becomes nonlinear in the symmetry-breaking regime.
        The results in panels (c) and (d) correspond to macroscopic predictions obtained using the Macroscopic Fluctuation Theory (MFT) as in \cite{perez-espigares13a}.
        In all panels, the global density is $\rho_0=1/3$ and $E=10>E_c$.
    }
    \label{fig:DPT_WASEPclosed}
\end{figure}


Interestingly, the closed WASEP also presents a symmetry-breaking DPT when the system is biased towards currents well below its typical value and in the presence of a strong enough field $\E$ \cite{bodineau05a,perez-espigares13a,hurtado-gutierrez20a}.
As in the open WASEP, the time-extensive current $\intcurrent$ is defined as the number jumps to the right minus the jumps to the left per site during a trajectory of duration $\traj$; so that the time-intensive one is just $\current = \intcurrent/\trajduration$.
When this DPT kicks in, the particles cluster together to form a jammed rotating condensate, see Figs.~\ref{fig:DPT_WASEPclosed}(a)-\ref{fig:DPT_WASEPclosed}(b).
This condensate gives rise to regions of either high or low density, characterized by low mobility $\mobility(\dens)$, thus hindering particle motion and facilitating the low current fluctuation.
This condensate spontaneously breaks the translation symmetry $\symmop_{T}$ and, whenever $\rho_0\ne 1/2$, travels at constant velocity along the lattice, thus breaking time-translation symmetry as well \cite{bodineau05a,perez-espigares13a}.
These features are the fingerprint of the recently discovered time-crystal phase of matter \cite{wilczek12a,shapere12a,zakrzewski12a,moessner17a,richerme17a,yao18a,sacha18a,hurtado-gutierrez20a}.

In order to study the current fluctuations giving rise to this DPT, we introduce the corresponding tilted generator according to Eq.~\eqref{eq:genmatT_theory}, 
\begin{equation}
\begin{split}
    \genmatT
    =
    \sum_{k=1}^{\lattsize} \Big[\; &\jumprateR \big( \text{e}^{+\currenttilt/\lattsize}\creationop_{k+1} \destructop_{k} - \occupop_{k  } (\identityop_{k+1} - \occupop_{k+1} )\big) \\ 
    + &\jumprateL (\text{e}^{-\currenttilt/\lattsize} \creationop_{k  } \destructop_{k+1} - \occupop_{k+1} (\identityop_{k} - \occupop_{k  } )\big)\Big] \, , \nonumber
\end{split}
\label{eq:WASEPclosed_tilted_gen}
\end{equation}
where $\currenttilt$ is the biasing field conjugated to the current.
\todo[inline]{Habría que mencionar que en este caso el generador tiltado hereda la simetría}

Specifically, the aforementioned DPT appears for external fields $\abs{\field} > E_c=\pi/\sqrt{\dens_0(1-\dens_0)}$ and for currents $\abs{\current} \le \crit{\current}=\dens_0(1-\dens_0)\sqrt{E^2-E_c^2}$, which correspond to biasing fields $\currenttiltcritmin < \currenttilt < \currenttiltcritmax$, with $\currenttiltcritminmax = -\field \pm\sqrt{\field^2 - \crit{\field}^2}$  \cite{bodineau05a,perez-espigares13a}.
For $\lambda$ outside this regime, the typical density field sustaining the fluctuation is just flat, structureless [Fig.~\ref{fig:DPT_WASEPclosed}(a)], while within the critical region a density wave [Fig.~\ref{fig:DPT_WASEPclosed}(b)] with a highly nonlinear profile develops.
Figure \ref{fig:DPT_WASEPclosed}(c) shows the density profiles of the resulting jammed condensate for different values of $\currenttilt$ in the macroscopic limit, while the inset to Fig.~\ref{fig:DPT_WASEPclosed}(d) displays the how the current drops as $\currenttilt$ decreases, developing a nonlinear behavior inside the critical regime.

A suitable way to characterize this DPT consists in measuring the packing of the particles in the 1D ring.
Therefore, a good order parameter for the DPT will be the position of the center of mass of the system in the two-dimensional plane.
For a configuration $\conf=\{\occup_k(\conf)\}_{k=1,\ldots,\lattsize}$, with $\occup_k(\conf)=0,1$ the occupation number of site $k$ at $\conf$, this is given by the \emph{packing order parameter} $\KDopar (\conf)$, defined as 
\begin{equation}
    \KDopar(\conf)
    =
    \frac{1}{N} \sum_{k=1}^{\lattsize} \occup_k(\conf) \, \text{e}^{i 2\pi k/\lattsize} = |\KDopar(\conf)| \text{e}^{i \varphi(\conf)} \, .
    \label{coher}
\end{equation}
The magnitude $|\KDopar(\conf)|$ of this packing parameter is close to zero for any homogeneous distribution of particles in the ring, but increases significantly for condensed configurations, while its complex phase $\varphi(\conf)$ signals the angular position of the condensate's center of mass.
In this way, we expect $|\KDopar(\conf)|$ to increase from zero when the condensate first appears at the DPT.
This is in fact confirmed in Fig.~\ref{fig:DPT_WASEPclosed}(d), which shows the evolution of $\avg{|\KDopar|}_\currenttilt$ as a function of the biasing field. 
Note that this order parameter corresponds to the well-known Kuramoto order parameters of synchronization \cite{kuramoto84a,kuramoto87a,pikovsky03a,acebron05a}.
\todo[inline]{Me gustaría hacer un diagrama ilustrando esto}

\section{Spectral analysis of the time-crystal phase} 

\begin{figure}
    \centering
    \vspace{-4mm}
    \includegraphics[width=0.78\linewidth]{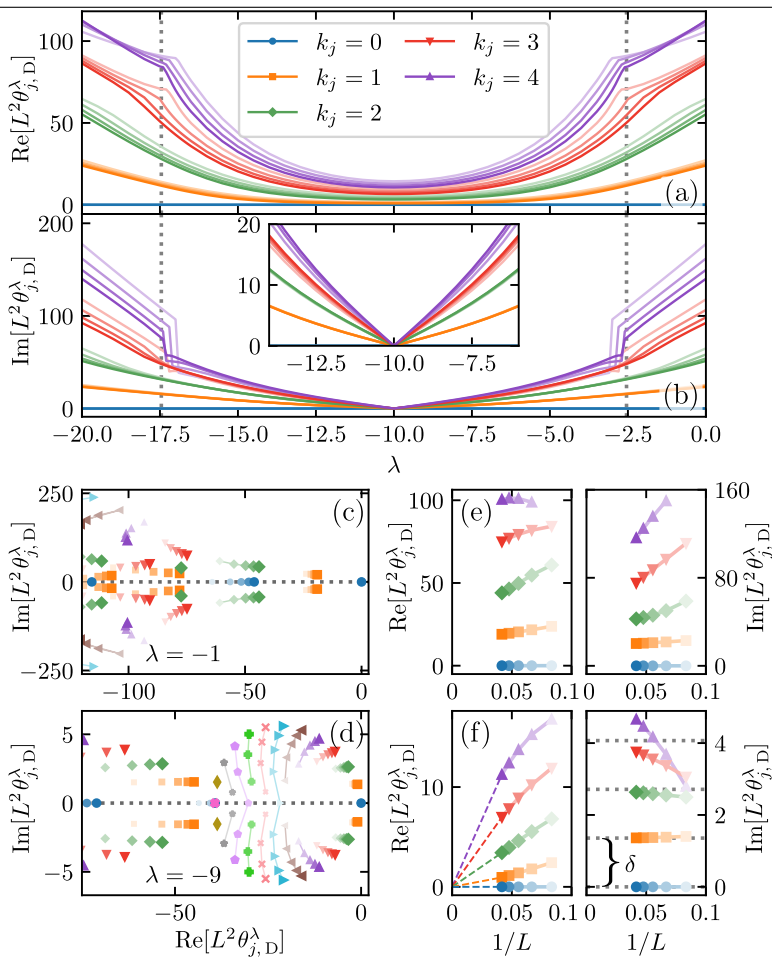}
    \vspace{-2mm}
    \caption{Diffusively scaled eigenvalues of the Doob's generator $\genmatD$ in the closed WASEP with $\rho_0=1/3$, $E=10>E_c$ and lattice sizes $L=9, 12, 15, 18, 21, 24$.
    The different colors and markers denote the symmetry eigenvalue $\symmeigval{j} = e^{i 2\pi k_j/\lattsize}$ corresponding to each $\eigvalD{j}$.
    Top panels show the evolution with $\lambda$ of the real (a) and imaginary (b) parts of the first few leading eigenvalues of $\genmatD$ for increasing values of $L$, denoted by increasing color intensity. 
    More specifically, they show the largest $\Delta^{\lambda}_j$ corresponding to each $k_j$.
    Bottom panels show the spectrum in the complex plane in the homogeneous phase for $\lambda=-1$ (c), and the condensate phase for $\lambda=-9$ (d).
    The size of each marker indicates the lattice size.
    The colors and markers beyond the ones in the legend correspond to $k_j=[5\isep 12]$, appearing in order in panel (d).
    Panels (e)-(f) show the finite-size scaling analysis for the real and imaginary parts of the leading eigenvalues in the homogeneous (e) and condensate (f) phases. In the condensate phase, the real parts converge to zero as a power law of $1/L$, while the imaginary parts exhibit a clear band structure with constant frequency spacing $\delta$, proportional to the condensate velocity.
    \todo[inline]{Me gustaría hacer algo para que se entendiera mejor}
    \todo[inline]{En las figuras (c) y (d) se está mostrando en realidad $k_j=+-1, 2,...$}
    \todo[inline]{Esta figura se podría mejorar añadiendo en (a) y (b) lineas verticales correspondientes a los plots de abajo}
    }
    \label{fig:eigvalues_closed}
\end{figure}

According to the theory developed in Chapter~\ref{chap:SBDPTs}, and the examples studied in Chapter~\ref{chap:DPTs_models}, we expect that the DPT in the closed WASEP will be accompanied by the emergence of a degenerate stationary subspace in the Doob generator, spanned by multiple eigenvectors with vanishing spectral gaps in the thermodynamic limit.
However, in stark contrast with previous examples, in this model the number of contributing eigenvectors to the stationary state is not fixed but increases linearly with the system size.
This is confirmed in Fig.~\ref{fig:eigvalues_closed}, which shows the diffusively scaled eigenvalues of the Doob stochastic generator $\genmatD$ for the model with $\rho_0=1/3$ and $E=10>E_c$.
In particular, Figs.~\ref{fig:eigvalues_closed}(a)-(b) show the evolution with $\lambda$ of the real (a) and imaginary (b) parts of the first few leading eigenvalues of $\genmatD$ for different system sizes $L$.
As expected, a clear change of behavior is apparent at $\lambda_c^\pm$.
Outside the region $[\currenttiltcritmin, \currenttiltcritmax]$, the spectrum is clearly gapped for any lattice size, while inside we can appreciate how the real part of the first eigenvalues vanish as $\lattsize$ increases.
This can be further seen in the rest of the panels, displaying the evolution of the eigenvalues with $\lattsize$ in the complex plane and their finite-size scaling for a $\currenttilt$ in the gapped phase corresponding to the flat density profile [panels (c) and (e)] and in the condensate one [panels (d) and (f)].
In particular, panel (f) clearly shows that the real part of the eigenvalues in the condensate phase decay to zero linearly with $1/\lattsize$.

An important observation from Fig.~\ref{fig:eigvalues_closed}(d) is that, as $\lattsize$ increases, so does the number of eigenvalues expected to contribute to the stationary state according to Chapter~\ref{chap:SBDPTs}.
Therefore, even though the real part of the first $\lattsize$ eigenvalues decay as the size of the system increases, for any given $\lattsize$, the real part of the last of these eigenvalues will be significantly different from zero.
So, although we are breaking a $\Zsymm{\lattsize}$, for any large but finite $\lattsize$ only a macroscopic fraction $\mathcal{O}(\lattsize)$ of the eigenvalues will have a negible real part.
Moreover, Fig.~\ref{fig:eigvalues_closed} shows that the imaginary parts of the gap-closing eigenvalues in the condensate phase are non-zero (except for the leading one) and exhibit an emerging band structure with a constant frequency spacing $\delta$ [see dashed horizontal lines in Fig.~\ref{fig:eigvalues_closed}(f)].
We will show below that this band structure in the imaginary axis can be directly linked with the velocity $v$ of the moving condensate.


In view of these spectral properties, we naturally expect a unique steady state given by the leading Doob eigenvector $\probTvectst = \eigvectDR{0}$ in the region where the spectrum is gapped, i.e. when $\lambda>\lambda_c^+$ or $\lambda<\lambda_c^-$.
This steady state, corresponding to the flat profile, remains invariant under the translation operator $\hat{S}_{\textrm{T}}$.
Conversely, in the gapless regime $\lambda_c^-<\lambda<\lambda_c^+$, the Doob \emph{stationary} subspace will be ${\cal O}(L)$-degenerate, and the resulting Doob \emph{steady state} will be in fact time-dependent and approximately equal to\todo{eso es mentiraa!}
\begin{equation}
    \probTvectst(t) \approx \eigvectDR{0} + \sum_{j=1}^{L-1} \text{e}^{t\Re(\eigvalD{j}) +i t \Im(\theta_{j,\text{D}}^\lambda)}\eigvectDR{j} \braket{\eigvectDLnk{j}}{\probvecttimenk{0}}\, ,
\end{equation}
see Eq.~\eqref{Psstime} in Section~\ref{ssecDeg} and the associated discussion.
For large values of $\lattsize$, we expect that the real part of $\eigvalD{j}$ for a fraction $\mathcal{O}(\lattsize)$ of the first $\lattsize$ will be negligible.
In contrast, the subsequent eigenvalues will exhibit an increasingly significant decay rate.
Therefore,  we will disregard the eigenvectors with non-negligible real parts to obtain the simpler expression
\begin{equation}
    \probTvectst(t) \approx \eigvectDR{0} + \sum_{j=1}^{\macrofraction} \text{e}^{+i t \Im(\theta_{j,\text{D}}^\lambda)}\eigvectDR{j} \braket{\eigvectDLnk{j}}{\probvecttimenk{0}}\, .
    \label{PssTC}
\end{equation}
While this expression is not strictly true, for large enough $\lattsize$ and for times $t$ such that the majority of eigenvectors with nonzero real part of $\eigvalD{j}$ have decayed, the behavior of $\probTvectst(t)$ will be dominated by the first $\macrofraction$ eigenvectors, and the previous expression will serve as an adequate approximation.
Still, for any finite $L$ the leading spectral gaps will not be completely closed, in fact they decay as $1/L$ in a hierarchical manner, see Fig.~\ref{fig:eigvalues_closed}(d) and Fig.~\ref{fig:eigvalues_closed}(f), and the Doob stationary subspace will be \emph{quasi}-degenerate \cite{gaveau98a,gaveau06a,minganti18a}.
As expected, the Doob steady state in this quasi-degenerate regime breaks spontaneously the translation symmetry, 
so $\hat{S}_{\textrm{T}}\probTvectst(t) \ne \probTvectst(t)$, see below.

\todo[inline]{No todos pertenecen a la estructura de bandas}
Assuming now $L$ to be odd for simplicity (all results can be trivially generalized to even $L$), and recalling that the complex eigenvalues and eigenvectors of $\genmatD$ come in complex-conjugate pairs, so $\Im(\theta_{2k+1,\text{D}}^\lambda)=-\Im(\theta_{2k,\text{D}}^\lambda)$, the band structure with constant spacing $\delta$ observed in the imaginary parts of the gap-closing eigenvalues implies that
\be
\Im(\theta_{j,\text{D}}^\lambda) =
\begin{cases}
+ j \delta/2 \, , & j=2,4,\ldots  \\
-(j+1) \delta/2 \, , & j=1,3,\ldots
\end{cases} 
\ee
and hence asymptotically
\be
\probTvectst (t) \approx \eigvectDR{0} + 2 \sum_{\substack{j=2 \\j\, \text{even}}}^{\macrofraction} \Re\left[ \text{e}^{+i t \frac{j\delta}{2}}\eigvectDR{j} \braket{\eigvectDLnk{j}}{\probvecttimenk{0}} \right]\, .
\label{PssTC2}
\ee
Similarly, the symmetry eigenvalues $\phi_j$ of the different Doob eigenvectors, such that $\hat{S}_\text{T} \eigvectDR{j}=\phi_j\eigvectDR{j}$, come in complex-conjugate pairs and obey
\be
\phi_j =
\begin{cases}
\text{e}^{+i \pi j/L} \, , & j=2,4,\ldots  \\
\text{e}^{-i \pi (j+1)/L} \, , & j=1,3,\ldots
\end{cases} 
\ee
such that $\phi_{2k+1}=\phi_{2k}^*$.
In this way, we can easily see that
\begin{equation}
\begin{split}
\label{PssTC3}
    \hat{S}_\text{T} \probTvectst (t) & \approx  \eigvectDR{0} + 2 \sum_{\substack{j=2\\ j\, \text{even}}}^{\macrofraction} \Re\Big[ \text{e}^{+i (t+\frac{2\pi}{L\delta}) \frac{j\delta}{2}}\eigvectDR{j} 
    \braket{\eigvectDLnk{j}}{\probvecttimenk{0}} \Big]
    \\
    &= \probTvectst(t+\frac{2\pi}{L\delta}) \, . 
\end{split}
\end{equation}
This shows that, in the quasi-degenerate phase $\lambda_c^-<\lambda<\lambda_c^+$, (i) the symmetry is spontaneously broken, $\hat{S}_\text{T} \probTvectst (t)\ne \probTvectst (t)$, but (ii) spatial translation and time evolution are two sides of the same coin in this regime. In particular, we have shown that a spatial translation of a unit lattice site is equivalent to a temporal evolution of time $2\pi/L\delta$, i.e. $\hat{S}_\text{T} \probTvectst (t) =\probTvectst (t+\frac{2\pi}{L\delta})$, leading to a time-periodic motion of period $2\pi/\delta$ or equivalently a density wave of velocity $v=L\delta/2\pi$.

For the phase probability vectors in the symmetry-broken regime, Eq.~\eqref{eq:probphasevect} implies that
\be
\ket{\Pi_l^\lambda} = \eigvectDR{0} + 2 \sum_{\substack{j=2\\ j\, \text{even}}}^{\macrofraction} \Re\Big[\text{e}^{+i \frac{\pi l}{L}j}\eigvectDR{j} \Big] \, ,
\label{PssTC4}
\ee
such that $\hat{S}_\text{T}\ket{\Pi_l^\lambda}=\ket{\Pi_{l+1}^\lambda}$.
As we will see when analyzing the reduced order parameter space, the dominant configurations in each phase probability vector correspond to static particle condensates localized around a different lattice site.
We will show that these \emph{localized} $\ket{\Pi_l^\lambda}$ are built as linear superpositions of the different \emph{delocalized} eigenvectors $\eigvectDR{j}$, shifted appropriately according to their symmetry eigenvalues, $(\phi_j)^l$ [this localization mechanism is sketched in Fig.~\ref{polar}(f), which will be discussed later].
However, notice that since only a fraction $\macrofraction$ of the eigenvectors join the quasi-degenerate subspace, we will not be able to define $\lattsize$ linearly independent $\ket{\Pi_l^\lambda}$ phase vectors.
Instead of $\lattsize$ phases perfectly localized in each lattice site, we will have a smaller number $\macrofraction$ of smeared-out phases.
Still, by increasing the lattice size we can localize such phases in an arbitrarily small interval.

We can write the time-dependent \emph{stationary} Doob state in terms of the static phase probability vectors as
\be
\probTvectst (t) \approx \sum_{l=0}^{\macrofraction} w_l(t) \ket{\Pi_l^\lambda} \, ,
\label{PssTC5}
\ee
where the different phase weights $w_l(t)$ are now time-dependent, see Eq.~\eqref{coeft},
\be
w_l (t)= \frac{1}{L} + \frac{2}{L}\sum_{\substack{j=2\\ j\, \text{even}}}^{\macrofraction}  \Re\Big[ \text{e}^{+i (\frac{t\delta}{2}-\frac{\pi l}{L})j} \braket{\eigvectDLnk{j}}{\probvecttimenk{0}} \Big] \, .
\label{PssTC6}
\ee
The periodicity of the resulting symmetry-broken state is reflected in the fact that $w_l(t+2\pi/\delta)=w_l(t)$ $\forall l\in[0\isep L-1]$.

\begin{figure}
    \includegraphics[width=1\linewidth]{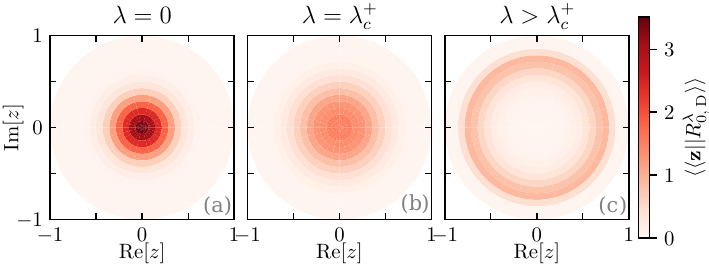}
    \caption{{\bf The leading eigenvector across the DPT in the closed WASEP.} Structure of $\bbraket{\KDopar}{R_{0,\text{D}}^\lambda}$ in the complex $\KDopar$-plane for different values of $\currenttilt$ across the DPT for $\rho_0=1/3$, $E=10>E_c$ and $L=24$. From left to right: (a) $\currenttilt = 0$ (symmetry-preserving phase, before the DPT), (b) $\currenttilt = -2.5 \approx \lambda_c^+$, (c) $\currenttilt = -9$ (symmetry-broken phase, after the DPT). Note the transition from unimodal$\bbraket{\KDopar}{R_{0,\text{D}}^\lambda}$ peaked around $|\KDopar|\approx 0$ for $\lambda>\lambda_c^+$ to the inverted Mexican-hat structure with a steep ridge around $|\KDopar|\approx 0.7$ for $\lambda_c^-<\lambda<\lambda_c^+$.}
    \label{fig:eigvect0_closed}
\end{figure}

\begin{figure}
    \centering
    \includegraphics[width=0.82\linewidth]{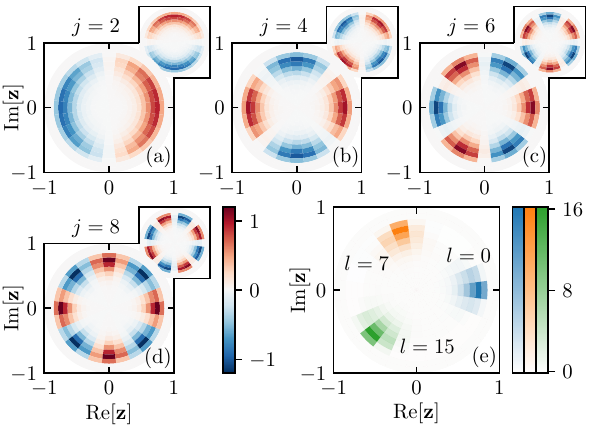}
    \includegraphics[width=0.78\linewidth]{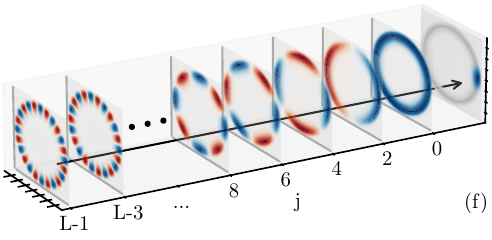}
    \caption{{Quasi-degenerate reduced eigenvectors in the closed WASEP and condensate localization.} (a)-(d) First few reduced eigenvectors of the Doob degenerate subspace in the closed WASEP. In particular, the real and imaginary parts of $\bbraket{\KDopar}{R_{j,\text{D}}^\lambda}$ are displayed for $\lambda_c^-<\lambda=-9<\lambda_c^+$ and different (even) values of $j=2,4,6,8$. Recall that the complex eigenvalues and eigenvectors of $\genmatD$ come in complex-conjugate pairs, so $\bbraket{\KDopar}{R_{2k+1,\text{D}}^\lambda}={\bbraket{\KDopar}{R_{2k,\text{D}}^\lambda}}^*$. The main panels show the real parts, while the insets display the imaginary parts for each $j$. Note the $j$-fold symmetry of reduced eigenvectors, and that non-negligible structure appears in all cases in the region $|\KDopar|\approx 0.7$, as expected in the symmetry-broken dynamical phase. (e) Sample of the resulting localized reduced phase probability vectors $\bbraket{\KDopar}{\Pi_l^\lambda}$ for $l=0,7,15$. (f) Sketch of the spectral localization mechanism that gives rise to a compact condensate. Each slice shows $\Re[\phi_j^l\bbraket{\KDopar}{R_{j,\text{D}}^\lambda}]$ for the corresponding $j$.}
    \todo[inline]{break into two images so that it fits}
    \label{polar}
\end{figure}

The spectral structure of the Doob stationary subspace is better explored in the reduced Hilbert space, introduced in Section~\ref{ssecOparam}, associated with the packing order parameter $\KDopar$ of Eq.~\eqref{coher}.
Figure~\ref{fig:eigvect0_closed} shows the structure of the leading reduced eigenvector $\bbraket{\KDopar}{R_{0,\text{D}}^\lambda}$ in the complex $\KDopar$-plane for varying biasing fields $\lambda$ across the DPT.
As observed in the examples of Chapter~\ref{chap:DPTs_models}, before the DPT occurs (i.e. for $\lambda>\lambda_c^+$ or $\lambda<\lambda_c^-$, when the spectrum of $\genmatD$ is gapped) the real distribution $\bbraket{\KDopar}{R_{0,\text{D}}^\lambda}$ is unimodal and peaked around $|\KDopar|\approx 0$.
This indicates that the configurations contributing to this phase symmetry-preserving phase are flat ($|\KDopar(\conf)|\approx 0$), as we expected.
As $\currenttilt$ approaches the critical point, $\bbraket{\KDopar}{R_{0,\text{D}}^\lambda}$ flattens and spreads over the unit complex circle, see Fig.~\ref{fig:eigvect0_closed}(b) for $\lambda\approx\lambda_c^+$, while deep inside the critical regime $\lambda_c^-<\lambda<\lambda_c^+$ the distribution $\bbraket{\KDopar}{R_{0,\text{D}}^\lambda}$ develops an inverted Mexican-hat shape, see Fig.~\ref{fig:eigvect0_closed}(c), with a steep ridge around $|\KDopar|\approx 0.7$ but homogeneous angular distribution.
Therefore, even if the typical configurations contributing to $\ket{R_{0,\text{D}}}$ correspond to symmetry-broken condensate configurations ($|\KDopar|\ne 0$), 
the resulting reduced eigenvector preserves its invariance under the reduced symmetry operator, $\hat{S}_\KDopar \kket{R_{0,\text{D}}^\lambda}=\kket{R_{0,\text{D}}^\lambda}$, where $\hat{S}_{\KDopar}$ corresponds now to a rotation of $2\pi/L$ radians in the complex $\KDopar$-plane.

As in the previous examples, the subleading eigenvectors spanning the (quasi-)degenerate subspace cooperate to break the symmetry, in this case by localizing the condensate at a particular point in the lattice. Figure \ref{polar} shows the structure of the real and imaginary parts of the first few subleading reduced eigenvectors in the closed WASEP, $\bbraket{\KDopar}{R_{j,\text{D}}^\lambda}$ for $j=2,4,6,8$.
As expected from its eigenvalue under the symmetry operator\todo{creo que esto no queda muy claro}, the $j$th-order ($j$ even) reduced eigenvector exhibits a clear 
$(j/2)$-fold angular symmetry in the complex plane [i.e. invariance under rotations of angles $2\pi/(j/2)$], with a structure around $|\KDopar|\approx 0.7\ne 0$ for the particular case $\lambda=-9$. 
All (quasi-) degenerate eigenvectors hence exhibit some degree of angular symmetry but their superposition, weighted by their symmetry eigenvalues $(\phi_j)^l$,
cooperates to produce a compact condensate localized at site $l$ and captured by the reduced phase probability vector $\kket{\Pi_l^\lambda}$, see also Eq.~\eqref{PssTC4}.
A sample of the resulting reduced phase probability vectors is shown in Fig.~\ref{polar}(e), which as expected are localized around different angular positions along the ring. Finally, Fig.~\ref{polar}(f) shows a sketch of the spectral localization mechanism that gives rise to a compact localized condensate from the superposition of multiple delocalized reduced eigenvectors in the degenerate subspace.
As described above, the time dependence introduced by the imaginary parts of the gap-closing eigenvalues, together with their imaginary band structure, leads to the motion of the condensate at constant velocity.

\begin{figure}
    \centering
    \includegraphics[width=0.8\linewidth]{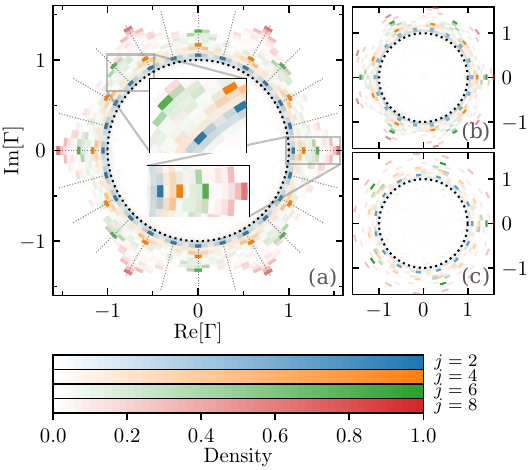}
    \caption{{\bf Structure of the degenerate subspace in the closed WASEP.} Density plot of $\Gamma_j=\braket{C}{\eigvectDRnk{j}}/\braket{C}{\eigvectDRnk{0}}$ in the complex $\Gamma$-plane, with $j=2,4,6,8$, obtained for a large set of configurations $\ket{C}$ sampled from the Doob \emph{stationary} distribution in the symmetry-broken regime for $\lambda=-9$, $E=10>E_c$ and (a) $L=24$, (b) $L=18$ and (c) $L=15$. Different colors correspond to different values of index $j$. The insets show zooms on compact regions around the complex unit circle to better appreciate the emerging structure.}
    \label{fig:eigvect_WASEPclosed_complete}
\end{figure}

Coming back to the original eigenvectors, according to Eq.~\eqref{eq:eigvectsfromeigvect0}, in the symmetry-broken regime we should expect a tight relation between the eigenvectors spanning the degenerate Doob subspace.
In particular, we expect that
\be
\braket{C}{\eigvectDRnk{j}} \approx \text{e}^{-i \frac{\pi j}{L}\ell_C} \braket{C}{\eigvectDRnk{0}} 
\label{PssTC7}
\ee
for the statistically-relevant configurations $\ket{C}$ in the Doob \emph{stationary} state belonging to the basin of attraction of phase $\ell_C\in[0,L-1]$, see the associated discussion in Section~\ref{ssecSpec}. To investigate this relation, we now plot in Fig.~\ref{fig:eigvect_WASEPclosed_complete} a density map in the complex plane for the quotients $\Gamma_j(C)=\braket{C}{\eigvectDRnk{j}}/\braket{C}{\eigvectDRnk{0}}$ for $j=2,4,6,8$ obtained from a large sample of configurations drawn from the Doob \emph{stationary} distribution $\probTvectst (t)$. As expected from Eq.~\eqref{PssTC7}, we observe a condensation of points around the complex unit circle, with high-density regions at nodal angles multiple of $\varphi_j=\pi j /L$. For instance, for $j=2$ we expect to observe sharp peaks in the density plot of $\Gamma_2(C)$ in the complex unit circle at angles $2\pi k/L$, $k\in[0\isep L-1]$, as confirmed in Fig.~\ref{fig:eigvect_WASEPclosed_complete}(a) for $L=24$.
The convergence to the complex unit circle improves as the system size $L$ increases, see Figs.~\ref{fig:eigvect_WASEPclosed_complete}(b) and \ref{fig:eigvect_WASEPclosed_complete}(c), and for fixed $L$ this convergence is better for smaller spectral index $j$, i.e. for the eigenvectors $\eigvectDRnk{j}$ whose (finite-size) spectral gap $\Delta_j^\lambda(L)$ is closer to zero, see Fig.~\ref{fig:eigvalues_closed}(f).
Note also that, when the (even) spectral index $j$ is commensurate with $2L$, we expect $2L/j$ nodal accumulation points in the density plot for $\Gamma_j(C)$, e.g. 
for $j=6$ and $L=24$ we expect 8($=2\times 24/6$) nodal points, 
as seen in Fig.~\ref{fig:eigvect_WASEPclosed_complete}(a). 
For an (even) spectral index $j$ incommensurate with $2L$ one should observe just $L$ nodal points, as shown in Fig.~\ref{fig:eigvect_WASEPclosed_complete}(b) for $j=4$ and $L=15$. Overall, this analysis confirms the tight structural relation imposed by the $\mathbb{Z}_L$ symmetry on the eigenvectors spanning the Doob \emph{stationary} subspace, confirming along the way that statistically relevant configurations in the Doob \emph{stationary} state for the closed WASEP can be classified into different symmetry classes.

\begin{figure}
    \centering
    \includegraphics[width=\linewidth]{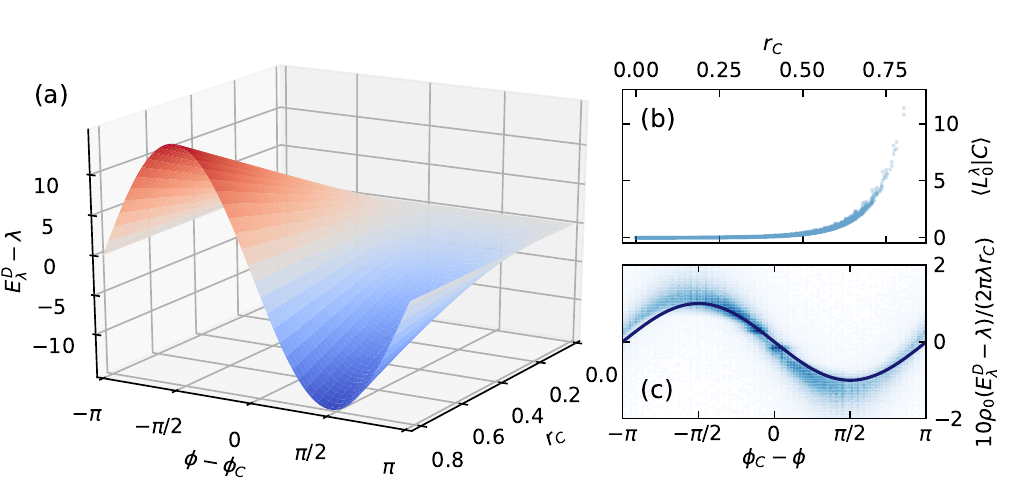}
    \caption{(a) Auxiliary field for $\rho_0=1/3$ and $\lambda=-9$ as a function of packing order parameter $r_C$ and the angular distance to the center-of-mass position. (b) $\braket{L_0^\lambda}{C}$ vs the packing order parameter $r_C$ for $L=24$, $\rho_0=1/3$, $E=10$, $\lambda=-9$ (condensate phase) and a large sample of microscopic configurations. (c) Angular dependence of the auxiliary field with respect to the center-of-mass angular location for a large sample of microscopic configurations and the same parameters, together with the $\sin(\phi_k-\phi_C)$ prediction (line).}
    \todo[inline]{Arreglar phis}
    \label{fig:smart_Doob_field}
\end{figure}

\section{Identifying the packing field mechanism in the Doob dynamics}
{
\newcommand{\jumpdir}{q}

\newcommand{\smartE}{E^{\text{D}}_\lambda}
\newcommand{\ctoc}{{\conf_j \to \conf_i}}

In the previous sections, we have found the fingerprints of a time-crystal phase in the DPT of the WASEP and we have analyzed the role of the quasi-degenerated eigenvectors in the emergence of a traveling-wave periodic condensate.
However, despite this understanding and the ability to compute the Doob dynamics leading to the DPT, we still lack an intuitive grasp of the dynamics within $\genmatD$.
To better understand the underlying physics in $\genmatD$, we will now shift our focus to the study of its rates.
In particular, we will rewrite the rates of the Doob's generator in terms of the original WASEP rates supplemented by an auxiliary field $\smartE$, i.e., we impose 
\begin{equation}
    (\genmatD)_{\ctoc}
    =
    (\genmat)_{\ctoc}
    \expsup{\jumpdir_{C_j \to C_i } (\smartE)_{\ctoc}/L}
\end{equation}
with $({\mathbb W}^{\lambda}_{\text{D}})_{\ctoc}=\bra{C_i}{\mathbb W}^{\lambda}_{\text{D}}\ket{C_j}$, $\jumpdir_{\ctoc}=\pm 1$ the direction of the particle jump in the transition $C_j\to C_i$ and $(\smartE)_{\ctoc}$ the value of the auxiliary field in the transition.
Together with the definition of ${\mathbb W}^{\lambda}_{\text{D}}$, this leads to
\begin{equation}
    (E^{\text{D}}_\lambda)_{\ctoc}
    =
    \lambda + \jumpdir_{\ctoc}L \ln \left( \frac{\braket{L_0^\lambda}{C_i}}{\braket{L_0^\lambda}{C_j}} \right) \, .
\label{EDoob}
\end{equation}
Therefore, $E^{\text{D}}_\lambda$ can be interpreted as the external field needed to make typical a rare event of bias field $\lambda$.
In order to disentangle the nonlocal complexity of the Doob's dynamics, we scrutinize the dependence of the left eigenvector $\eigvectTL{0}$ with the packing parameter $\KDopar(\conf)$.
In particular, since we know that \todo{esto lo sé yo, pero no sé si queda claro} this eigenvector is invariant under the rotations, it will not depend on the complex argument of the order parameter, $\varphi(C) = \arg(\KDopar(\conf))$ marking the angular position of the center of mass of a given configuration.
Therefore, we just need to study the relation between $\eigvectTL{0}$ and the magnitude of the order parameter $r_C \equiv \abs{\KDopar(\conf)}$.
In particular, Fig.~\ref{fig:smart_Doob_field}(b) plots the projections $\braket{\eigvectTLnk{0}}{\conf}$ vs the packing parameter $r_C$ for a large sample of microscopic configurations $C$, as obtained for $\E=10$, $\rho_0=1/3$, $L=24$ and $\lambda=-9$ (condensate phase).
Interestingly, this shows that $\braket{L_0^\lambda}{C}\simeq f_{\lambda,L}(r_C)$ to a high degree of accuracy, with $f_{\lambda,L}(r)$ some unknown $\lambda$- and $L$-dependent function of the packing parameter.

This means in particular that the auxiliary field $(E^{\text{D}}_\lambda)_{ij}$ depends essentially on the packing parameter of configurations $C_i$ and $C_j$, a radical simplification.
Moreover, as elementary transitions involve just a local particle jump, the resulting change on the packing parameter is perturbatively small for large enough $L$.
If $C'(k)$ is the configuration that results from $C$ after a particle jump at site $k\in[1,L]$, we have, up to order $1/\lattsize$
\begin{equation}
\label{eq:packfield_derivation_buildingtc}
    r_{C'(k)}
    \simeq
    r_C + \frac{2\pi \jumpdir_{C \to C'(k)}}{\rho_0 L^2} \sin(\varphi(C)-\varphi_k)
    ,
\end{equation}
with $\varphi_k\equiv 2\pi k/L$ the angle corresponding to site $k$.
Therefore, we obtain that the auxiliary field for the transition $\conf\to\conf'(k)$ is given by
\begin{equation}
    (E^{\text{D}}_\lambda)_{C \to C'_k}
    \simeq
    \lambda + 2\pi (\rho_0 L)^{-1} g_{\lambda,L}(r_C) \sin(\varphi(C)-\varphi_k)   
    ,
\end{equation}
with $g_{\lambda,L}(r)\equiv f'_{\lambda,L}(r)/f_{\lambda,L}(r)$, and we empirically \todo{esto hubiera estado bien mostrarlo} find an approximately linear dependence
\begin{equation}
    g_{\lambda,L}(r)
    \approx
    -\lambda L r/10   
\end{equation}
near the critical point $\lambda_c^+$.
Therefore, we expect the following relation by inverting Eq.~\eqref{eq:packfield_derivation_buildingtc}
\begin{equation}
    \sin(\varphi(C)-\varphi_k)
    \approx
    \frac{10\rho_0}{(2\pi \lambda r_C)} [(E^{\text{D}}_\lambda)_{C\to C'(k)}-\lambda]
    .
\end{equation}
Figure~\ref{fig:smart_Doob_field}(c) tests this relation by plotting the right-hand-side of the previous equation for a large sample of connected configurations $C\to C'_k$, as a function of $\varphi(C)-\varphi_k$. \todo{no estoy seguro que quede claro cómo calcular el rhs}
The agreement between the numerical data and our prediction is remarkable, especially when comparing the simplicity of our approximation with the complexity of exact Doob dynamics.
In this way, the nonconstant part of the auxiliary field $(E^{\text{D}}_\lambda-\lambda)$ acts as a \emph{packing field} on a given configuration $C$, pushing particles that lag behind the center of mass while restraining those moving ahead, see Fig. \ref{fig:smart_Doob_field}(a), with an amplitude proportional to the packing parameter $\lambda$ and $r_C$. 
The dependence on the order parameter $r_C$, measuring the packing of the particles around its center of mass, induces a nonlinear feedback mechanism that competes with the tendency of the diffusion to flatten profiles and the pushing constant field, amplifies the naturally occurring fluctuations of the packing parameter, leading eventually to emergence of a traveling-wave condensate displaying the features of a time-crystal phase.
Similar effective potentials for atypical fluctuations have been found in other driven systems \cite{kaviani20a,popkov10a}.

\section{Engineering time periodic behavior: the time-crystal lattice gas}

\begin{figure}
    \centering
    \includegraphics[width=1\linewidth]{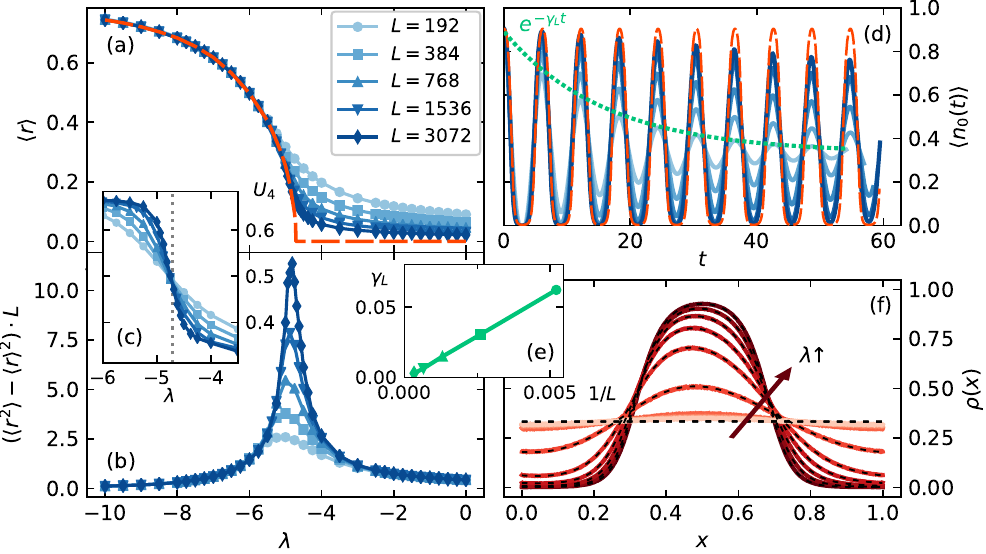}
    \vspace{-0.25cm}
    \caption{Numerics for the time-crystal lattice gas. Average packing order parameter (a), its fluctuations (b) and Binder's cumulant (c) measured for $\rho_0=1/3$, $E=10$ and different $L$. (d) Local density as a function of time and different $L$'s in the time-crystal phase ($\lambda=-9$). Note the persistent oscillations typical of time crystals. (e) Decay of the oscillations damping rate as $L\to\infty$, a clear sign of the rigidity of the time-crystal phase in the thermodynamic limit. (f) Average density profile of the condensate for $L=3072$ and varying $\lambda$. Dashed lines correspond to hydrodynamic predictions.}
    \label{fig4}
\end{figure}
}
Inspired by the results of the previous analysis, we now simplify the auxiliary field to introduce the time-crystal lattice gas (tcLG).
This is a variant of the $1d$ WASEP where a particle at site $k$ hops stochastically under a configuration-dependent packing field 
\begin{equation}
    E_\lambda(C;k)= E + \lambda + 2 \lambda r_C \sin(\phi_k-\phi_C)   
    ,
\end{equation}
with $E$ being a constant external field and $\lambda$ now a control parameter.
We note that this auxiliary field can be also written as a Kuramoto-like long-range interaction term
\begin{equation}
    E_\lambda(C;k)
    =
    E+\lambda+\frac{2\lambda}{N}\sum_{j\neq k} \sin(\phi_k-\phi_j),   
\end{equation}
highlighting the link between the tcLG and the Kuramoto model of synchronization \cite{kuramoto84a, kuramoto87a, pikovsky03a, acebron05a}.
However, we stress that this link is only formal, as the Kuramoto model lacks any particle transport in real space.

According to the discussion in the previous section, we expect this lattice gas to display a putative steady-state phase transition to a time-crystal phase with a rotating condensate at some critical $\lambda_c$ as $L \to \infty$. \todo{tenía puesto esto: (due to the Perron-Frobenius theorem) ?????}
To test this picture, we performed extensive Monte Carlo simulations and a finite-size scaling analysis of the tcLG at density $\rho_0=1/3$, which is shown in Fig.~\ref{fig4}.
As expected, the average packing parameter $\avg{r}$ increases steeply but continuously for parameters below a critical point $\lambda< \lambda_c=-\pi/(1-\rho_0)\approx -4.7$ see Fig.~\ref{fig4}(a), converging toward the macroscopic hydrodynamic prediction (which is commented below) as $L\to\infty$.
Moreover, the associated susceptibility, as measured by the packing fluctuations $\avg{r^2} - \avg{r}^2$, exhibits a well-defined peak around the same $\lambda_c$ which sharpens as $L$ grows and is compatible with a divergence in the thermodynamic limit [Fig.~\ref{fig4}(b)].
The critical point location can also be inferred from the crossing of the finite-size Binder cumulants $\binder(L)=1-\avg{r^4}/(3\avg{r^2})$ for different $L$'s, see Fig.~\ref{fig4}(c), which again agrees with the hydrodynamic prediction for $\lambda_c$.
Interestingly, the average density at a given point exhibits persistent oscillations as a function of time with period $v^{-1}$ (in the diffusive timescale), see Fig.~\ref{fig4}(d), with $v$ the condensate velocity, a universal feature of time crystals \cite{yao18a, wilczek12a, shapere12a, zakrzewski12a, moessner17a, richerme17a, yao18a, sacha18a, bruno13a, nozieres13a, watanabe15a, khemani16a, keyserlingk16a, else16a, gambetta19a, yao17a, yao17a, zhang17a, choi17a, nakatsugawa17a, lazarides17a, gong18a, tucker18a, osullivan18a, lazarides19a, yao18b, gambetta19b, iemini18a, medenjak20a}, and converges toward the hydrodynamic (undamped) periodic prediction as $L\to \infty$. Indeed the finite-size damping rate of oscillations, $\gamma_L$, obtained from an exponential fit to the envelope of $\avg{n_0(t)}$, decays to zero in the thermodynamic limit [Fig.~\ref{fig4}(e)], a clear signature of the rigidity of the long-range spatio-temporal order emerging in the time crystal phase of tcLG. We also measured the average density profile of the moving condensate for different values of $\lambda$, see Fig.~\ref{fig4}(f), which becomes highly nonlinear deep into the time-crystal phase. 

In the macroscopic limit, one can show using a local equilibrium approximation \cite{prados12a, hurtado13a, lasanta15a, lasanta16a, manacorda16a, gutierrez-ariza19a} that the tcLG is described by the hydrodynamic equation (\ref{eq:hydroWASEP_buildingtc}) with a $\rho$-dependent local field 
\begin{equation}
    E_\lambda(\rho;x)= E + \lambda + 2 \lambda r_\rho \sin(2\pi x-\varphi_\rho)
    ,
\end{equation}
with $r_\rho=|z_\rho|$, $\varphi_\rho=\arg(z_\rho)$ and 
\begin{equation}
    z_\rho=\rho_0^{-1}\int_0^1 dx \rho(x) \text{e}^{i2\pi x}
\end{equation}
 the density field generalization of our order parameter.
A local stability analysis then shows \cite{hurtado11a,hurtado14a,tizon-escamilla17b} that the homogeneous solution $\rho(x,t)=\rho_0$ becomes unstable at 
\begin{equation}
    \lambda_c=-2\pi\rho_0D(\rho_0)/\sigma(\rho_0)=-\pi/(1-\rho_0)   
    ,
\end{equation}
 where a traveling-wave condensate emerges.
 As we observed before, the hydrodynamic predictions fully coincide with the microscopic simulations in Fig.~\ref{fig4}. 

\section{Conclusion}

This chapter has been dedicated to the examination of the DPT present in the periodic WASEP when conditioned to sustain currents well below its typical value.
As we have seen, this DPT is characterized by the emergence of a traveling particle condensate, which exhibits the features of a time crystal phase.
Leveraging on the tools developed in previous chapters, we have performed a throughout analysis of the spectral properties of this phase, which has allowed us to determine the role played by the eigenvectors in the quasi-degenerated subspace.
Next, we have explored the details of the dynamics within Doob's generator $\genmatD$, and we have identified the packing field mechanism giving rise to the appearance of the traveling-wave condensate.
With this insight, we have explored with great success the possibility of engineering time-crystalline phases by applying this packing field directly to the WASEP.

The modern experimental control of colloidal fluids trapped in quasi-$1d$ periodic structures, such as circular channels \cite{wei00a,lutz04a} or optical traps based e.g. on Bessel rings or optical vortices \cite{ladavac05a,roichman07a,roichman07b}, together with feedback-control force protocols to implement the nonlinear packing field $E_\lambda(C;k)$ using optical tweezers \cite{kumar18a,grier97a,ortiz14a,martinez15a,martinez17a,rodrigo18a}, may allow the engineering and direct observation of this time-crystal phase, opening the door to further experimental advances in this active field.
Moreover, the ideas developed in this chapter can be further explored in other models or greater dimensions $d>1$, where DPTs exhibit a much richer phenomenology \cite{tizon-escamilla17b,tizon-escamilla17a}.
This may lead, via the Doob's transform pathway here described, to materials with a rich phase diagram composed of multiple spacetime-crystalline phases.
In fact, the next chapter will be dedicated to further analysis of the mechanism described here.
    }

    \chapter{Generalizing the packing field mechanism to engineer complex time-crystal phases}
    \chaptermark{Generalizing the packing field mechanism}
    \label{chap:genpackfield}
    {
\newcommand{\Epg}[1]{\mathcal{E}_{\pos}^{(#1)}}
\newcommand{\Epm}{\Epg{\Eorder}}

\newcommand{\loccenterang}[2]{\phi_{#1}^{(#2)}}
\newcommand{\loccenterangm}{\loccenterang{m}{j}}
\newcommand{\denstwt}{\tilde{\denstw{}}}
\newcommand{\denstwtprime}{\tilde{\denstw{}'}}
\newcommand{\twargt}{\tilde{\twarg}}
\newcommand{\damprateL}{\gamma_\lattsize}
\newcommand{\meancurrent}{J}
\newcommand{\meancurrentflat}{\meancurrent_{\mathrm{flat}}}

\newcommand{\occupvect}{\mathbf{\occup}}
\newcommand{\kp}{k^{\prime}}

\todo[inline]{
    ABSTRACT://
    Time crystals are many-body systems that spontaneously break time-translation symmetry, and thus exhibit long-range spatiotemporal order and robust periodic motion. We have recently shown how an external time-independent packing field coupled with density fluctuations of a driven diffusive system can induce the spontaneous emergence of time-crystalline particle waves. Here we generalize this mechanism to engineer and control, on demand, continuous time-crystal phases characterized by a single or multiple rotating condensates. 
    We show that there is a critical coupling to the packing field leading to the time-crystal phase, and elucidate how such a critical point depends on the transport coefficients of the system. Further, we demonstrate that the shape of the higher-order traveling condensates is obtained by rescaling the parameters of the first-order condensate. We illustrate our findings by solving the hydrodynamic equations for several representative diffusive systems. These systems exhibit either continuous or first-order transitions toward the time crystalline phase, contingent upon the specific transport coefficients involved.
}

\begin{figure}
    \centering
    \includegraphics[width=0.7\linewidth]{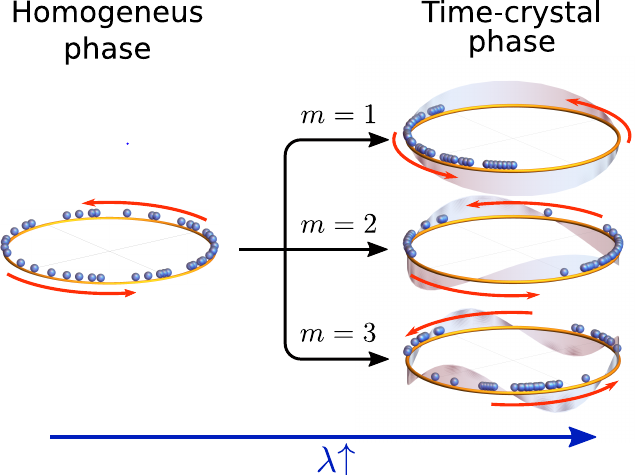}
    \caption{
        The packing-field route to time crystals.
        (Left) A stochastic particle fluid in the presence of a constant driving field sustains a net current of particles with a homogeneous density structure on average subject to small density fluctuations.
        (Right) By switching on a $\Eorder$-order packing field with coupling beyond a critical value, an instability is triggered that leads to a complex time-crystal phase characterized by the emergence of $\Eorder$ rotating particle condensates with a velocity controlled by the driving field.
        \todo[inline]{CAMBIAR signo al campo y sentido flecha roja -> SEGURO?}
    }
    \label{fig1_largo}
\end{figure}

\todo[inline]{Le falta un poco de prosa a esto}
In this chapter, we will step away from the study of rare events and we will delve into the analysis of the packing field mechanism introduced in the previous chapter.
The core idea behind this mechanism was to take a system of particles arranged in a periodic lattice, subjected to a constant driving field, and introduce an additional packing field coupled to density fluctuations.
This field amplified the spontaneous fluctuations of the density around the center of mass of the system, giving rise to the emergence of a particle condensate that periodically traveled around the ring---a time crystal.

Here, we will generalize this mechanism to yield more complex time-crystal phases characterized by an arbitrary number of symmetric traveling-wave condensates, as illustrated in Fig.~\ref{fig1_largo}.
This generalized mechanism will be studied in detail, obtaining the dependence of the phase transition on the transport coefficients of the selected model and establishing a connection between the dynamics corresponding to different numbers of traveling condensates.
These findings will be illustrated and verified in various representative driven diffusive systems.

\todo[inline]{Esto se podría cerrar mejor}


\section{Introducing the generalized packing field mechanism}
\todo[inline]{cambiar el título}

\begin{figure}
   \centering
   \includegraphics[width=0.85\linewidth]{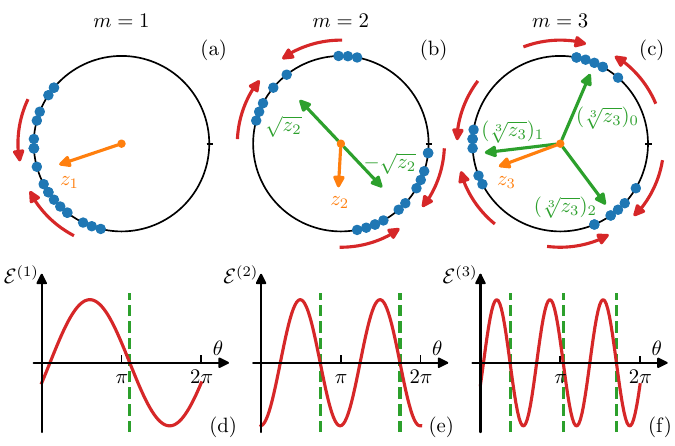}
   \caption{
        Sketch showing the mechanism of the packing field defined in Eq.~\eqref{eq:packingfield_genpackfield} for different packing orders $\Eorder = 1,2,3$ in a model of particle transport.
        The top row (a-c) shows possible lattice configurations in the time-crystal phase for varying $m$, together with the corresponding complex Kuramoto-Daido order parameter $\KDoparm$.
        The angular positions of the localization centers are marked by the complex argument of the $\Eorder$th roots of the order parameter, 
        $(\sqrt[\Eorder]{\KDoparm})_j \equiv \sqrt[\Eorder]{\KDoparm}\textrm{e}^{\textrm{i}2\pi j/\Eorder}$, with the magnitude of $\KDoparm$ measuring the packing around such centers.
        The arrows in each plot signal the local direction of the packing field, which hinders particle motion ahead of each localization point and pushes particles lagging behind, promoting in this way particle packing. 
        This is further seen in the row (d-f), which displays the value of the packing field as a function of the angular position along the ring.
        Dashed vertical lines mark the angular position of the localization centers.
    }
    \label{fig:packing_field_sketch_largo}
\end{figure}

Our starting point is the hydrodynamic equation describing the evolution of a one-dimensional diffusive system in a ring geometry under the action of a local external field, given by
\todo{Me falta una cita}
\begin{equation}
    \label{eq:hydro_field_genpackfield}
    \partial_t \dens
    =
    - \partial_{\pos} \Big[-\diffcoef(\dens) \partial_{\pos} \dens + \mobility(\rho) \E_{\pos}[\dens] \Big]
\end{equation}
with $\pos \in [0, 1]$, and where $\diffcoef(\dens)$ and $\mobility(\dens)$ are the diffusivity and the mobility of the system, respectively, and the brackets in the RHS correspond to the local current $j(x,t)$.
Similarly to the packing field introduced in the previous chapter, the local external field takes the form 
\begin{equation}
    \label{eq:packingfielddriving_genpackfield}
    \E_{\pos}[\dens] = \Ed +  \Epampl \Epm[\rho],
\end{equation}
with a constant driving $\Ed$ and a packing field $\Epm[\rho]$ whose intensity is controlled by the field coupling $\Epampl>0$.
However, in this case, the packing field takes an additional parameter $\Eorder$, so that its equation is
\begin{equation}
    \label{eq:packingfield_genpackfield}
    \Epm[\rho]
    =
    \abs{\KDoparm[\dens]} \sin(\KDoparargm[\dens] - 2\pi\pos \Eorder)
\end{equation}
with $2\pi \pos \equiv \theta(x)$ the angular position along the ring and $\KDoparm[\dens]$ the $\Eorder$th-order {Kuramoto-Daido parameter} \cite{daido1992}, defined by
\begin{equation}
\label{eq:z_field}
    \KDoparm[\dens]
    =
    \frac{1}{\meandens} \int_0^1 \dd\pos \dens(\pos,t) \expsup{\textrm{i}2\pi\Eorder\pos} \equiv \abs{\KDoparm[\dens]} \expsup{\textrm{i} \KDoparargm[\dens]}\, ,    
\end{equation}
and where $\meandens=\int_0^1 \dens(\pos) \dd \pos$ is the average conserved density.
We will refer to this new parameter $\Eorder$ as the packing field \emph{order}, which, as it shall show below, controls the number of symmetric particle condensates that appear in the time-crystal phase when $\Epampl$ is above the critical threshold.\todo{aquí habría que explicar qué es kuramoto daido}
\todo[inline,size=\normalsize]
{
    This field may also be written as 
    \begin{equation}
    \label{eq:extfield_pairs_field}
        \Epm(\dens) = \frac{1}{\meandens} \int_0^1 \dd\tilde{\pos} \dens(\tilde{\pos}) \sin(2\pi \Eorder (\tilde{\pos}-\pos) ) \, ,
    \end{equation}
}

Each term of the external field in Eqs.~\eqref{eq:packingfielddriving_genpackfield},\eqref{eq:packingfield_genpackfield} plays a different role in the emergence of the time-crystal phase, as we shall prove in the next sections.
On the one hand, the constant driving field $\epsilon$ gives rise to a net particle current in the desired direction and controls the velocity of the resulting particle condensates, as well as the asymmetry of the associated density profiles.
On the other, the packing field $\Epm[\dens]$ pushes particles locally towards the $\Eorder$ rotationally\todo{is it understood?} symmetric points where particles are most clustered. 
These are the \emph{localization centers}, whose angular position is given by $\loccenterangm = (\KDoparargm + 2\pi j)/\Eorder$, with $j\in[0,\Eorder-1]$, i.e., the complex arguments of the $\Eorder$th roots of the Kuramoto-Daido parameter, $(\sqrt[\Eorder]{\KDoparm})_j \equiv \sqrt[\Eorder]{\KDoparm}\expsup{\textrm{i}2\pi j/\Eorder}$ [see Fig.~\ref{fig:packing_field_sketch_largo}(a-c)].
The packing field mechanism works by restraining the motion of particles ahead of the closest localization center, i.e. for nearby positions $x$ where $\Epm[\rho]<0$, while enhancing such motion for particles lagging behind this point (where $\Epm[\dens]>0$), as illustrated in Fig.~\ref{fig:packing_field_sketch_largo}(d-f).
In addition, the strength of the packing field is proportional to the instantaneous magnitude of the Kuramoto-Daido parameter $\abs{\KDoparm}$, which measures the density packing around the \todo{dejo lo de emergent?} localization centers and thus plays the role of the order parameter.
In this way, as in the previous chapter, the packing field leads to a nonlinear feedback mechanism that amplifies the packing fluctuations naturally present in the system, resulting eventually in a phase transition to a time-crystal phase for large enough values of $\Epampl$.

Note that, for $\Eorder=1$, this equation resembles the hydrodynamic evolution equation for the oscillator density in the mean-field Kuramoto model with white noise \cite{acebron05a}.
Nevertheless, the Kuramoto model considers coupled oscillators, while here we are studying diffusive models displaying particle or energy transport in real space.
Therefore, in our case, we will deal with different---including non-linear---transport coefficients depending on the considered model, which will give rise to different and more intricate phenomenologies.
\todo{esta discusión puede que esté repetida en el capítulo anterior}

\section{Hydrodynamic instability in the time-crystal phase transition}
\label{sec:criticalpoint_genpackfield}

The model introduced above exhibits a continuous phase transition from a time-translation symmetric phase characterized by homogeneous density profiles to a complex time-crystal phase characterized by multiple traveling condensates.
This transition is triggered for strong enough packing fields, as controlled by the coupling $\lambda$.
Under this situation, the packing field induces a nonlinear feedback mechanism that amplifies the naturally occurring fluctuations of the particles' clustering around $\Eorder$ symmetric points (as quantified by $|\KDoparm|$).
After the coupling constant $\Epampl$ exceeds a critical threshold $\lambda_c$, these fluctuations become unstable and $m$ traveling condensates emerge.

To determine this critical threshold, we first note that for any value of $\Epampl$ the homogeneous density profile $\densxt = \meandens$ is a solution of the hydrodynamic equation \eqref{eq:hydro_field_genpackfield}.
Therefore, a linear stability analysis of this solution will allow us to find the critical value $\Epampl_c^{(\Eorder)}$.
We hence consider a small perturbation over the flat profile, $\densxt = \meandens + \denspxt$, with $\int_0^{1} d\pos \denspxt = 0$ so as to conserve the global density of the system, $\rho_0$.
Plugging this perturbation into Eq.~\eqref{eq:hydro_field_genpackfield} and linearizing it to first order in $\denspxt$ we obtain
\begin{eqnarray}
    \label{eq:hydro_linearized_genpackfield}
    \partial_t \delta\rho &=& -\partial_x \Big[-D(\rho_0) \partial_x\delta\rho + \epsilon \sigma'(\rho_0) \delta\rho \nonumber \\
    &+& \sigma(\rho_0) \big(\epsilon+\lambda \abs{\KDoparm[\delta\dens]} \sin(\varphi_m[\delta\dens] - 2\pi\pos \Eorder) \big) \Big] \, ,
\end{eqnarray}
where $\mobility'(\rho_0)$ stands for the derivative of the mobility $\mobility(\dens)$ with respect to its argument evaluated at $\rho_0$, and where we have used that $\abs{\KDoparm}$ is already first-order in $\delta\rho$, see Eq. \eqref{eq:z_field}. The periodicity of the lattice can be used to expand the density field perturbation in Fourier modes,
\begin{equation}
    \denspxt = \sum_{j=-\infty}^{\infty} \fcoeft{j} e^{i 2\pi\pos j} \, ,
\end{equation}
where the $j$-th Fourier coefficient is given by $\fcoeft{j} = \int_0^{1} \dd x\denspxt \textrm{e}^{-\textrm{i} 2\pi\pos j}$. \todo{transición brusca}
Note that the Kuramoto-Daido parameter is proportional to the $(-m)$-th Fourier coefficient in this expansion, i.e., $\KDoparm[\delta\dens] = \fcoeft{-\Eorder} / \meandens$.
Replacing the Fourier expansion in the linearized Eq.~\eqref{eq:hydro_linearized_genpackfield}, we thus obtain
\begin{equation}
\sum_{j=-\infty}^{\infty}\Big(\partial_t C_j(t) + \zeta_j C_j(t) \Big) \textrm{e}^{\textrm{i} 2\pi\pos j } = 0 \, ,
\label{Fourier1}
\end{equation}
where we have defined
\begin{eqnarray}
\label{Fourier2}
    \zeta_j 
    &\equiv&
    \left(2\pi j\right)^2 \diffcoef (\meandens)  + \textrm{i} 2\pi j \mobility^{\prime}(\meandens)\Ed  \nonumber
    \\
    &-& \Epampl\frac{\mobility(\meandens)}{2\meandens} 2\pi\Eorder (\delta_{j, \Eorder} + \delta_{j,-\Eorder}) \, ,
\end{eqnarray}
and $\delta_{j, \Eorder}$ and $\delta_{j,-\Eorder}$ are Kronecker deltas.
As the different complex exponentials in Eq. \eqref{Fourier1} are linearly independent, each term in the sum must be zero, so the solution for the different Fourier coefficients is just $\fcoeft{j} = C_j(0) \textrm{e}^{-\zeta_j \tv}$, with $C_j(0)$ the coefficients associated with the initial perturbation $\delta\dens(x,0)$.
The stability of the different Fourier modes is then controlled by the real part of $\zeta_j$, for which we have to consider two distinct cases: $\abs{j} \neq \Eorder$ and $\abs{j} = \Eorder$. In the first case $\abs{j} \neq \Eorder$, we have 
$
    \Re(\zeta_j)
    =
    \diffcoef(\meandens) \left(2\pi j\right)^2>0\; \forall j
$, 
so that these Fourier modes will always decay.
On the other hand, when $\abs{j} = \Eorder$, the decay rate involves a competition between the diffusion term and the packing field,
\begin{equation}
    \label{Recrit}
    \Re(\zeta_{\pm m})
    =
    \left(2\pi \Eorder\right)^2\left( \diffcoef(\meandens)  - \Epampl\frac{\mobility(\meandens)}{4\pi \Eorder\meandens}\right) \,.
\end{equation}
The critical value of $\Epampl$ is reached whenever $\Re(\zeta_{\pm m}) = 0$, and reads
\begin{equation}
    \label{eq:clambda_genpackfield}
    \Epamplcrit{\Eorder}
    =
    4\pi\Eorder \frac{\diffcoef(\meandens)\meandens}{\mobility(\meandens)} \, .
\end{equation}
In this way we expect the homogeneous density solution $\densxt = \meandens$ to become unstable for $\lambda>\lambda_c^{(\Eorder)}$, leading to non-flat a density field solution with a more complex spatiotemporal structure. 
In particular, right after the instability takes place, it seems reasonable to
expect from the previous analysis that the main non-flat contributions to the density profile will be the $\pm\Eorder$ order Fourier modes. 
Therefore, in this regime we expect a traveling-wave profile close to $\dens(x,t)=\meandens + A\cos(\omega t - 2\pi\Eorder\pos)$ with $A$ small and where the angular velocity $\omega$ is obtained from the imaginary part of Eq.~\eqref{Fourier2}.
\begin{equation}
\label{eq:omega_twperturbative_genpackfield}
    \omega
    =
    2\pi \Eorder \mobility'(\meandens)\Ed
\end{equation}
This suggests that beyond the instability, the homogeneous density turns into a single ($m=1$) or multiple ($m>1$) condensates periodically moving at a constant velocity proportional to $\mobility'(\meandens)$ and $\Ed$, thus giving rise to a continuous time crystal.
In addition, the average current $\meancurrent=\int_0^1 \dd\pos j(x,t)$ for this profile can be calculated from the local current in the linearized equation \eqref{eq:hydro_linearized_genpackfield}, resulting in
\begin{equation}
\label{eq:J_twperturbative_genpackfield}
    \meancurrent
    =
    \meancurrentflat+ A^2 \mobility''(\meandens) \Ed/4
    ,
\end{equation}
where $\meancurrentflat = \mobility(\meandens) \Ed$ is the averaged current in the homogeneous phase.
While we only expect these equations to hold true close to $\Epampl = \Epamplcrit{\Eorder}$, they highlight the relevance of the transport coefficients in the behavior of the model under the packing field.
Depending on the slope and convexity of the mobility of each model, the packing field will enhance or lower the current and speed of the traveling wave.

\todo[inline, size=\normalsize]{
    Finally, regarding Eq.~\eqref{eq:clambda_genpackfield}, it is worth commenting on the dependence of the critical packing-field coupling $\lambda_c^{(\Eorder)}$ with the packing order $m$.
    This admits a simple interpretation in terms of the competition between diffusion and the effect of the packing field: While the effect of diffusion, which tends to destroy the $m$ particle condensates, scales with $m^2$ in Eq. \eqref{Recrit}, the action of the packing field promoting the condensates scales with $m$.
    Therefore a stronger packing-field coupling, $\lambda_c^{(\Eorder)}$, is needed as $m$ increases to destabilize the homogeneous solution.
}

\section{Mapping traveling-wave profiles across different packing orders}
\sectionmark{Mapping density profiles across different packing orders}
\label{sec:mappingorders_genpackfield}
    
Next, we will explore traveling-wave solutions to Eq.~\eqref{eq:hydro_field_genpackfield} and investigate the relationship between solutions corresponding to different packing orders $\Eorder$.
Specifically, we set out to show that, provided that a $2\pi/m$-periodic traveling-wave solution exists for packing order $\Eorder$, this solution can be  
built by gluing together $\Eorder$ copies of the solution of the $m=1$ hydrodynamic equation with properly rescaled parameters.
To prove this we start by using a traveling-wave ansatz $\dens_{\Eorder}(\pos,\tv) = \denstw{\Eorder}(\omega \tv - 2\pi \pos)$ in the hydrodynamic equation with $\Eorder$-th order packing, where $\denstw{\Eorder}$ is a generic $2\pi$-periodic function, and  
$\omega$ denotes the traveling-wave velocity. Substituting this into Eq. \eqref{eq:hydro_field_genpackfield} we get
\begin{equation}
\label{eq:hydrotw_genpackfield}
\begin{split}
    \omega \denstw{\Eorder}^{\prime}(\twarg)
    &=
    2\pi \dv{\twarg} \Big\{ D(\denstw{\Eorder}) 2\pi \denstw{\Eorder}^{\prime}(\twarg) 
    \\
    &+ \mobility(\denstw{\Eorder}) \big[\Ed + \Epampl \abs{\KDoparm} \sin(\KDoparargmini + \Eorder\twarg)\big]\Big\}\, ,
\end{split}
\end{equation}
where we have introduced the variable $\twarg = \omega \tv - 2\pi \pos$.
In addition, we have used that under the traveling-wave ansatz the magnitude of the order parameter is constant and its complex phase increases linearly in time, i.e., $\KDoparm[\denstw{\Eorder}] = \abs{\KDoparm} \expsup{i(\KDoparargmini + \Eorder\omega t)}$ and thus $\KDoparargm[\denstw{\Eorder}]=\KDoparargmini + \Eorder\omega t$.

Since we expect the formation of $m$ equivalent particle condensates once the homogeneous density profile becomes unstable, it is reasonable to further assume that the resulting traveling density wave will exhibit $(2\pi/\Eorder)$-periodic behavior, i.e. $\denstw{\Eorder}(\twarg) = \denstwt(\Eorder\twarg)$ with $\denstwt$ a new $2\pi$-periodic function.
Under this additional assumption, the initial $\Eorder$-th order Kuramoto-Daido parameter reads
\begin{equation}
\begin{split}
    \KDoparm(0)
    &=
    \frac{1}{\meandens}
        \int_0^{1} \dd \pos \denstw{\Eorder}(- 2\pi \pos) \mathrm{e}^{i 2\pi\Eorder\pos}
    \\
    &=
    \frac{1}{\meandens}
        \int_0^{1} \dd \pos \denstwt(- 2\pi \Eorder \pos) \mathrm{e}^{i 2\pi\Eorder\pos}
    \\
    &=
    \frac{1}{\meandens}
        \int_0^{1} \dd \tilde{\pos} \denstwt(- 2\pi \tilde{\pos}) \mathrm{e}^{i 2\pi\tilde{\pos}}
    \equiv
    \tilde{\KDopar}_1(0) \, ,
\end{split}
\end{equation}
where we haven taken into account the periodicity of $\denstwt$ and where $\tilde{\KDopar}_1(0)$ is defined as the initial $m=1$ Kuramoto-Daido parameter of $\denstwt(\twarg)$.
Using this result we can rewrite Eq.~\eqref{eq:hydrotw_genpackfield} in terms of $\denstwt$ and a new variable $\tilde{\twarg} = \Eorder \twarg$,
\begin{eqnarray}
    \frac{\omega}{\Eorder} \denstwtprime(\twargt)
    &=&
    2\pi \dv{\twargt} \Big\{D(\denstwt) 2\pi \denstwtprime(\twargt)  \nonumber
    \\
    &+& \mobility(\denstwt) \big[\frac{\Ed}{\Eorder} + \frac{\Epampl}{\Eorder} \abs{\tilde{\KDopar}_1} \sin(\tilde{\varphi}_{1,0} - \twargt)\big]\Big\} \, , 
    \label{wave2}
\end{eqnarray}
where we have used that $\denstw{\Eorder}^{\prime}(\twarg) = \Eorder \denstwtprime(\twargt)$.
This equation 
is nothing but the original equation for the traveling wave Eq.~\eqref{eq:hydrotw_genpackfield} with a first-order packing $\Eorder=1$ and rescaled parameters
\begin{equation}
\label{eq:params_rescaling_profile_equivalence}
    \tilde{\omega}
    =
    \frac{\omega}{\Eorder}  \, , \qquad  \tilde{\Ed} = \frac{\Ed}{\Eorder} \, , \qquad \tilde{\Epampl}=\frac{\Epampl}{\Eorder} \, .
\end{equation}
In this way, if $\denstwt(\tilde{\omega} \tv - 2\pi\pos)$ is a traveling-wave solution of the hydrodynamic equation \eqref{eq:hydro_field_genpackfield} with $m=1$ and parameters $\tilde{\Ed}$ and $\tilde{\Epampl}$, we have just proved that $\dens_{\Eorder}(\pos, \tv) = \denstwt(\Eorder^2\tilde{\omega} \tv - \Eorder2\pi\pos)$ is a solution of the corresponding hydrodynamic equation with $\Eorder$-th order packing and parameters $\Ed=\Eorder\tilde{\Ed}$ and $\Epampl = \Eorder \tilde{\Epampl}$.
This exact mapping between the traveling-wave solutions for different packing orders will be fully confirmed below in simulations of the stochastic model.

In the Kuramoto model, the presence of linear transport coefficients can be exploited to prove a full dynamical equivalence between the first-order and higher-order couplings \cite{delabays2019}.
However such equivalence cannot be extended to our case since our transport coefficients may be, in general, nonlinear.

\todo[inline]{No sé si merece la pena alguna imagen aquí para ilustrar el concepto}
\todo[inline]{Se puede demostrar la equivalencia de la dinámica de las perturbaciones lineales sobre la solución estacionaria, pero no sé si merece la pena}
\todo[inline]{Se puede comentar que esto es una diferencia con respecto al modelo de Kuramoto con muchos modos, donde la solución no tiene porqué ser $m$-simétrica.}

\section[Time-crystals in particular models of diffusive transport]{Multicondensate time-crystal phases in particular models of diffusive transport}
\todo[inline]{Cambia el título}
\label{sec:models_hydro_genpackfield}

To illustrate these results\todo{le falta brillo}, we will analyze the hydrodynamic behavior of several lattice gas models under the action of the packing field of Eq.~\eqref{eq:packingfield_genpackfield}.
This behavior is completely specified by the hydrodynamic equation \eqref{eq:hydro_field_genpackfield} and the transport coefficients of the considered models, which are:
\begin{itemize}
    \item
    the Weakly Asymmetric Simple Exclusion Process (WASEP) \cite{gartner87a,de-masi89a,}, 
    \begin{equation}
        \diffcoef(\dens)=1/2
        ,\quad
        \mobility(\dens)=\dens (1-\dens)
        ;
    \end{equation}

    \item
    the Kipnis-Marchioro-Presutti (KMP) model \cite{kipnis82a}, 
    \begin{equation}
        D(\rho)=1/2
        ,\quad
        \sigma(\rho)=\rho^2
        ;
    \end{equation}

    \item
    the random walk (RW)\todo{añadir alguna cita de esto}, 
    \begin{equation}
        D(\rho)=1/2
        ,\quad
        \sigma(\rho)=\rho
        ;
    \end{equation}
    
    \item
    and the Katz-Lebowitz-Spohn (KLS) model \cite{katz84,hager01a,Krapivsky13a} with parameters $\nu = 0.9$ and $\delta = 0$, which presents nonlinear diffusion coefficient and mobility (see Appendix~\ref{app:KLScoeffs} for their explicit expressions).
\end{itemize}
The mobilities of such models are illustrated in Fig.~\ref{fig:observableshydro_models_genpackfield}(a).

Studying the dynamics of these models involves solving the nonlinear second-order traveling-wave equation \eqref{eq:hydrotw_genpackfield}.
This equation, with periodic boundaries and where the Kuramoto-Daido parameter introduces a dependence on the integral of the solution, poses a challenge for standard numerical methods.
To address this, we devised an alternative approach in which we transformed this equation into an ordinary first-order differential equation supplemented by several self-consistence relations.
This method, whose details are provided in Appendix~\ref{app:selfconsisthydro}, allowed us to calculate the traveling-wave behavior of the presented models, which is shown in Fig.~\ref{fig:observableshydro_models_genpackfield}(b-d) and Fig.~\ref{fig:profileshydro_models_genpackfield}.

In Fig.~\ref{fig:observableshydro_models_genpackfield}, we consider the previous models with different combinations of the average density $\meandens$: the WASEP with densities $\meandens = 1/3,\, 3/5$, the KMP model with $\meandens = 1/3$, the random walk with $\meandens = 1/3$ and the KLS with $\meandens = 1/3$ and $1/2$.
In all cases, the packing order is $\Eorder=1$ and the driving in the packing field is $\Ed = 10$.
As we have anticipated above, after the critical value of the coupling $\lambda_c$ given by Eq.~\eqref{eq:clambda_genpackfield} the flat profile $\dens(x,t) = \meandens$ becomes inestable and a phase transition to a time-crystal phase in the form of a traveling density profile takes place.
Such transition is well captured by the magnitude of the Kuramoto-Daido order parameter $|\KDopar_{1}|$, which becomes non-zero after $\Epamplcritnom$, as displayed in Fig.~\ref{fig:observableshydro_models_genpackfield}(b).
The exception is the KLS with $\meandens = 1/2$, where the traveling-wave phase starts before $\Epamplcritnom$, which will be discussed later. \todo{se me hace un poco rara la frase}

For the RW, KMP and WASEP the transition is continuous for any value of the total density $\rho_0$, though we just display the results for $\rho_0=1/3$ (and additionally $\rho_0=3/5$ in the WASEP).
In these cases, since the condensates emerge continuously, their velocities right after the instability follow the perturbative prediction of Eq.~\eqref{eq:omega_twperturbative_genpackfield}, i.e. are proportional to $\mobility'(\meandens)$ and $\Ed$, as illustrated in Fig.~\ref{fig:observableshydro_models_genpackfield}(c).
The WASEP model with $\meandens = 1/2$ has a negative derivative of the mobility, $\mobility'(\meandens)$, and thus displays a negative $\omega$, while the other cases show $\mobility'(\meandens) > 0$ and thus $\omega > 0$.

\begin{figure}
    \centering
    \includegraphics[width=\linewidth]{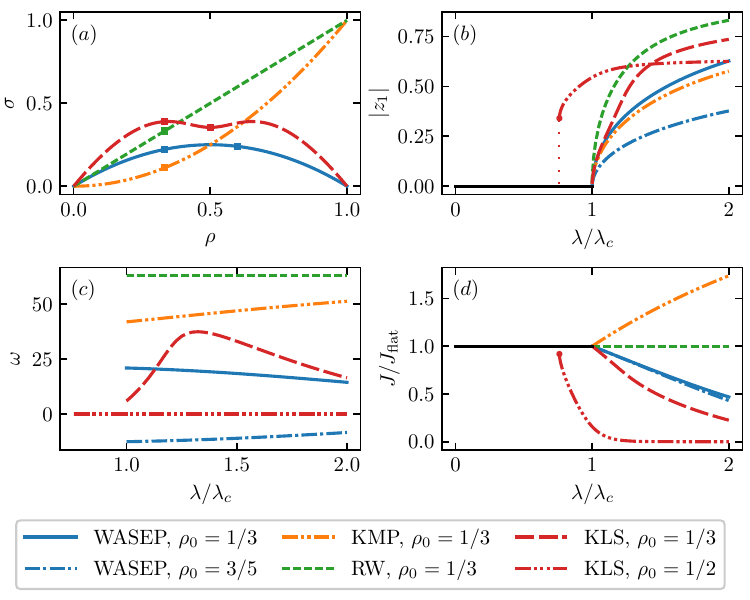}
    \caption{
        Application of the packing field with $\Eorder=1$ to the random walk (RW), WASEP, KMP and KLS models.
        (a) Mobility of the models as a function of the density in the range $\dens = [0,1]$.
        The square markers display the mobility mobility of each model evaluated on its mean density.
        (b) Magnitude of the Kuramoto-Daido parameter $\KDoparm$ as a function of the coupling $\Epampl$ to the packing field.
        Most models present a continuous phase transition to a traveling-wave profile (marked by $\abs{\KDoparm}>0$) when the flat profile becomes unstable at $\Epampl = \Epamplcritnom$.
        However, the KLS model with $\dens=1/2$ displays a discontinuous phase transition accompanied by a region of bistability from $\Epampl/\Epamplcritnom \approx 0.76$ to $\Epampl/\Epamplcritnom = 1$.
        (c) Angular velocity of the traveling wave profile against $\Epampl/\Epamplcritnom$.
        (d) Mean current $\meancurrent/\meancurrentflat$ relative to the one of the flat profile against $\Epampl/\Epamplcritnom$. 
        In this case, the KLS model with $\dens=1/2$ also presents a small discontinuity.
        \todo[inline]{does it feel strange to have only 4 curves in panel (a)?}
        \todo[inline]{I would change $\meandens$ values in the WASEP to make them more different}
    }
    \label{fig:observableshydro_models_genpackfield}
\end{figure}

\begin{figure}
    \centering
    \includegraphics[width=0.8\linewidth]{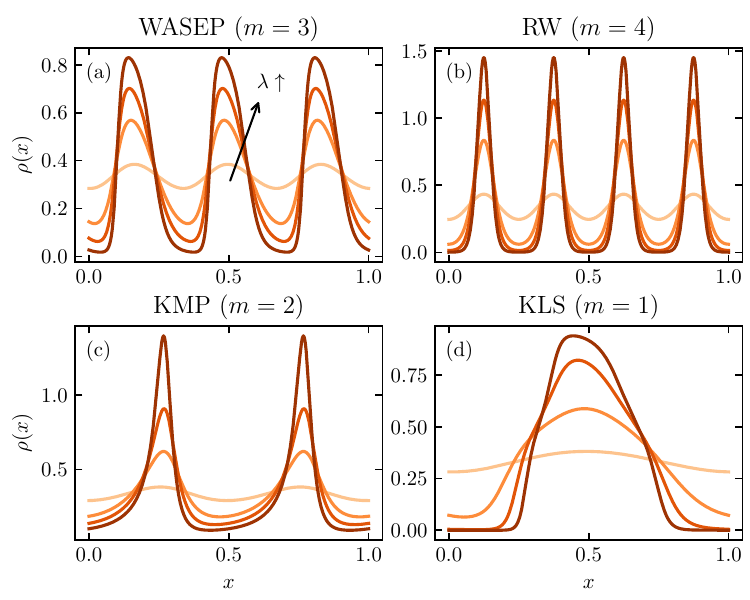} 
    \caption{
        Traveling-wave profiles dependence on the coupling $\Epampl$.
        Each panel corresponds to a different model: (a) WASEP with $\Eorder=3$, (b) random walk with $\Eorder=4$, (c) KMP model with $\Eorder=2$ and (d) KLS models with $\Eorder=1$.
        In all cases, the mean density is $\meandens = 1/3$, the driving term of the field is $\Ed=10\Eorder$ and the couplings are $\Epampl/\Epamplcrit{\Eorder} = 1.01,\, 1.2,\, 1.5,$ and $2$, with larger values shown in increasing color density. 
    }
\label{fig:profileshydro_models_genpackfield}
\end{figure}

The current also manifests different behaviors right after the critical point, since it depends on the value of $\mobility''(\meandens)$ [see Eq.~\eqref{eq:J_twperturbative_genpackfield}].
This is shown in Fig.~\ref{fig:observableshydro_models_genpackfield}(d).
The RW model does not change its current across the transition since the mobility is linear, and thus $\mobility''(\meandens)=0$. \todo{actually in the RW the mean current does not depend on the density profile at all, only in $\meandens$}
In the WASEP model for both mean densities, we have that $\sigma''(\rho_0)<0$ so the current is reduced in the traveling-wave phase.
This is due to the packing-field mechanism giving rise to regions of either high or low density, therefore leading to lower local mobilities. \todo{exlusion in the responsible of low mobilities at high densities}
The opposite happens in the KMP model, where the higher densities promoted by the packing mechanism increase the average mobility and the current.

The KLS model, on the other hand, exhibits a richer behavior due to the additional complexity of its transport coefficients.
When $\meandens=1/3$, with $\mobility'(\meandens)>0$ and $\mobility''(\meandens)<0$, it displays continuous transition obeying the perturbative predictions for $\omega$ and $\meancurrent$.
Still, a different behavior compared to the other models is reflected in the non-monotonous behavior of $\omega$.
For $\meandens=1/2$, with $\mobility'(\meandens)=0$ and $\mobility''(\meandens)>0$, it presents an abrupt change of both the order parameter and the current before the critical point, resembling the explosive synchronization observed for some oscillators \cite{leyva12a,Hu14a}. \todo{creo que la velocidad es nula porque tengo PH symmetry}
This signals the onset of a first-order phase transition, presenting bistability in the region from $\Epampl/\Epamplcritnom \approx 0.76$ to $\Epampl/\Epamplcritnom = 1$. \todo{do all first-order phase transitions have bistable regimes?}
While we will delve much more into this model, its behavior highlights the richness found in the packing field mechanism.

Finally, Fig.~\ref{fig:profileshydro_models_genpackfield} shows how the traveling-wave profile emerges as $\Epampl$ increases for several choices of the models and parameters.
The distinction between models incorporating exclusion, such as the WASEP and KLS, and those without is evident.
In the former, as $\Epampl$ increases, the condensates tend to adopt a shape reminiscent of a square pulse, since the density can not go beyond $\dens = 1$.
Conversely, in the absence of exclusion, the concentration around their respective peaks simply keeps increasing.
Another interesting observation is that, except for the random walk, where there is no interaction between particles, each condensate within the traveling wave exhibits asymmetry around its maximum.
This is further explored in Fig.~\ref{fig:twprofile_vs_a_largo} in the WASEP model with $\meandens=1/3$, where the traveling-wave profiles for increasing values of the driving field $\Ed$ are shown.
We observe that the stronger the driving field $\Ed$, the more important this difference between the leading and rear fronts is, resulting in an increasing asymmetry.
In the inset to the figure, we also explore the effect of the driving field on the condensate velocity $\omega$, which increases with $\Ed$ as expected.
Interestingly, for large couplings $\Epampl$, this increase in velocity is superlinear for weak and moderate driving fields ($\Ed\lesssim 10$), becoming progressively linear for $\Ed> 10$.

\begin{figure}
    \centering
    \includegraphics[width=0.7\linewidth]{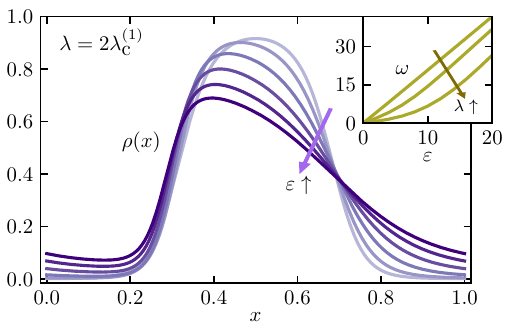}
    \caption{{Condensate asymmetry and velocity.} 
        Main: Hydrodynamic solution of the traveling wave in the WASEP with $\meandens=1/3$, $\Eorder=1$, $\Epampl=2\lambda_c^{(1)}$ and driving fields $\Ed=0, 2, 4, 6, 9, 10$. The asymmetry of the condensate density profile increases as $\Ed$ grows. Inset: Traveling-wave velocity as a function of the driving field $\Ed$ for couplings $\Epampl/\Epamplcritnom = 1.05, 2, 3$.
    }
    \label{fig:twprofile_vs_a_largo}
\end{figure}

\section{Monte Carlo microscopic analysis in the closed WASEP}
\label{sec:numerics}
\todo[inline]{TCEP o WASEP en el título}
\todo[inline]{Habría que especificar los tiempos de relajación}
\todo[inline]{Habría que especificar el método de simulación}

In this section, we will perform further analysis of the packing field mechanism by performing simulations of the microscopic dynamics of the WASEP.
In particular, we will verify the instability prediction and mapping across different orders of Sections \ref{sec:criticalpoint_genpackfield} and \ref{sec:mappingorders_genpackfield}, we will study the rigidity of the time-crystal phase and we will discuss the universality class of the phase transition. 

As in Chapter~\ref{chap:buildingtc}, the packing field mechanism is implemented in the microscopic WASEP 
by replacing the constant external field controlling the asymmetry in the rates with the one introduced in Eq.~\eqref{eq:packingfielddriving_genpackfield}.
If we consider a lattice with $\lattsize$ sites, the system state is given by the vector $\occupvect\equiv \{n_k, k\in[1,L]\}$, with $n_k = 0,1$ the occupation of the $k$th site. \todo{no me gusta la conexión}
The external field at the $k$th site is then given by
\begin{equation}
    \label{eq:packfieldmicro_genpackfield}
    \mathcal{E}^{(m)}_k(\occupvect)
    =
    \abs{\KDoparm(\occupvect)} \sin(\KDoparargm(\occupvect) - \frac{2\pi \Eorder k}{\lattsize}) \, ,
\end{equation}
where the $\Eorder$th-order Kuramoto-Daido parameter is calculated as
\begin{equation}
\label{eq:KDoparammicro_genpackfield}
    \KDoparm(\occupvect)
    =
    \frac{1}{\nprtcls}{\sum_{\kp=1}^\lattsize \occup_{\kp} e^{i 2\pi \Eorder \kp/\lattsize}} \equiv \abs{\KDoparm(\occupvect)} \textrm{e}^{\textrm{i}\KDoparargm(\occupvect)} 
    \, .
\end{equation}

In particular, we carried out extensive discrete-time Monte Carlo simulations \cite{bortz1975,gillespie1977,newman99a,binder2010} of the 
closed WASEP under different packing fields ($\Eorder = 1,2,3$) at a constant global density $\meandens=\nprtcls/\lattsize=1/3$ and lattice sizes of $\lattsize\in[480,2880]$, so as to perform a detailed finite-size scaling analysis of the predicted phase transition.
Unless otherwise specified, we used a driving field $\Ed(\Eorder) = 2.5 \Eorder$. \todo{en realidad en esta sección siempre se utiliza esto}
\todo[inline]{
    Concerning the simulation of the stochastic gas trajectories, we used a discrete-time acceptance-rejection method instead of the usual continuous-time one (also known as Kinetic Monte Carlo or Gillespie \cite{bortz1975,gillespie1977}). This allows us to reduce the computational effort significantly, as explained in Appendix~\ref{sec:discrete-time-method}, where the method details are discussed.
    This allows us to avoid recalculating all the jump rates in each timestep (they all change since $\KDoparm$ does) for the computation of the escape rate required to sample the time for the next transition, which significantly reduces the computational effort.
    The time-dependent observables can be compared with the continuous-time results by rescaling the time by measuring the activity in the model and the ``theoretical'' escape rate.
    The details of the algorithm are explained in the Appendix.
}

\begin{figure*}
            \includegraphics[width=\linewidth]{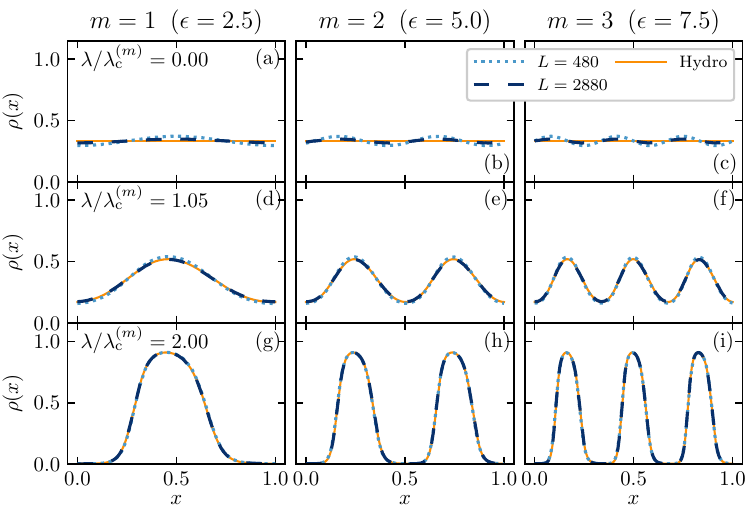}
    \caption{
        Measured density profiles for packing coupling orders $\Eorder=1$ (left column), $\Eorder=2$ (central column), and $\Eorder=3$ (right column), as obtained from Monte Carlo simulations of the WASEP for $\rho_0=1/3$, driving field $\Ed(\Eorder) = 2.5 \Eorder$, and different values of $L$ (see legend). Panels (a)-(c) show data obtained for packing field coupling $\lambda=0$, corresponding to the homogeneous density phase, panels (d)-(f) display measurements right above the critical point $\lambda_c^{(m)}$, and panels (g)-(i) show data deep inside the time-crystal phase. 
        Full lines in panels (d)-(i) correspond to the solutions of the traveling wave Eq.~\eqref{eq:hydrotw_genpackfield} for the different parameters, while lines in panels (a)-(c) represent the constant density solution $\rho_0$.
        }
    \label{fig:density_profiles_phases_largo}
\end{figure*}

The phase transition is most evident at the configurational level, so we measured the average particle density profiles along the $1d$ ring for different values of the packing field coupling $\lambda$ across the predicted critical point $\lambda_c^{(m)}$, different packing coupling orders $m=1,~2,~3$, and varying $L$, as shown in Fig.~\ref{fig:density_profiles_phases_largo}.
In order to capture the shape of the traveling wave, we repeatedly measured the instantaneous occupation vector of the lattice, we centered the vectors around the angular position given by the argument of the order parameter, i.e, we shifted them so that $\KDoparargm\to 0$, and we then performed the average. \todo{no estoy seguro de que se entienda}
The results are shown in Fig.~\ref{fig:density_profiles_phases_largo}.
Note that this procedure leads to a spurious weak structure in the flat phase, $\Epampl<\Epamplcritnom$, due to the way the fluctuations of the particles' packing are averaged in this method [see Figs.~\ref{fig:density_profiles_phases_largo}(a)-(c)].
This spurious structure disappears as we go toward larger $\Epampl$.
Conversely, supercritical ($\lambda>\lambda_c^{(m)}$) density fields exhibit a much-pronounced structure resulting from the appearance of traveling particle condensates, see Figs.~\ref{fig:density_profiles_phases_largo}(d)-(i).
In all cases, the correspondence between the hydrodynamic solution and the microscopic results, even for $\lattsize=480$, is remarkable.

\begin{figure}
    \centering
    \includegraphics[width=0.6\linewidth]{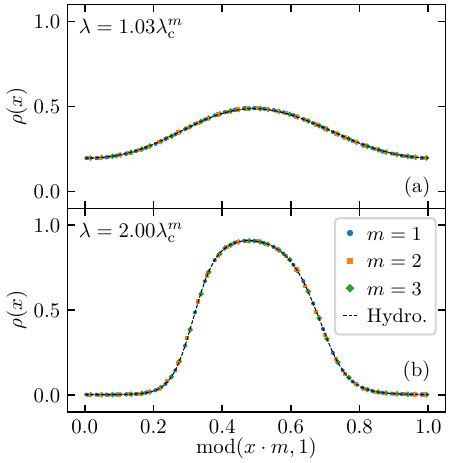}
    \caption{Mapping between density profiles of different orders.
             Collapse of the measured condensate density profiles in the WASEP for different packing coupling orders $\Eorder = 1,~2,~3$, for $\lambda=1.03\lambda_c^{(m)}$ (a) and $\lambda=2\lambda_c^{(m)}$ (b).
             The different markers correspond to simulations of different packing orders $\Eorder$ and $\lattsize = 1920$, while the dashed line corresponds to the hydrodynamic solution.
             The driving field for each packing coupling order is $\Ed = 2.5\Eorder$, and the global density is $\rho_0=1/3$.
        }
    \label{fig:density_profiles_collapse_largo}
\end{figure}

Regarding these profiles, it remains to be checked whether, under the proper choice of the parameters, the shape of each of the emergent particle condensates remains constant for the different packing orders $m$, as it is predicted 
in Section~\ref{sec:mappingorders_genpackfield}.
To address this question, we measured the density profiles in two cases $\lambda/\lambda_c^{(m)} = 1.03$ and $2$ inside the time-crystal phase for $m=1,~2,~3$ and with parameters $\Ed$ and $\Epampl$ scaled according to Eq.~\eqref{eq:params_rescaling_profile_equivalence}.
This is shown in Fig.~\ref{fig:density_profiles_collapse_largo}.
As predicted by the hydrodynamic theory, the measured condensate profiles collapse neatly on top of each other, $f_m(x)=f_1(mx)$, once the spatial coordinate $x$ is stretched by the corresponding packing order $\Eorder$ and then remapped to the $x \in [0,1[$ interval using the transformation $m x \to \textrm{mod}(m x,1)$.
Moreover, the collapsed shape perfectly agrees with the hydrodynamic curve, see dashed lines in Fig.~\ref{fig:density_profiles_collapse_largo}.

\begin{figure}
    \centering
    \vspace{-5mm}
    \includegraphics{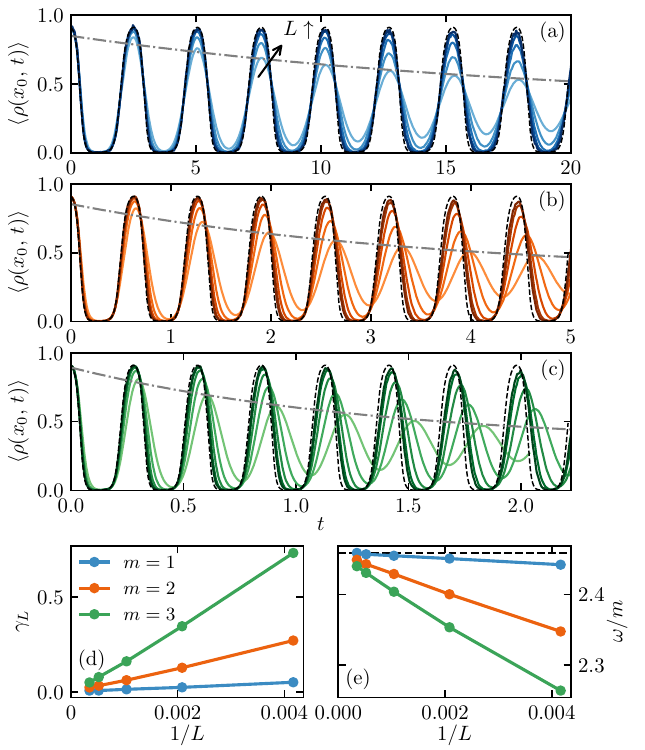}
    \vspace{-2mm}
    \caption{
            Rigidity of the time-crystal phase in the WASEP with packing fields of orders $\Eorder=1,\,2,\,3$ and parameters $\Ed=2.5\Eorder$ and $\Epampl = 2\Epamplcrit{\Eorder}$.
            Panels (a-c) show the oscillations of the average density at a given lattice site $\langle \dens(x_0,t)\rangle$ for packing orders $\Eorder = 1,\,2,\,3$, respectively, and for different lattice sizes $\lattsize = 240,\,480,\,960,\,1920,\,2880$ (drawn in increasing color intensity).
            We observe that the oscillations' amplitude is damped in time for finite lattice sizes.
            The black dashed line shows the hydrodynamic solution, while the gray dash-dotted line displays the exponential envelope of the $\lattsize=240$ curve.
            Panel (d) shows the dependence of the damping rate $\damprateL$ of the oscillations on the lattice size, displaying how it vanishes in the macroscopic limit.
            Panel (e) illustrates the convergence of the finite-size angular speed $\omega$ of the traveling wave to the hydrodynamic one, shown as a black horizontal dashed line.
            Notice that, for the choice of the parameters, Eq.~\eqref{eq:params_rescaling_profile_equivalence} dictates that the different hydrodynamical $\omega$'s are simply related by a factor $\Eorder$.
            In this panel, the error bars are not shown because they are smaller than the markers.
            For all choices of the parameters, the results were obtained from the average of at least 1000 runs of the simulations.
            \todo[inline]{se podría pintar la varianza de la omega también}
        }
    \label{fig:oscillations_WASEP}
\end{figure}

{
\newcommand{\ta}{t}
\newcommand{\tb}{t^{\prime}}

The rigidity of the time-crystal phase can be assessed by monitoring the average density at a given point along the ring, $\avg{\dens(x_0, t)}$, as a function of time. \todo{should I specify the parameters in the main text?}
In particular, this is measured using the following procedure:
\begin{enumerate}
    \item 
    First, once in the traveling-wave regime, the profile is shifted so that its maximum aligns with $x=0$, and time is set to $t=0$.

    \item
    Next, the evolution of the profile $\dens(x, t)$ is measured up to a specified time $T$.

    \item
    Finally, the process is iterated by returning to the first step, and this cycle is repeated until the necessary statistics are obtained.
    The density profiles at each time step are averaged across different realizations to obtain $\avg{\dens(x_0, t)}$.
\end{enumerate}
Applying this approach to our model, we find persistent oscillations as a function of time, see Fig.~\ref{fig:oscillations_WASEP}(a-c), a universal signature of time-crystalline order \cite{yao18a, wilczek12a, shapere12a, zakrzewski12a, moessner17a, richerme17a, sacha18a, bruno13a, nozieres13a, watanabe15a, khemani16a, keyserlingk16a, else16a, gambetta19a, yao17a, yao17a, zhang17a, choi17a, nakatsugawa17a, lazarides17a, gong18a, tucker18a, osullivan18a, lazarides19a, yao18b, gambetta19b, iemini18a, medenjak20a}.
These oscillations are characterized by a period $2\pi\omega^{-1}$ in the diffusive scale, with $\omega$ the condensate angular velocity.
In addition, these panels show that, for any finite lattice size $L$, the oscillations' amplitude diminishes over time.
This is due to the deviations from perfectly periodic traveling waves when $\lattsize$ is finite. \todo{I don't like the sentence}
Consequently, as time $t$ progresses, the positions of the condensates in different realizations become more dispersed, so that when the average is taken the structure of the traveling wave becomes increasingly blurred.
For long times, the amplitude of the oscillation is expected to follow an exponential decay.
This, alongside the traveling-wave ansatz, means that we expect the following relation between the averaged density profiles at two different times $\tb>\ta$:
\begin{equation}
    \avg{\dens(x, \tb)}
    =
    \meandens + \expsup{-\damprateL(\tb-\ta)} \left(\avg{\dens(x - \frac{\omega}{2\pi}(\tb-\ta), \ta)} - \meandens \right),
\end{equation}
where $\damprateL$ is the damping rate of the amplitude, which can be measured using the following procedure. 
For all pairs $\ta, \tb$, we can compare the measured $\avg{\dens(x, \tb)}$ with the one estimated with the previous equation using $\avg{\dens(x, \ta)}$, and then we can adjust $\damprateL$ to minimize their differences.
The damping rates estimated using this method in the previous simulations are shown in Fig.~\ref{fig:oscillations_WASEP}(d), in which we used the profiles measured in the last tenth of the time interval in order to have a compromise between approaching the exponential regime and having enough statistics.
We observe that the measured $\damprateL$ decays to zero in the thermodynamic $L\to \infty$ limit, meaning that the observed local density oscillations converge toward the hydrodynamic (undamped) periodic prediction (dashed black line in Figs. (a-c)).
This clearly signals the rigidity of the long-range spatio-temporal order as we approach the thermodynamic limit.
}
Finally, Fig.~\ref{fig:oscillations_WASEP}(e) displays the measured angular speed $\omega$ in the simulations for different lattice sizes and packing orders.
In all cases, we observe a clear convergence to the hydrodynamical prediction.
It is worth mentioning that the convergence to the hydrodynamic solution is slower for larger packing orders $\Eorder$. This can be attributed to the decrease in the number of particles per condensate as $\Eorder$ increases.

\begin{figure}
    \includegraphics[width=\linewidth]{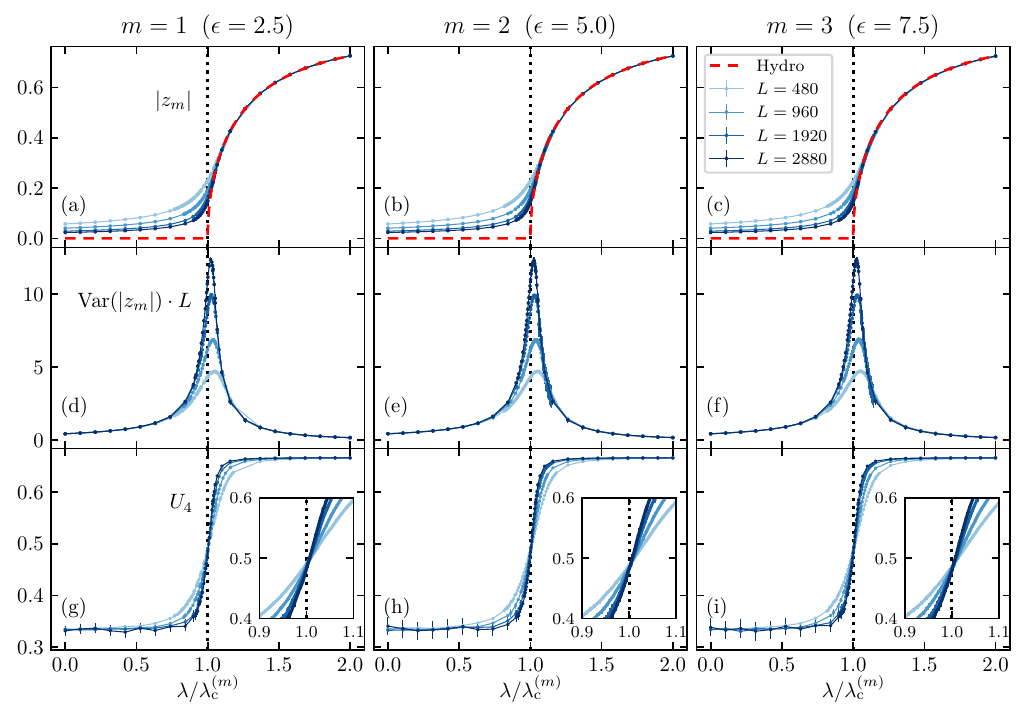}
    \caption{
        Characterizing the phase transition.
        Monte Carlo results for the order parameter $\langle |\KDopar_{\Eorder}|\rangle$ [panels (a)-(c)], its scaled fluctuations $\lattsize \cdot \Var(\abs{\KDoparm})$ [panels (d)-(f)], and the Binder parameter $U_4$ [panels (g)-(i)], as a function of the packing field coupling $\lambda$ and measured for packing coupling orders $\Eorder=1$ (left column), $\Eorder=2$ (central column), and $\Eorder=3$ (right column), global density $\rho_0=1/3$, driving field $\Ed(\Eorder) = 2.5 \Eorder$, and different values of $L$ (see legend). The red dashed lines in panels (a)-(c) correspond to the hydrodynamic prediction for the $\lambda$-dependence of the order parameter, while the vertical dashed lines in all panels signal the predicted critical point $\lambda_c^{(m)}$, see Eq.~\eqref{eq:clambda_genpackfield}.
        }
    \label{fig:critical_point_sim_largo}
\end{figure}

Now we turn to the confirmation of the critical point in Section~\ref{sec:criticalpoint_genpackfield}.
Figures~\ref{fig:critical_point_sim_largo}.(a)-(c) show the measured $\langle |\KDopar_{\Eorder}|\rangle$ as a function of the packing field coupling $\lambda$ for $\rho_0=1/3$, $m=1,~2,~3$ and different system sizes $L$.
As expected, the order parameter remains small for $\lambda<\lambda_c^{(m)}$, i.e. before the predicted critical point, 
approaching zero in the thermodynamic $L\to\infty$ limit. 
Upon crossing $\lambda_c^{(m)}$, the value of $\langle |\KDopar_{\Eorder}|\rangle$ increases sharply but in a continuous way, as expected for a second-order continuous phase transition.
Remarkably, even for moderate system sizes, inside the time-crystal phase the measured $\langle |\KDopar_{\Eorder}|\rangle(\lambda)$ aligns closely with the hydrodynamic predictions.
We also measured the susceptibility across the transition as captured by the order parameter scaled fluctuations, 
$
\lattsize \cdot \Var(|z_m|) \equiv \lattsize\left(\avg{|\KDopar_{\Eorder}|^2} - \avg{|\KDopar_{\Eorder}|}^2\right)
$, see Figs.~\ref{fig:critical_point_sim_largo}(d-f).
This observable displays a pronounced peak around $\Epamplcrit{\Eorder}$.
The sharpness of this peak increases with the system size $\lattsize$, strongly indicating a divergence of the scaled order parameter fluctuations at the critical point $\Epamplcrit{\Eorder}$ in the thermodynamic limit, as expected for a continuous phase transition.
It is also worth mentioning that, for each $\lattsize$, the finite-size peak in the scaled variance is always located slightly after the critical point, $\Epampl > \Epamplcrit{\Eorder}$, an observation in contrast to the behavior reported in the Kuramoto model of synchronization \cite{daido1990, hong2015} and probably linked to the particle exclusion interactions that characterize our model.
In order to further determine the critical packing field coupling $\lambda_c^{(m)}$ from simulations, we also measured the Binder cumulant $\binder=1 - \avg{|\KDopar_{\Eorder}|^4}/(3\avg{|\KDopar_{\Eorder}|^2}^2)$ as a function of $\lambda$ \cite{binder2010}, see Figs.~\ref{fig:critical_point_sim_largo}.(g)-(i).
It is well known that, due to the finite-size scaling properties of this observable, the curves $U_4(\lambda)$ measured for different values of $L$ should cross at a $L$-independent value of $\lambda$ that identifies the infinite-size critical point $\lambda_c^{(m)}$.
This is indeed the case in Figs.~\ref{fig:critical_point_sim_largo}(g)-(i), where we can verify that the numerically measured $\Epamplcrit{\Eorder}$ agrees remarkably well with the hydrodynamic predictions for all the packing orders.

\begin{figure}
    \includegraphics[width=\linewidth]{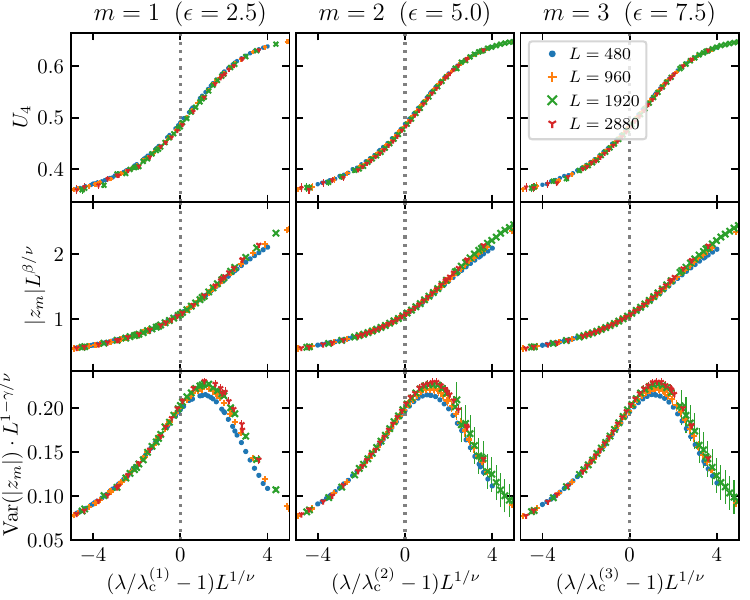}
    \caption{
        Data collapses around the critical region.
        Scaling plots for the Binder cumulant $U_4$ (top row), the order parameter $\langle |\KDopar_{\Eorder}|\rangle$ (middle row), and its scaled fluctuations $\lattsize \cdot \Var(\abs{\KDoparm})$ (bottom row), using all data shown in Fig.~\ref{fig:critical_point_sim_largo} (corresponding to $\rho_0=1/3$, $L\in[480,2880]$, $m=1,~2,~3$ and $\Ed(\Eorder) = 2.5 \Eorder$) and the critical exponents $\nu=2$, $\gamma=1$, $\beta=1/2$ corresponding to the mean field universality class. Vertical dashed lines signal the critical point.
        \todo[inline]{Ahora mismo las barras de error corresponden a 1 sigma}
        }
        \label{fig:critical_exponents_collapse_largo}
\end{figure}

Finally, we performed extensive measurements around the critical region to determine numerically the critical exponents characterizing the universality class of the observed phase transition to this time-crystal phase.
Due to the scale invariance underlying any continuous phase transitions, the relevant observables near the critical point are expected to be \emph{homogeneous functions} of the two length scales characterizing the problem at hand, namely the correlation length $\xi$ and the system size $L$ \cite{barber1983,binder2010}.
This immediately implies the following scaling relations near the critical point,
\begin{eqnarray}
    |\KDopar_{\Eorder}|(\Epampl, \lattsize) &=& \lattsize^{-\KDoparexp/\clengthexp} \mathcal{F}\left[(\Epampl/\lambda_c^{(m)}-1)\lattsize^{1/\clengthexp}\right]
    \, ,
    \\
    \Var(\abs{\KDoparm})
    (\Epampl, \lattsize) &=& \lattsize^{\suscepexp/\clengthexp} \mathcal{G}\left[(\Epampl/\lambda_c^{(m)} - 1)\lattsize^{1/\clengthexp}\right]
    \, , 
    \\
    \binder(\Epampl, \lattsize) &=& \mathcal{H}\left[(\Epampl/\lambda_c^{(m)}-1)\lattsize^{1/\clengthexp}\right] \, , 
\end{eqnarray}
where 
$\KDoparexp$, $\suscepexp$ and $\clengthexp$ are the critical exponents characterizing the power-law behavior near the critical point of the order parameter, the scaled fluctuations, and the correlation length, respectively.
The functions $\mathcal{F}(\cdot)$, $\mathcal{G}(\cdot)$ and $\mathcal{H}(\cdot)$ are the master curves or scaling functions for this universality class.
These scaling relations are tested for our model in Fig.~\ref{fig:critical_exponents_collapse_largo}. 
This figure shows the data collapses of the order parameter $\langle |\KDoparm|\rangle$, its scaled fluctuations $\lattsize\cdot\Var({\KDoparm})$, and the Binder cumulant $\binder$ when using the mean-field exponents $\KDoparexp = 1/2$, $\suscepexp$ = 1, and $\clengthexp=2$, which were the ones providing the best collapse.
Such exponents coincide with the expected mean-field values
\cite{botet_size_1982,hong2015}.
All scaling plots display a remarkable collapse for different lattice sizes.
Note however that, while the collapse for the order parameter scaled fluctuations is excellent for $\Epampl<\Epamplcritm$, it falls off slightly for $\Epampl>\Epamplcritm$. 
\todo[inline]{a scaling behavior already observed in some models of synchronization <- Creo que esto lo había visto en Daido 1990. Pero claro, el es que tiene exponentes diferentes a ambos lados del punto crítico}
Our results thus strongly suggest that the universal scaling properties around the transition to the complex time-crystal phase do not depend on the number $m$ of emergent particle condensates induced by the packing field or other details of the driving mechanism.

\section{Decorated time crystals}

\begin{figure}
    \centering
    \includegraphics[width=0.7\linewidth]{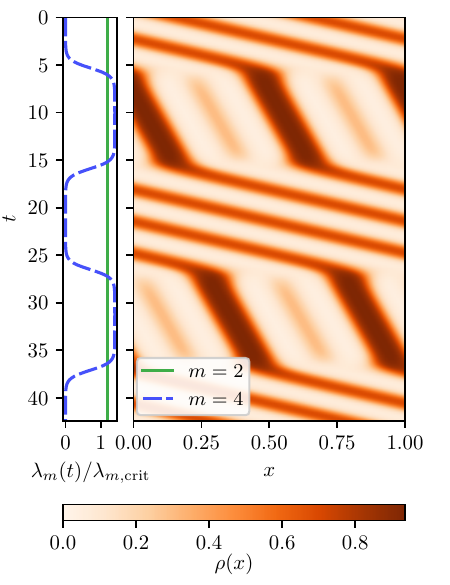}
    \caption{Engineering complex time-crystal phases in the hydrodynamic WASEP with $\meandens=1/3$ and $\Ed=0.5$.
        (Left panel) Time modulation of the packing field couplings.
        The full line corresponds to the (constant and supercritical) primary packing field coupling $\lambda_2=1.2\Epamplcrit{2}$, whereas the dashed line is the secondary time-dependent coupling $\lambda_{4}(t)$, which oscillates between 0 and $\lambda_{4}^\textrm{(max)}=1.4\lambda_c^{(4)}$ with period $T=21.2$ time units.
        (Right panel) Raster plot of the spatiotemporal evolution of the density field $\rho(x,t)$ solution of the hydrodynamic equation~\eqref{eq:hydro_field_genpackfield} subject to the time-modulated external field corresponding to Eq.~\eqref{eq:amplit}. Note the decorated pattern that emerges.
    }
    \label{fig:raster}
\end{figure}

As we have shown throughout this chapter, the general mechanism introduced in this chapter 
allows for the possibility of engineering different time-crystal phases with multiple rotating condensates.
These time-crystal phases can be further ``decorated'' with additional matter waves by introducing extra higher-order packing fields modulated in time.
As a proof of concept, let us consider a time model with an external field given by 
\begin{equation}
\label{eq:amplit}
    \tilde{E}_\pos[\rho] = \epsilon + \lambda_m \mathcal{E}^{(m)}_{\pos}[\rho] + \lambda_{2m}(t) \mathcal{E}^{(2m)}_{\pos}[\rho],
\end{equation}
where $\lambda_m$ is a constant coupling for the packing field $\mathcal{E}^{(m)}_k(\mathbf{\rho})$ [given by Eq.~\eqref{eq:packingfield_genpackfield}], and $\lambda_{2m}(t)$ is a secondary time-dependent coupling of an additional packing field of order $2m$, $\Epg{2\Eorder}[\dens]$.
For its simplicity, we will consider the case $m=2$ in the hydrodynamic WASEP model with $\meandens=1/3$ and $\Ed=0.5$, though one can imagine a myriad of potentially interesting combinations.
We will set the primary $m=2$ coupling beyond its critical value, $\lambda_2 = 1.2\lambda_c^{(2)}$, 
and we will introduce a periodic temporal modulation in the secondary coupling $\lambda_{4}(t)$ switching between zero and $1.4\Epamplcrit{4}$.
This time dependence of the couplings is depicted in the left panel of Fig.~\ref{fig:raster}, with the corresponding spatiotemporal evolution of the density field shown in its right panel.
Interestingly, a time-dependent and decorated pattern emerges, switching in phase with the coupling $\lambda_{2m}(t)$ between a time-crystal phase with $m=2$ traveling condensates when $\lambda_{2m}(t)\approx 0$, and a phase with $m=4$ matter waves when $\lambda_{2m}(t)$ is close to its maximum.
This simple example shows the huge potential of the packing-field route to engineer and control time-crystal phases in stochastic fluids, opening new avenues of future research. 

\section{Conclusion}

In this chapter, we have further explored the packing-field route to time crystals devised in Chapter~\ref{chap:buildingtc}.
We have generalized this packing field mechanism in order to trigger a new phase transition to a time crystal characterized by an arbitrary number $m$ of traveling-wave condensates.

We've characterized the instability leading to the phase transition and established a connection between traveling-wave profiles for different packing orders $\Eorder$ (with properly rescaled field parameters). 
The mechanism has been tested in the hydrodynamic equation of several lattice gases, revealing a diverse range of behaviors depending on the model's transport coefficients.
Microscopic Monte Carlo simulations of the WASEP have fully confirmed our predictions, both for the instability threshold and the equivalence of the density profiles for different packing orders.
Additionally, this microscopic approach has facilitated a numerical characterization of the critical behavior of the model through a finite-size scaling analysis, allowing us to determine its critical exponents and scaling functions.
This suggests that the critical exponents of the WASEP model under the packing field correspond to those of mean-field models.
Lastly, we have demonstrated the potential for developing more complex phases by combining several packing fields of different orders and introducing time-dependent couplings.

Altogether, these results highlight the potential and versatility of the packing-field mechanism in creating and studying new models that exhibit time-crystalline behavior.
In this direction, the exploration of combining several packing fields and including time-dependent couplings hinted at in the last section requires further investigation.
Other open venues include the detailed characterization of the discontinuous phase transition in models such as the KLS, or the determination of the critical exponents in other models beyond the WASEP.

    }

    \chapter{Conclusions}
    {
%
%
%
%
%
%

In this thesis, we have conducted a comprehensive analysis of the large fluctuations or rare events associated with time-integrated dynamical observables in stochastic processes. 
Specifically, we have focused on fluctuations giving rise to dynamical phase transitions (DPTs) that involve spontaneous symmetry breaking.
These transitions 
often involve dramatic changes in the dynamics of system trajectories\todo{buscar frase bonita}. 
Through the application of large deviation theory, we have 
characterized the general microscopic mechanism that generates these phase transitions, by means of deeper comprehension of their spectral properties\todo{resaltar la relevancia de esto}.
In addition, insights gained from studying the specific DPT in the closed WASEP have allowed us to introduce a new mechanism for designing transport models that exhibit phase transitions into time crystals in their typical dynamics.
In the following, we detail the results that have been obtained in this thesis.


First, we provided a review of the main concepts underpinning this thesis.
We began with a comprehensive analysis of Markov processes, which model the stochastic systems that we have considered.
Then we explored large deviation theory, which quantifies the occurrence of rare events in the system dynamics, and its applications to the statistics of trajectories.
In addition, we delved into the biased ensemble and the Doob transform, the main tools we have used in the analysis of rare events.
\todo{cerrar un poco mejor}

In Chapter~\ref{chap:SBDPTs}, we set our focus on symmetry-breaking DPTs.
We analyzed the rigid structure imposed by the presence of a $\Zsymm{n}$ symmetry in the generators of systems undergoing such DPTs.
Specifically, we explored how the properties of a symmetry-breaking phase transition constrain the general structure of the eigenvectors of the Doob-transformed generator, which describe the different broken-symmetry phases.
We found that, for the eigenvectors contributing to the steady state, their elements---associated with the different configurations of the system---are approximately equal in magnitude.
On the other hand, the corresponding complex phases of such elements are fully determined by the associated eigenvalues under the symmetry operator and the particular phase to which they belong.
Such insights led us to unveil the phase selection mechanism governing the dynamics of the system and the particular role played by the eigenvectors.
Additionally, we investigated a reduction of the vector space to one defined by the possible values of the order parameter.
This greatly simplified the analysis to understand the underlying symmetry-breaking mechanism.
\todo{no me encanta esta frase}

Building on the concepts established earlier, Chapter~\ref{chap:DPTs_models} delves into applying these principles to four distinct one-dimensional lattice models.
This exploration began with the boundary-driven WASEP, a model displaying a particle-hole symmetry-breaking DPT for current fluctuations.
We then studied the Potts model with three and four spin orientations, observing discrete rotational symmetry-breaking DPTs in energy fluctuations.
The closed WASEP presented a particularly compelling case, exhibiting a continuous symmetry-breaking DPT in the limit of infinite lattice size ($\lattsize\to\infty$) combined with the breaking of the time-translation symmetry, giving rise to a time-crystal phase with a rotating density wave.
Our spectral analysis of symmetry-breaking DPTs found concrete validation in these models, offering novel insights into spontaneous symmetry breaking at the fluctuating level.
The open WASEP demonstrated the breaking of the $\Zsymm{2}$ particle-hole symmetry. 
When the system was conditioned to sustain a current way below its typical value, the lattice either filled or got emptied, decreasing the mobility and thus increasing the probability of the fluctuation.
Regarding its spectrum, the first eigenvector $\eigvectDR{0}$ illustrated the transition from an unimodal to a bimodal stationary probability distribution of lattice occupation.
As expected, the second eigenvector $\eigvectDR1$ echoed the structure of $\eigvectDR0$ but with a different sign in each phase, thus acting as the selector of the phase.
The Potts model exemplified the extension of this mechanism to symmetries beyond $\Zsymm{2}$.
Specifically, we analyzed the three- and four-state Potts model, which exhibited $\Zsymm{3}$ and $\Zsymm{4}$ symmetries under the rotation of their spins.
The observed DPTs were from a paramagnetic in its typical behavior to a ferromagnetic phase when sustaining low energy fluctuations.
In the three-state model, $\eigvectDR{0}$ incorporated the probability distribution for the three symmetry-broken phases, with $\eigvectDR{1}$and $\eigvectDR{2}$ selectively favoring one phase through their complex phase interplay.
The four-state model's mechanism was more intricate, involving four eigenvectors for phase selection and distribution.
In summary, the examination of these models underscored the efficacy of our theoretical approach as a powerful analytical tool for understanding DPTs.

Chapter~\ref{chap:buildingtc} focuses in the analysis of the DPT found in the closed WASEP and its use to devise new models of time crystals.
When conditioned to sustain low particle currents well below its typical value, a DPT appears in the closed WASEP in which a periodic traveling particle condensate emerges. 
This results in the breaking of both time and space-translation symmetries, marking the onset of a traveling-wave phase.
After applying the tools of Chapter~\ref{chap:SBDPTs}, we obtained that in this model each phase corresponded to a condensate position, with the system moving in periodic motion between them, a phenomenon linked to the appearance of a band structure in the imaginary part of the eigenvalues associated with the stationary eigenvectors.
In this model, the phase selection mechanism involved a macroscopic number of eigenvectors, which collaborate to localize the position of the condensate. 
In addition to studying the spectral properties of the time-crystal phase, we focused on dissecting the intricate dynamics induced by the Doob-transformed generator and the mechanism giving rise to the time-crystal phase.
Despite the complex structure of the generator, key elements crucial to the phase transition were identified.
Remarkably, it was discovered that the dynamics under the Doob transform could be approximately understood as the original dynamics supplemented with an auxiliary packing field.
This field functioned by accelerating particles lagging behind the condensate's center of mass while restraining those ahead, thereby amplifying naturally occurring density fluctuations and leading to the formation of a periodic traveling wave or time-crystal phase.
This understanding facilitated the proposal of a novel stochastic lattice gas model incorporating a similar packing field.
In contrast to the closed WASEP, this new model presented a phase transition to a time-crystal phase, not in the fluctuations on the current, but in its typical behavior when changing its coupling to the packing field.
An extensive numerical analysis of this new model was undertaken to characterize the nature of the phase transition, which confirmed the theoretical predictions.


Finally, in Chapter~\ref{chap:genpackfield}, we left fluctuations to focus on the packing-field mechanism to create time crystals introduced in the previous chapter.
Here, we presented an extended version of the packing field mechanism, designed in this case to generate an arbitrary number $\Eorder$ of traveling-wave condensates\todo{resaltar que se puede aplicar a muchos modelos}.
This advancement was achieved by generalizing the packing field from the previous chapter to amplify fluctuations in the $\Eorder$-th Fourier mode of the density profile.
Regarding this new packing field, we established several general results.
In particular, a local stability analysis of the hydrodynamic equations of diffusive models under the packing field was performed, finding the critical coupling to the field in terms of the transport coefficients of the considered model\todo{hay que decir antes que esto vale para cualquier modelo}.
Moreover, we established a direct relationship between the steady-state traveling-wave solutions of the hydrodynamic equation appearing for different values of the number of condensates $\Eorder$\todo{un poco largo}.
Then we explored the application of the field using two different approaches.
First, using a hydrodynamic description, the mechanism was explored in several models of transport, displaying different behaviors in the time-crystal phase depending on their transport coefficients.
Finally, we performed a detailed microscopic analysis of the WASEP.
Extensive Monte Carlo simulations validated our theoretical predictions regarding the instability threshold and the equivalence between the traveling-wave profiles for different values of $\Eorder$\todo{esto no queda muy claro}.
In addition, the simulations provided an in-depth view of the critical behavior, enabling us to evaluate critical exponents and scaling functions\todo{no se ha comentado el tema de la asimetría}.

    }

    \begin{appendices}
        \chapter{Transport coefficients in Katz-Lebowitz-Spohn model}
        \chaptermark{Transport coefficients in the KLS model}
        \label{app:KLScoeffs}
        {
            \todo[inline]{Habría que poner alguna figura aquí ilutrando el comportamiento del modelo}

Below we provide the expressions for the calculation of the transport coefficients of the Katz-Lebowitz-Spohn (KLS) model as given in \cite{baek15a}.
In contrast to the other microscopic transport models presented in this thesis, the richer dynamics of the KLS model give rise to a more convoluted expression for their coefficients.

Specifically, the diffusion coefficient is obtained in terms of the quotient
\begin{align}
    D(\rho) = \frac{\mathcal{J}(\rho)}{\chi(\rho)} \,,
\end{align}
where $\mathcal{J}(\rho)$ is the average current in the totally asymmetric version of the model and $\chi(\rho)$ is its compressibility.
The first is given by
\begin{equation}
    \mathcal{J}(\rho)
    =
    \frac{\nu [1+\delta(1-2\rho)]-\eta\sqrt{4\rho(1-\rho)}}{\nu^3} \,,
\end{equation}
while the second obeys,
\begin{equation}
    \chi(\rho) = \rho(1-\rho)\sqrt{(2\rho-1)^2 + 4\rho(1-\rho)e^{-4\beta}}
    .
\end{equation}
In turn, $\nu$ and $\beta$ are determined from the expressions
\begin{equation}
    \nu \equiv \frac{1+ \sqrt{(2\rho-1)^2 + 4\rho(1-\rho)e^{-4\beta}}}{\sqrt{4\rho(1-\rho)}} 
    ,
\end{equation}
and
\begin{equation}
    \quad e^{4\beta} \equiv \frac{1+\eta}{1-\eta}
    \,.
\end{equation}

Finally the mobility coefficient $\sigma(\rho)$ is obtained from the diffusion coefficient and the compressibility using the Einstein relation
\begin{align}
    \sigma(\rho) = 2 D(\rho) \, \chi(\rho)
    \, .
\end{align}

The mobility $\mobility(\dens)$ and the diffusion coefficient $\diffcoef(\dens)$ for the parameters $\nu=0.9$ and $\delta=0$ (the ones used in Section~\ref{sec:models_hydro_genpackfield}) are displayed in Fig.~\ref{fig:mobdiffcoeff_KLScoeffs}
\begin{figure}
    \centering
    \includegraphics[width=0.9\linewidth]{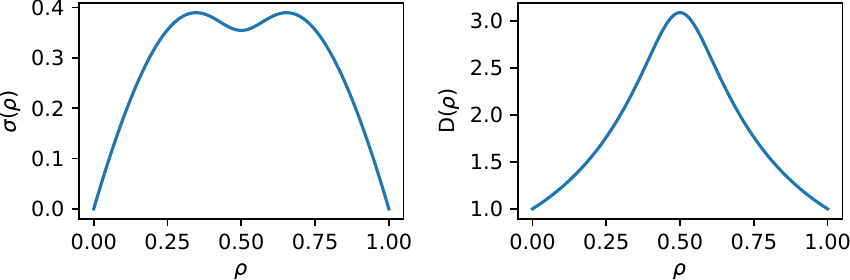}
    \caption{
        Mobility $\mobility(\dens)$ and diffusion coefficient $\diffcoef(\dens)$ of the KLS model for the parameters $\nu=0.9$ and $\delta=0$.
    }
    \label{fig:mobdiffcoeff_KLScoeffs}
\end{figure}

        }
        
        \chapter{Numerical approach to the traveling-wave hydrodynamic equation with a packing field}
        \chaptermark{Solving the traveling-wave hydrodynamic equation}
        \label{app:selfconsisthydro}
        {
            \todo[inline]{Añadir referencias a dónde aparecen estas ecuaciones}

\newcommand{\denstwt}{\tilde{\denstw{}}}
\newcommand{\twargt}{\tilde{\twarg}}
\newcommand{\inicond}{y_0}
\newcommand{\odeconst}{C}

As hinted in Chapters~\ref{chap:buildingtc} and \ref{chap:genpackfield}, the numerical treatment of the equation governing the behavior of diffusive systems under the action of a packing field is quite challenging.
In this appendix, we will detail the method used in this Thesis to overcome this problem.

Indeed, the hydrodynamical equation
\begin{align}
\label{eq:hydroeq_selfconsisthydro}
    \partial_t \dens = - \partial_x \Big[-\diffcoef(\rho) \partial_x \dens + \mobility(\dens)
    \Big(\Ed +  \Epampl \abs{\KDoparm[\dens]} \sin(\KDoparargm[\dens] - 2\pi\pos \Eorder)\Big)
    \Big],
\end{align}
with $\pos \in [0, 1]$, is a nonlinear second-order differential equation with periodic boundaries, in which the Kuramoto-Daido parameter given by 
\begin{align}
\label{eq:opar_selfconsisthydro}
    \KDoparm[\dens]
    &=
    \frac{1}{\meandens} \int_0^1 \dd x \rho(x,t) \expsup{\textrm{i}2\pi \Eorder x}
    \equiv
    \abs{\KDoparm[\dens]} \expsup{\textrm{i} \KDoparargm[\dens]}
    \, ,
\end{align}
introduces a dependence on the integral of the solution.
While this equation can be solved using standard techniques for partial differential equations---such as finite difference methods---, reaching the traveling-wave regime with enough precision becomes increasingly difficult when the system is close to the critical point or deep into the nonlinear regime. 
\todo{is the meaning of this nonlinear regime (for large lambda/lambda critical) clear?}
An alternative approach is to consider the hydrodynamic equation with a traveling-wave ansatz, $\dens(x,t) = \denstw{\Eorder}(\omega t - 2\pi x)$, to obtain an ordinary second-order differential equation depending only on the variable $\twarg = \omega t - 2\pi x$,
\begin{eqnarray}
    \label{eq:hydrotw_selfconsisthydro}
    \omega \denstw{\Eorder}^{\prime}(\twarg)
    &=&
    2\pi \dv{\twarg} \Big\{ \diffcoef(\denstw{\Eorder}) 2\pi \denstw{\Eorder}^{\prime}(\twarg) 
    \\
    &+& \mobility(\denstw{\Eorder}) \big[\Ed + \Epampl \abs{\KDoparm} \sin(\KDoparargmini + \Eorder\twarg)\big]\Big\} \nonumber
    \, .
\end{eqnarray}
Here we have used that, under this ansatz, the magnitude of the order parameter is constant and its complex phase increases linearly in time, i.e., $\KDoparm[\denstw{\Eorder}] = \abs{\KDoparm} \expsup{i(\KDoparargmini + \Eorder\omega t)}$ with $\KDoparargmini$ the complex argument at $t=0$.

This simplifies the equation, but it makes it harder to deal with the order parameter.
In the original partial differential equation, given the density profile at a particular time step, we just needed to evaluate $\KDoparm(\dens)$ in order to obtain the next one.
However, in the traveling-wave version of Eq.~\eqref{eq:hydrotw_selfconsisthydro}, in order to compute the differential equation we need the integral Eq.~\eqref{eq:opar_selfconsisthydro} of its solution, which renders usual differential equations methods invalid.

To overcome this issue, we will first integrate Eq.~\eqref{eq:hydrotw_selfconsisthydro} to obtain a first-order differential equation easier to tackle numerically,
\begin{eqnarray}
\label{eq:hydrotwfirst_selfconsisthydro}
    \omega \denstw{\Eorder}(\twarg)
    &=&
    \wnumber \Big\{ D(\denstw{\Eorder}) \wnumber \denstw{\Eorder}^{\prime}(\twarg)
    \\
    &+& \mobility(\denstw{\Eorder}) \big[\Ed + \Epampl \abs{\KDoparm} \sin(\Eorder\twarg)\big]\Big\} \nonumber
        + \odeconst\, ,
\end{eqnarray}
where $C$ is the integration constant and where we have chosen $\KDoparargmini = 0$ without loss of generality (we have set the origin of $\twarg$ to the angular position given by $\KDoparargmini$, i.e., $\twarg \mapsto \twarg - \KDoparargmini$).

Now, the key is to consider $|\KDoparm|$ as a parameter instead of an integral of the solution, thus transforming the previous equation into a standard ordinary differential equation depending on three parameters: $|\KDoparm|$, $\omega$, and $\odeconst$.
Such a differential equation can now be solved using the initial condition at the left boundary $\denstw{\Eorder}(0) = \inicond$ to obtain the solution $\denstw{\Eorder}(\twarg; \inicond, |\KDoparm|, \omega, \odeconst)$.
Nevertheless, this function will not be, in general, a solution to the original problem.
The parameters $\inicond$, $|\KDoparm|$, $\omega$, and $\odeconst$, must be carefully chosen to ensure the compatibility of the solution with the specifications of the original problem: the periodicity of the solution, its average density $\meandens$ and the consistency between the chosen $|\KDoparm|$ and the value calculated from $\denstw{\Eorder}(\twarg; \inicond, |\KDoparm|, \omega, \odeconst)$.
These conditions are captured in the following system of equations:
\begin{align}
    \denstw{\Eorder}(0; \inicond, |\KDoparm|, \omega, \odeconst)
    =
    \denstw{\Eorder}(2\pi; \inicond, |\KDoparm|, \omega, \odeconst)
    \,, \quad &\textrm{(periodicity)}
    \\
    \meandens
    =
    \int_0^1 \dd x \denstw{\Eorder}(-2\pi x; \inicond, |\KDoparm|, \omega, \odeconst)
    \,, \quad &\textrm{(total mass)}
    \\
    \abs{\KDoparm}
    =
    \int_0^{1} \dd x \denstw{\Eorder}(-2\pi x; \inicond, |\KDoparm|, \omega, \odeconst) \expsup{\textrm{i} 2\pi x \Eorder}
    \,, \quad
        &\hspace{-1,5mm}\begin{array}{c}
            \textrm{(order parameter}
            \\
            \textrm{at $t=0$)}
        \end{array}
\end{align}
which completes the set of tools required to solve the problem (we have two real equations and a complex one to determine four real parameters).
To determine the solution to the original problem, we define a function $G(\inicond, |\KDoparm|, \omega, \odeconst)$ which calculates the profile $\denstw{\Eorder}(0; \inicond, |\KDoparm|, \omega, \odeconst)$ using Eq.~\eqref{eq:hydrotwfirst_selfconsisthydro} and returns the squared sum of the errors in the previous self-consistent equations for this profile.
With this, the correct parameters $\inicond, |\KDoparm|, \omega, \odeconst$ can be found by performing a numerical optimization of $G(\inicond, |\KDoparm|, \omega, \odeconst)$, and the traveling wave profile $\denstw{\Eorder}$ corresponding to such parameters will be the solution of the original problem.
This approach is reminiscent of the shooting method \cite{press2007} used to solve two-point boundary value problems.

\todo[inline]{
    In practice, the convergence of the numerical solution was faster when the initial condition for the density field was obtained from a numerical simulation of the full hydrodynamic equation.
}
        }
    \end{appendices}

\backmatter

\printbibliography[heading=bibintoc]

\end{document}